\documentclass[11pt,notitlepage,aps,prd,preprintnumbers,superscriptaddress,nofootinbib,floatfix]{revtex4-2}

\usepackage{iftex}
\ifPDFTeX
  \pdfoutput=1
\fi
\usepackage{graphicx}
\usepackage{dcolumn}
\usepackage{bm}
\usepackage{extarrows}
\usepackage{multirow,array}
\usepackage{amsmath,amssymb,physics}
\usepackage{subfigure}
\usepackage{enumitem}
\setlist{nolistsep}
\usepackage{url}
\usepackage{slashed}
\usepackage{booktabs}

\usepackage{orcidlink}

\usepackage{xcolor}

\definecolor{nicered}{rgb}{0.5,0.,0.}
\definecolor{nicegreen}{rgb}{0.,0.5,0.}
\definecolor{niceblue}{rgb}{0.,0.,0.5}

\usepackage{braket}
\usepackage{colortbl}
\usepackage{gensymb}
\usepackage{comment}
\usepackage{setspace}
\usepackage[normalem]{ulem}

\usepackage{hyperref}
\usepackage[capitalise]{cleveref}
\hypersetup{colorlinks,citecolor=nicegreen,linkcolor=nicered,urlcolor=niceblue}

\def\q{\textbf{q}}

\def\0{\textbf{0}}

\def\v{\textbf{v}}

\begin{document}
\title{Scalar inelastic scattering of pseudo-Majorana dark matter\\ through a radion portal}
\author{Ruifeng Leng}
\email{lengrf@mail.sysu.edu.cn}
\affiliation{School of Physics, Sun Yat-Sen University, Guangzhou 510275, China}
\author{Kirtimaan A. Mohan}
\email{kamohan@msu.edu}
\affiliation{Department of Physics and Astronomy, Michigan State University, East Lansing, Michigan 48824, USA}
\author{Keping Xie}
\email{kepingxie@sjtu.edu.cn}
\affiliation{State Key Laboratory of Dark Matter Physics \& Key Laboratory for Particle Astrophysics and Cosmology (MOE) \& Shanghai Key Laboratory for Particle Physics and Cosmology, Tsung-Dao Lee Institute \& School of Physics and Astronomy, Shanghai Jiao Tong University, Shanghai 201210, China\looseness=-1}
\author{Hong-Hao Zhang}
\email{zhh98@mail.sysu.edu.cn}
\affiliation{School of Physics, Sun Yat-Sen University, Guangzhou 510275, China}

\singlespacing	
\begin{abstract}

We investigate a radion portal for pseudo-Majorana dark matter as an interpretation of the high-recoil event LZ230616.
In contrast to the familiar pseudo-Dirac regime, the pseudo-Majorana structure allows a dominant scalar
transition while suppressing elastic scattering.
Endothermic kinematics remove low-energy recoils, while the nuclear response shapes the surviving high-recoil spectrum.
Combining the accepted LZ event count with the Born-level thermal relic abundance yields a reference solution with
$m_{\chi}=1\,\mathrm{TeV}$, $\delta=340\,\mathrm{keV}$, and $m_{r}=30\,\mathrm{GeV}$.
This solution lies within the published LZ spin-independent $\mathcal{O}_{1}^{s}$ interval and remains compatible
with existing OPAL and CMS constraints, while a simplified HL-LHC diphoton projection could probe it.
Solar neutrino searches offer a complementary test.
The benchmark's energetic-daughter proxy rates are comparable in magnitude to the rate scales reconstructed from IceCube's direct $b\bar{b}$ and $\tau^{+}\tau^{-}$ templates. An illustrative IceCube-Gen2 rescaling suggests future sensitivity, while a model-specific assessment requires the radion-cascade neutrino spectra and detector response.
\end{abstract}

\maketitle
\tableofcontents{}

\section{Introduction}
\label{sec:intro}

The LUX-ZEPLIN (LZ) Collaboration recently reported an unusual event, LZ230616, in its extended nuclear-recoil (NR) analysis based on a 2.84 tonne-year exposure~\cite{LZ:2026axp}. Under the NR hypothesis, its reconstructed energy and corresponding momentum transfer are
\begin{equation}
 E_{R}^{\rm obs}=248\pm23_{\rm stat}\pm23_{\rm sys}\,\mathrm{keV},
 \qquad
 Q_{\rm obs}\simeq\sqrt{2m_{\rm Xe}E_{R}^{\rm obs}}\simeq246\,\mathrm{MeV}.
 \label{eq:event}
\end{equation}
Here $E_{R}$ denotes the kinetic energy of the recoiling xenon nucleus. The event lies in a region where the expected LZ background is small and at a recoil energy above the range where conventional spin-independent dark matter (DM) spectra are concentrated. Among the tested signal models, the maximum local significance is $3.4\sigma$, reduced to $2.6\sigma$ globally after the look-elsewhere correction~\cite{LZ:2026axp}.

Although LZ investigated several rare background processes and detector effects, none was identified as a likely explanation for
the event~\cite{LZ:2026axp}.
This motivates asking whether a DM interaction can produce a recoil near $250\,\mathrm{keV}$ without generating the much larger population of lower-energy recoils typically expected from conventional elastic scattering. Increasing the DM mass can make such a high recoil kinematically accessible, but does not by itself remove the strong preference of an ordinary coherent elastic spectrum for lower recoil energies.
Endothermic scattering, $\chi_{1} A\to\chi_{2} A$, provides a natural kinematic filter. Producing the heavier state requires an excitation energy $\delta=m_{\chi_{2}}-m_{\chi_{1}}>0$~\cite{Tucker-Smith:2001myb,Tucker-Smith:2004jv,Barello:2014uda,Bramante:2016rdh}. Energy and momentum conservation imply that, for a nucleus of mass $m_{A}$, the minimum incident speed grows as $v_{\min,A}\sim\delta/\sqrt{2m_{A}E_{R}}$ when $E_{R}\to0$. Since the Galactic halo has a finite maximum speed, sufficiently low-energy recoils become kinematically inaccessible. A large mass splitting can therefore remove the low-recoil population while restricting the surviving scattering events to the fastest halo particles.

\begin{figure}[t]
\centering
\includegraphics[width=0.45\textwidth]{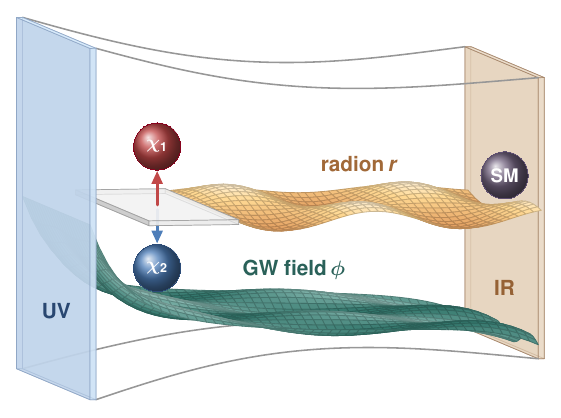}
\caption{Extra-dimensional realization of the pseudo-Majorana radion portal.
Two bulk Dirac fermions with opposite $\mathcal{S}^{1}/\mathbb{Z}_{2}$ parities furnish complementary light Weyl modes that mix into $\chi_{1,2}$, while the SM resides on the IR brane.
Metric and GW scalar contributions to the radion couplings can compete, permitting opposite-sign responses to positive Majorana masses and a dominant scalar inelastic transition.}
\label{fig:rad_portal}
\end{figure}

We study this process through a radion portal to a nearly degenerate Majorana pair. 
The radion is the scalar fluctuation of the interbrane separation in a warped extra dimension stabilized by the Goldberger-Wise (GW) field~\cite{Randall:1999ee,Goldberger:1999uk,Goldberger:1999wh,Csaki:2000zn}. 
The Standard Model (SM), localized on the infrared (IR) brane, couples through its energy-momentum trace, while bulk-fermion radion couplings also depend on their extra-dimensional profiles~\cite{Azatov:2008vm}. In a stabilized $\mathcal{S}^{1}/\mathbb{Z}_{2}$ orbifold, two bulk Dirac fermions with opposite parity assignments furnish complementary light modes. 
We formulate their mass-proportional interactions in a four-dimensional (4D) effective theory and exhibit an illustrative 5D realization in Appendix~\ref{app:orbifold-fixed-metric}.
The dark-sector mass spectrum is dominated by nearly degenerate Majorana masses, while the small Dirac mixing and their mass difference determine the physical splitting. 
We refer to the Majorana-dominated hierarchy as the \emph{pseudo-Majorana regime}, using the term only to distinguish it from the familiar pseudo-Dirac hierarchy~\cite{Tucker-Smith:2001myb}.
Fig.~\ref{fig:rad_portal} illustrates the stabilized-orbifold realization.

The mass hierarchy alone does not ensure that scattering is predominantly inelastic.
We identify a region in which the radion coupling coefficients for the two Majorana mass terms are approximately opposite.
At equal Majorana masses, these couplings generate a scalar transition in the physical basis, whereas the small Dirac mixing gives diagonal scalar couplings proportional to the splitting.
This realizes the required interaction within a real, mass-proportional two-field effective field theory (EFT), without introducing an independent scalar-transition operator or complex couplings.
Matching the visible energy-momentum tensor onto nucleons gives a coherent scalar response, modulated at finite momentum transfer by isotope-dependent Helm form factors~\cite{Lewin:1995rx}.
We calculate the recoil spectrum in true NR energy using the published LZ NR signal efficiency.
The endothermic threshold and finite halo endpoint determine the allowed recoil range, while the nuclear response shapes the surviving spectrum through diffraction minima.
We use kinematic diffraction to describe this interplay of inelastic kinematics, the halo endpoint, and the nuclear response.
The absolute event count is the direct detection observable used in our parameter analysis.
Following the benchmark construction logic of recent particle models of the event~\cite{Unwin:2026rdp}, we combine the one-event LZ count with the central thermal relic abundance evaluated at Born level.
We retain the original radion parameters rather than treating the two vertices as a single coupling.
Both Majorana states contribute to freeze-out, while the direct detection calculation assumes that the present halo is dominated by the lighter state.
The joint conditions determine the radion scale and transition strength for chosen DM and radion masses.
We further examine the LZ spin-independent $\mathcal{O}_{1}^{s}$ interval, OPAL and CMS collider constraints, and the projected HL-LHC diphoton sensitivity.
Solar capture provides an additional probe through exothermic conversion, possible orbital re-ejection, and Sommerfeld-enhanced annihilation, followed by channel-dependent comparisons with the ten-year IceCube results and illustrative IceCube-Gen2 projections.

This paper is organized as follows.
Section~\ref{sec:framework} defines the radion portal and dominant scalar transition.
Section~\ref{sec:rate} presents the recoil calculation, LZ event count and likelihood support, and Born-level thermal relic abundance.
Section~\ref{sec:results} presents the joint parameter analysis, collider constraints, and solar capture and neutrino phenomenology.
We conclude in Section~\ref{sec:conclusions}.
The appendices provide supporting derivations, an explicit 5D realization, collider constraints, the HL-LHC sensitivity projection, and details of the solar analysis.

\section{Radion and pseudo-Majorana dark matter}
\label{sec:framework}

In this section, we formulate the radion portal for pseudo-Majorana DM used in the recoil analysis. 
We first introduce the effective radion coupling,
then construct the nearly degenerate Majorana spectrum and derive the scalar transition that dominates the DM scattering process.

\subsection{Visible sector and SM-radion coupling}
\label{sec:visible-radion}

A stabilized warped dimension provides a geometric motivation for a massive radion $r$~\cite{Randall:1999ee,Goldberger:1999uk,
Goldberger:1999wh,Csaki:2000zn}. The interbrane separation is a dynamical modulus, and its stabilization leaves a scalar fluctuation identified with the radion. In the 4D effective theory, we characterize the radion by its mass $m_{r}$ and interaction scale $\Lambda_{r}$.

For the visible sector, we adopt the radion coupling to the SM energy-momentum trace,
\begin{equation}
 \mathcal{L}_{r,\rm SM}
 =-\frac{r}{\Lambda_{r}}T^{\mu}{}_{\mu,\rm SM}.
 \label{eq:rsm}
\end{equation}
This normalization corresponds to the conventional trace coupling for a visible sector localized on the IR brane in a warped realization. 
The visible trace contains both the mass terms and the quantum trace anomaly,
\begin{equation}
\begin{aligned}
T^{\mu}{}_{\mu,\rm SM}
={}&
\sum_{f}m_{f}\bar f f
-2m_{W}^{2}W_{\mu}^{+}W^{-\mu}
-m_{Z}^{2}Z_{\mu}Z^{\mu}
\\
&-\partial_{\mu}h\,\partial^{\mu}h
+2m_{h}^{2}h^{2}
+\frac{\beta_{3}(g_{s})}{2g_{s}}
G_{\mu\nu}^{a}G^{a\mu\nu}
+\frac{\beta_{e}(e)}{2e}
F_{\mu\nu}F^{\mu\nu}.
\end{aligned}
\label{eq:smtrace}
\end{equation}
We neglect Higgs-radion mixing and treat $r$ as the physical radion field throughout the effective description.
The gluon term is important because QCD generates most of the nucleon mass. Equation~\eqref{eq:rsm} therefore implies a direct radion coupling to nucleons, even though Eq.~\eqref{eq:smtrace} is expressed in terms of quark and gluon fields~\cite{Shifman:1978zn,Giudice:2000av,Rizzo:2002pq}.

\subsection{Pseudo-Majorana spectrum and dark-radion couplings}
\label{sec:mass-sector}

The dark sector contains two canonically normalized left-handed Weyl fields $\eta$ and $\xi$, both singlets under the SM. 
In the realization of Appendix~\ref{app:orbifold-fixed-metric}, they arise from complementary light modes of two bulk Dirac fermions with opposite orbifold parities~\cite{Grossman:1999ra,Gherghetta:2000qt}.

Both fields are odd under an exact dark $\mathbb Z_{2}$, while the radion and SM fields are even. 
This symmetry forbids operators containing a single dark fermion and stabilizes the lightest $\mathbb Z_{2}$-odd mass
eigenstate. The mass terms are
\begin{equation}
 \mathcal{L}_{\chi}\supset-M_{D}\eta\xi
 -\frac{\mu_{\eta}}{2}\eta\eta
 -\frac{\mu_{\xi}}{2}\xi\xi+\mathrm{h.c.}
 \label{eq:dark-mass-lagrangian}
\end{equation}
We take a positive Dirac mass $M_{D}>0$ and positive, nearly degenerate Majorana masses,
\begin{equation}
 \mu_{\chi}\equiv\frac{\mu_{\eta}+\mu_{\xi}}{2},
 \qquad
 \mu_{\eta}\simeq\mu_{\xi}\simeq\mu_{\chi}\gg M_{D}.
 \label{eq:main-pseudo-majorana-hierarchy}
\end{equation}
We refer to this Majorana-dominated hierarchy as the \emph{pseudo-Majorana regime}. Both physical mass eigenstates are
Majorana fermions.

For $\psi=(\eta,\xi)^{\text{T}}$, the symmetric mass matrix is
\begin{equation}
 M=\begin{bmatrix}
 \mu_{\eta}&M_{D}\\
 M_{D}&\mu_{\xi}
 \end{bmatrix}.
 \label{eq:main-mass-matrix}
\end{equation}
We define
\begin{equation}
 \Sigma=\mu_{\eta}+\mu_{\xi},\qquad
 \Delta=\mu_{\eta}-\mu_{\xi},\qquad
 D=\sqrt{4M_{D}^{2}+\Delta^{2}}.
\end{equation}
Here $\Sigma$ determines the overall Majorana mass scale, while $\Delta$ measures the difference between the two Majorana masses.
The quantity $D$ combines this difference with the Dirac mass $M_{D}$ and determines the physical level splitting. 
Thus the same small splitting can arise from different combinations of $\Delta$ and $M_{D}$.
In the pseudo-Majorana regime, both eigenvalues of $M$ are positive.
The two physical masses are
\begin{equation}
 m_{\chi_{1}}=\frac{\Sigma-D}{2},
 \qquad
 m_{\chi_{2}}=\frac{\Sigma+D}{2}.
 \label{eq:masses}
\end{equation}
For the nearly degenerate pair, it is useful to characterize the
overall DM mass scale by the average mass
\begin{equation}
 m_{\chi}\equiv\frac{m_{\chi_{1}}+m_{\chi_{2}}}{2}
 =\frac{\Sigma}{2}
 =\mu_{\chi},
 \label{eq:average-mass}
\end{equation}
which controls the overall mass scale of the dark states.
The second physical scale is the excitation energy
\begin{equation}
\delta \equiv m_{\chi_{2}}-m_{\chi_{1}}
=D =\sqrt{4M_{D}^{2}+\Delta^{2}}.
\label{eq:mass-splitting}
\end{equation}
Unlike $m_{\chi}$, the splitting $\delta$ is controlled by the Majorana-mass difference $\Delta$ and the Dirac mass $M_{D}$.
It is this much smaller scale that governs the endothermic transition $\chi_{1}A\to\chi_{2}A$.
Both $|\Delta|$ and $M_{D}$ must remain small compared with $\mu_{\chi}$,
while their relative size determines how a fixed splitting is realized in the original $(\eta,\xi)$ basis. 
A general parametrization of this freedom and the corresponding alignment conditions are given in
Appendix~\ref{sec:takagi-majorana}.

The radion coupling to the dark-sector mass terms is parametrized as
\begin{equation}
 \mathcal{L}_{r\chi}=-\frac{r}{\Lambda_{r}}
 \left(c_{D}M_{D}\eta\xi+\frac{c_{\eta}\mu_{\eta}}{2}\eta\eta
 +\frac{c_{\xi}\mu_{\xi}}{2}\xi\xi\right)+\mathrm{h.c.}
 \label{eq:realresponse}
\end{equation}
The real dimensionless coefficients $c_{D},c_{\eta},c_{\xi}$ specify the radion couplings associated with the corresponding dark mass terms.
We use the common radion scale $\Lambda_{r}$ for the visible and dark sectors, while allowing these coefficients to be nonuniversal among the dark mass operators and relative to the visible trace coupling. 
Each radion vertex is proportional to the corresponding mass term, and the $\mathbb Z_{2}$ symmetry remains exact.

We note that the two Majorana states $\chi_{1}$ and $\chi_{2}$ are not identified
with adjacent Kaluza-Klein (KK) modes. The sub-MeV splittings considered below satisfy $\delta/m_{\chi}\ll1$ and arise from the 4D mass matrix in Eq.~\eqref{eq:main-mass-matrix}, rather than from a KK level spacing.
The 5D theory also contains heavier fermion KK levels and gravitational and stabilization-sector excitations. Their effects beyond the radion and light dark pair are not included in the present phenomenological calculation~\cite{Chivukula:2021xod,Chivukula:2022tla}.

\subsection{Pseudo-Majorana scalar transition and elastic suppression}
\label{sec:scalar-transition}

Let $\zeta_{i}$ $(i=1,2)$ denote the left-handed two-component mass eigenstates obtained by diagonalizing the mass matrix in
Eq.~\eqref{eq:main-mass-matrix}. The corresponding four-component Majorana fields are defined in the chiral basis by
\begin{equation}
 \chi_{i}\equiv
 \begin{bmatrix}
  \zeta_{i\alpha}\\
  \zeta_{i}^{\dagger\dot\alpha}
 \end{bmatrix},
 \qquad
 \chi_{i}^{c}=\chi_{i},
 \qquad i=1,2.
 \label{eq:main-majorana-fields}
\end{equation}
In the mass basis, the radion interaction takes the form
\begin{equation}
 \mathcal{L}_{r\chi}
 =-\frac{r}{2}\sum_{i,j=1}^{2}
 \bar\chi_{i}
 \left(g_{ij}^{S}+i g_{ij}^{A}\gamma^{5}\right)
 \chi_{j}.
 \label{eq:generalcoupling}
\end{equation}
Here $g^{S}$ and $g^{A}$ are real dimensionless coupling matrices for the scalar and pseudoscalar bilinears, respectively. 
In the original Weyl basis, the interaction in Eq.~\eqref{eq:realresponse} is encoded by the coupling matrix
\begin{equation}
 G=\frac{1}{\Lambda_{r}}
 \begin{bmatrix}
  c_{\eta}\mu_{\eta}&c_{D}M_{D}\\
  c_{D}M_{D}&c_{\xi}\mu_{\xi}
 \end{bmatrix}.
 \label{eq:main-response-matrix}
\end{equation}

In the pseudo-Majorana regime, both eigenvalues of $M$ are positive, so the Takagi transformation reduces to the real orthogonal rotation $U=O$. The coupling matrix in the mass basis is therefore $\widehat G=O^{\text{T}}GO$, which is real, and
\begin{equation}
 g^{S}=\widehat G,\qquad g^{A}=0.
\end{equation}
The scalar transition coupling is
\begin{equation}
 g_{{12}}^{S}
 =\frac{M_{D}}{\Lambda_{r}D}
 \left[
 (c_{\eta}-c_{D})\mu_{\eta}
 -(c_{\xi}-c_{D})\mu_{\xi}
 \right].
 \label{eq:main-scalar-transition}
\end{equation}
Thus no pseudoscalar coupling is generated in this regime.
Before specializing to a particular configuration, we identify the general conditions for transition dominance. For fixed coupling coefficients and $\delta\ll m_{\chi}$, the leading scalar couplings can be organized as
\begin{align}
g_{{11}}^{S}+g_{{22}}^{S}
&=
\frac{m_{\chi}}{\Lambda_{r}} (c_{\eta}+c_{\xi}) +\mathcal{O} \left(\frac{\delta}{\Lambda_{r}}\right),
\\
g_{{22}}^{S}-g_{{11}}^{S}
&=
\frac{m_{\chi}}{\Lambda_{r}}(c_{\eta}-c_{\xi}) \frac{\mu_{\eta}-\mu_{\xi}}{\delta} 
+\mathcal{O}\left(\frac{\delta}{\Lambda_{r}}\right),
\\
g_{{12}}^{S}
&=
\frac{m_{\chi}}{2\Lambda_{r}}(c_{\eta}-c_{\xi}) \sqrt{1-\frac{(\mu_{\eta}-\mu_{\xi})^{2}}{\delta^{2}}}
+\mathcal{O} \left(\frac{\delta}{\Lambda_{r}}\right).
\end{align}
At leading order, the coefficient $c_{D}$ drops out and contributes only through terms suppressed by $\delta/m_{\chi}$. A dominant scalar transition with suppressed diagonal couplings is therefore obtained when
\begin{equation}
 \left|\frac{\mu_{\eta}-\mu_{\xi}}{\delta}\right|\ll1,
 \qquad
 \left|\frac{c_{\eta}+c_{\xi}}{c_{\eta}-c_{\xi}}\right|\ll1,
 \qquad
 c_{\eta}\ne c_{\xi}.
 \label{eq:main-transition-region}
\end{equation}
The first condition requires the two Majorana masses to be degenerate at the scale of $\delta$, while the second requires their radion coupling coefficients to be approximately opposite. The exact couplings and the quantitative alignment requirement are given in Appendix~\ref{sec:takagi-majorana}.

A simple central configuration satisfying these conditions is
\begin{equation}
 \mu_{\eta}=\mu_{\xi}=\mu_{\chi}=m_{\chi},\qquad
 M_{D}=\frac{\delta}{2},\qquad
 c_{\eta}=-c_{\xi}\equiv c_{M}\ne0.
 \label{eq:main-central-configuration}
\end{equation}
The equal Majorana masses correspond to a $\pi/4$ rotation of the two Weyl fields. In the mass basis, the coupling matrices become
\begin{equation}
 g^{S}=\frac{1}{\Lambda_{r}}
 \begin{bmatrix}
  -c_{D}\delta/2&c_{M}m_{\chi}\\
  c_{M}m_{\chi}&c_{D}\delta/2
 \end{bmatrix},
 \qquad
 g^{A}=0.
 \label{eq:main-central-response}
\end{equation}
The radion interaction is therefore
\begin{equation}
 \mathcal{L}_{r\chi}
 =-r\,\frac{c_{M}m_{\chi}}{\Lambda_{r}}
 \bar\chi_{2}\chi_{1}
 +\frac{r c_{D}\delta}{4\Lambda_{r}}
 \left(
 \bar\chi_{1}\chi_{1}
 -\bar\chi_{2}\chi_{2}
 \right).
 \label{eq:main-central-interaction}
\end{equation}
At this configuration, the opposite radion couplings associated with the two Majorana mass terms generate the scalar transition, while the coupling proportional to the small Dirac mass $M_{D}$ generates the diagonal interactions. Consequently,
\begin{equation}
 \left|\frac{g_{{11}}^{S}}{g_{{12}}^{S}}\right|
 =
 \left|\frac{g_{{22}}^{S}}{g_{{12}}^{S}}\right|
 =
 \left|\frac{c_{D}}{c_{M}}\right|
 \frac{\delta}{2m_{\chi}}.
 \label{eq:main-elastic-hierarchy}
\end{equation}
For $c_{D},c_{M}=\mathcal{O}(1)$, the transition coupling is $\mathcal{O}(m_{\chi}/\Lambda_{r})$, whereas the diagonal couplings are $\mathcal{O}(\delta/\Lambda_{r})$. They are therefore suppressed relative to the transition by $\delta/m_{\chi}$ without requiring $c_{D}\ll1$. Conversely, equal coefficients $c_{\eta}=c_{\xi}$ at equal Majorana masses eliminate the transition.
The conventional radion trace coupling suggests common responses to Majorana masses, so the alignment in Eq.~\eqref{eq:main-central-configuration} requires novel explanation. 
Appendix~\ref{app:orbifold-fixed-metric} realizes it in a stabilized $\mathcal{S}^{1}/\mathbb{Z}_{2}$ orbifold: two bulk Dirac fermions with opposite orbifold parities furnish the light Weyl modes, while their profiles and GW interactions lead to distinct metric and stabilizer contributions to the effective radion couplings~\cite{Azatov:2008vm}.
After dimensional reduction, these contributions can yield opposite signs even for $\mu_{\eta},\mu_{\xi}>0$.
An illustrative 5D benchmark reproduces $c_{\eta}\simeq-c_{\xi}\simeq3.19$ and the reference mass spectrum without introducing another mediator.
Orbifold parity makes the complementary modes available, but its radiative stability and higher-mode effects remain to be established.

For the recoil calculation, it is convenient to define the dimensionless scalar transition coefficient $c_{{12}}^{S}$ through
\begin{equation}
 g_{{12}}^{S}
 \equiv
 c_{{12}}^{S}\frac{m_{\chi}}{\Lambda_{r}}.
 \label{eq:scalartarget}
\end{equation}
Using Eq.~\eqref{eq:main-scalar-transition}, its general expression is
\begin{equation}
 c_{{12}}^{S}
 =
 \frac{M_{D}}{m_{\chi}D}
 \left[
 (c_{\eta}-c_{D})\mu_{\eta}
 -(c_{\xi}-c_{D})\mu_{\xi}
 \right].
 \label{eq:main-c12}
\end{equation}
Thus $c_{12}^{S}$ is determined by the underlying mass and radion coupling parameters rather than representing an additional independent interaction.
The recoil rates below are written in terms of $c_{12}^{S}$. The diagonal scalar channels are parametrically suppressed in the central configuration, where $c_{12}^{S}=c_{M}$.

For comparison, in the pseudo-Dirac regime $|\mu_{\eta,\xi}|\ll M_{D}$, one eigenvalue of the mass matrix is negative before the Takagi rephasing, so a relative Majorana phase is required to obtain positive physical masses. The resulting scalar
transition vanishes, $g_{{12}}^{S}=0$, while the off-diagonal coupling is pseudoscalar and controlled by the small Majorana masses. 
The diagonal scalar couplings remain $\mathcal{O}(M_{D}/\Lambda_{r})$, and the pseudoscalar scattering bilinear is additionally suppressed in the nonrelativistic limit. The complete derivation is given in Appendix~\ref{sec:takagi-pseudodirac}. Thus, within the same real mass-proportional coupling structure, the pseudo-Dirac regime does not
realize the dominant scalar transition considered here.

\subsection{Nucleon matching and xenon cross section}
\label{sec:matching}

The visible radion coupling in Eq.~\eqref{eq:rsm} is expressed in terms of the SM fields. 
For scattering on xenon, the energy-momentum trace must first be matched onto nucleon states. 
For a nucleon $N=p,n$, we define the dimensionless trace form factor $f_{N}(t)$ by
\begin{equation}
 \langle N(p')|T^{\mu}{}_{\mu,\rm SM}|N(p)\rangle
 =m_{N}f_{N}(t)\,\bar u_{N}(p')u_{N}(p),
 \qquad
 t=(p'-p)^{2}.
 \label{eq:nucleontrace}
\end{equation}
The nucleon mass matrix element of the full energy-momentum tensor fixes the forward normalization $f_{N}(0)=1$. 
At finite momentum transfer, the leading radius expansion gives $f_{N}(-Q^{2})\simeq 1-Q^{2}r_{\Theta}^{2}/6$.
Taking the nucleon trace radius $r_{\Theta}\simeq0.55\,\mathrm{fm}$~\cite{Kharzeev:2021qkd} gives
$f_{N}(-Q_{\mathrm{obs}}^{2})\simeq0.92$ at $Q_{\mathrm{obs}}\simeq246\,\mathrm{MeV}$.
We retain $f_{N}(t)\simeq1$ in the reference calculation,
while finite-momentum nucleon structure, nuclear binding, and scalar two-body currents contribute to the
theoretical uncertainty in the recoil rate.

Combining Eqs.~\eqref{eq:rsm}, \eqref{eq:scalartarget}, and \eqref{eq:nucleontrace} gives the radion-mediated transition
$\chi_{1}N\to\chi_{2}N$. For a nucleus with mass number $A$, the nucleon amplitudes add coherently in the long-wavelength limit, while at finite momentum transfer the coherent sum is modulated by the nuclear form factor $F_{A}(Q)$.
We denote the four-momentum transfer by $q^{\mu}$ and its spatial magnitude by $Q\equiv|\mathbf q|$, with $t=q^{2}$. 
For a nucleus initially at rest, the recoil energy $E_{R}$ is the kinetic energy of the outgoing nucleus,
\begin{equation}
 E_{R}\equiv k'^{0}-m_{A}
 =\sqrt{m_{A}^{2}+Q^{2}}-m_{A}.
 \label{eq:recoil-energy}
\end{equation}
The exact kinematic relations are
\begin{equation}
 Q^{2}=2m_{A}E_{R}+E_{R}^{2},
 \qquad
 t=-2m_{A}E_{R}.
 \label{eq:recoil-momentum-relations}
\end{equation}
In the nonrelativistic limit $E_{R}\ll m_{A}$, these reduce to
\begin{equation}
 E_{R}\simeq\frac{Q^{2}}{2m_{A}},
 \qquad
 Q^{2}\simeq-t.
 \label{eq:recoil-NR-relations}
\end{equation}
Within the one-body coherent approximation, the nuclear amplitude is
\begin{equation}
 \mathcal{M}_{A} =
 \frac{c_{{12}}^{S}m_{\chi}}{\Lambda_{r}}
 \frac{1}{Q^{2}+m_{r}^{2}}
 \frac{m_{N}f_{N}(t)}{\Lambda_{r}}
 A F_{A}(Q)\,
 (\bar u_{2}u_{1})(\bar u_{A}u_{A}).
 \label{eq:nuclearamplitude}
\end{equation}
The factor $A F_{A}(Q)$ arises from the coherent sum over nucleons and gives the characteristic $A^{2}F_{A}^{2}(Q)$ dependence of the cross section.
We define the reference zero-momentum scalar nucleon cross section
\begin{equation}
 \sigma_{N}^{0}
 =\frac{\mu_{\chi N}^{2}}{\pi}
 \left[
 \frac{c_{{12}}^{S}m_{\chi}f_{N}(0)m_{N}}
 {\Lambda_{r}^{2}m_{r}^{2}}
 \right]^{2},
 \qquad
 \mu_{\chi N}
 =\frac{m_{\chi}m_{N}}{m_{\chi}+m_{N}}.
 \label{eq:sigma0}
\end{equation}
With $f_{N}(t)\simeq f_{N}(0)=1$, the differential cross section for a xenon isotope $A$ is
\begin{equation}
 \frac{d\sigma_{12,A}}{dE_{R}}
 =
 \frac{m_{A}A^{2}}{2\mu_{\chi N}^{2}v^{2}}\,
 \sigma_{N}^{0}F_{A}^{2}(Q)
 \left(
 \frac{m_{r}^{2}}{m_{r}^{2}+Q^{2}}
 \right)^{2}.
 \label{eq:dsigma}
\end{equation}
For the xenon nuclear form factor in Eq.~\eqref{eq:dsigma}, we use the
isotope-dependent Helm parametrization in the Lewin-Smith convention~\cite{Lewin:1995rx},
\begin{equation}
F_{A}(Q)
=3\frac{j_{1}(QR_{1,A})}{QR_{1,A}}
\exp\left[-\frac{Q^{2}s^{2}}{2}\right],
\qquad
R_{1,A}^{2}
=c_{A}^{2}+\frac{7\pi^{2}a^{2}}{3}-5s^{2}.
\label{eq:helm}
\end{equation}
where
$c_{A}=(1.23A^{1/3}-0.60)\,\mathrm{fm}$, $j_{1}(x)$ is the spherical Bessel function of the first kind,
$a=0.52\,\mathrm{fm}$, $s=0.90\,\mathrm{fm}$.
The form factor is normalized as $F_{A}(0)=1$.
For natural xenon, the adopted Helm factor gives diffraction minima near $E_{R}\simeq95$ and $280\,\mathrm{keV}$, with a second-lobe maximum near $150\,\mathrm{keV}$, where $F_{A}^{2}(Q)\sim3\times10^{-3}$.

We use the average mass $m_{\chi}$ in the reduced masses and flux factors, neglecting relative corrections of
$\mathcal{O}(\delta/m_{\chi})$ while retaining the excitation energy in the inelastic kinematics.
Note that Eq.~\eqref{eq:dsigma} is understood on its kinematically allowed support, $v\ge v_{\min,A}(E_{R})$. The minimum speed and the endothermic condition are presented in Section~\ref{sec:rate}, while
Appendix~\ref{sec:cross-section-derivation} derives Eq.~\eqref{eq:dsigma} and shows how this support follows from the
energy-conservation delta function. Since $Q=0$ is not kinematically accessible for the endothermic process with $\delta>0$,
$\sigma_{N}^{0}$ serves as a reference normalization of the scalar interaction rather than a physical zero-momentum scattering point.

\section{Recoil kinematics and cosmological probes}
\label{sec:rate}

In this section, we formulate the observables used to constrain the radion portal for pseudo-Majorana DM. 
We first derive the endothermic recoil kinematics and Galactic recoil rate, then construct the LZ event-count
likelihood and the thermal relic abundance. 
We finally distinguish the total DM abundance from the relative populations of the two
Majorana states.

\subsection{Kinematic support and excitation ceiling}
\label{sec:kinematic-support}

For the endothermic transition $\chi_{1}A\to\chi_{2}A$ on a xenon isotope $A$, energy-momentum conservation gives the minimum incident speed required to produce a recoil energy $E_{R}$,
\begin{equation}
 v_{\min,A}(E_{R})
 =
 \frac{1}{\sqrt{2m_{A}E_{R}}}
 \left(
 \frac{m_{A}E_{R}}{\mu_{\chi A}}+\delta
 \right),
 \qquad
 \mu_{\chi A}
 =
 \frac{m_{\chi}m_{A}}{m_{\chi}+m_{A}},
 \label{eq:vmin}
\end{equation}
as derived in Appendix~\ref{sec:cross-section-derivation}.
The recoil contribution to $v_{\min,A}$ grows as $\sqrt{E_{R}}$,
whereas the excitation contribution decreases as $1/\sqrt{E_{R}}$.
Consequently, $v_{\min,A}(E_{R})$ reaches a minimum at
\begin{equation}
 E_{R,A}^{\star}
 =
 \frac{\mu_{\chi A}}{m_{A}}\delta,
 \qquad
 v_{\min,A}^{\star}
 =
 \sqrt{\frac{2\delta}{\mu_{\chi A}}}.
 \label{eq:star}
\end{equation}
A transition is therefore kinematically possible only if the incident speed satisfies $v\ge v_{\min,A}^{\star}$.
For a fixed incident speed above this threshold, the allowed recoil interval is
\begin{equation}
 E_{R,A}^{-}(v)
 \leq
 E_{R}
 \leq
 E_{R,A}^{+}(v),
\end{equation}
with
\begin{equation}
 E_{R,A}^{\pm}(v)
 =
 \frac{\mu_{\chi A}^{2}v^{2}}{2m_{A}}
 \left(
 1\pm
 \sqrt{
 1-\frac{2\delta}{\mu_{\chi A}v^{2}}
 }
 \right)^{2}.
 \label{eq:recoiledges}
\end{equation}
The two branches merge at $v=v_{\min,A}^{\star}$, where $E_{R,A}^{-}=E_{R,A}^{+}=E_{R,A}^{\star}$.

To determine whether the halo supplies particles above this threshold,
we specify the Galactic velocity inputs here. We adopt the truncated Standard Halo Model with
\begin{equation}
 \rho_{\chi}=0.30\,\mathrm{GeV\,cm^{-3}},
 \qquad
 v_{0}=238\,\mathrm{km/s},
 \qquad
 v_{\rm esc}=544\,\mathrm{km/s},
 \label{eq:halo-parameters}
\end{equation}
following Ref.~\cite{Baxter:2021pqo}. Here $v_{0}$ is the characteristic Maxwellian speed, which we identify in the adopted halo model with the magnitude of the Local Standard of Rest velocity $v_{\rm LSR}$, while $v_{\rm esc}$ is the Galactic escape speed at the Solar position. 
The Earth velocity in the Galactic rest frame is
\begin{equation}
 \mathbf v_{E}(t)
 =
 \mathbf v_{\rm LSR}
 +\mathbf v_{\rm pec}
 +\mathbf v_{\rm orb}(t),
 \label{eq:earth-velocity}
\end{equation}
with
\begin{equation}
 \mathbf v_{\rm LSR}=(0,v_{0},0),
 \qquad
 \mathbf v_{\rm pec}=(11.1,12.24,7.25)\,\mathrm{km/s},
 \label{eq:solar-velocity}
\end{equation}
where all velocity vectors are expressed in Galactic coordinates $(v_{r},v_{\phi},v_{\theta})$. 
$\mathbf v_{\rm pec}$ is the Sun's peculiar velocity relative to the Local Standard of Rest. 
The Earth's orbital velocity relative to the Sun is
\begin{equation}
 \mathbf v_{\rm orb}(t)
 =
 v_{\rm orb}
 \left[
 \boldsymbol{\epsilon}_{1}\cos\varphi(t)
 +\boldsymbol{\epsilon}_{2}\sin\varphi(t)
 \right],
 \label{eq:orbital-velocity}
\end{equation}
with
\begin{equation}
 \begin{gathered}
 v_{\rm orb}=29.8\,\mathrm{km/s},\\
 \boldsymbol{\epsilon}_{1}=(0.9941,0.1088,0.0042),
 \qquad
 \boldsymbol{\epsilon}_{2}=(-0.0504,0.4946,-0.8677),
 \end{gathered}
 \label{eq:orbital-inputs}
\end{equation}
and
\begin{equation}
 \varphi(t)=\omega(t-t_{\rm ref}),
 \qquad
 \omega=0.0172\,\mathrm{day}^{-1},
 \qquad
 t_{\rm ref}=\text{22 March 2018}.
 \label{eq:orbital-phase}
\end{equation}
Neglecting the Earth's daily rotation, the detector velocity in the Galactic rest frame is approximated by $\mathbf v_{E}(t)$\footnote{We neglect the detector velocity induced by the Earth's daily rotation, whose magnitude is below about $0.5\,\mathrm{km/s}$. It is subleading to the annual orbital motion.}.

For the adopted truncation, the DM velocity $\mathbf u$ in the Galactic rest frame satisfies
\begin{equation}
 u\equiv|\mathbf u|\le v_{\rm esc}.
\end{equation}
In the laboratory frame, the incident DM velocity is
$\mathbf v=\mathbf u-\mathbf v_{E}(t)$. The maximum incident speed is therefore obtained for $u=v_{\rm esc}$ with $\mathbf u$ opposite to $\mathbf v_{E}(t)$, giving
\begin{equation}
 v_{\max}(t)
 =
 v_{\rm esc}+v_{E}(t).
 \label{eq:halo-endpoint}
\end{equation}
Evaluating Eqs.~\eqref{eq:earth-velocity}--\eqref{eq:orbital-phase} over a complete orbital cycle gives
\begin{equation}
 v_{E}(t)\simeq237.4\text{--}266.5\,\mathrm{km/s},
 \qquad
 v_{\max}(t)\simeq781.4\text{--}810.5\,\mathrm{km/s}.
 \label{eq:halo-endpoint-range}
\end{equation}
Requiring $v_{\max}(t)\ge v_{\min,A}^{\star}$ gives the largest kinematically accessible excitation energy,
\begin{equation}
 \delta_{\max,A}(t)
 =
 \frac{1}{2}\mu_{\chi A}v_{\max}^{2}(t)
 =
 \frac{1}{2}\mu_{\chi A}
 \left[v_{\rm esc}+v_{E}(t)\right]^{2}.
 \label{eq:delta-max}
\end{equation}

Equation~\eqref{eq:delta-max} shows that the accessible splitting depends on both the DM mass and the target isotope.
As $v_{\min,A}^{\star}$ approaches $v_{\max}(t)$, the available phase space becomes increasingly restricted to the high-velocity tail and therefore more sensitive to the annual motion of the Earth and to the isotope dependence of $\mu_{\chi A}$.
To characterize the recoil energy region associated with LZ230616, we define
\begin{equation}
 \mathcal{B}= 248\pm32.5 \,\mathrm{keV},
 \label{eq:diagnostic-band}
\end{equation}
where the statistical and systematic uncertainties of the reported recoil energy are combined in quadrature.
At sufficiently low recoil energies, the excitation term causes $v_{\min,A}(E_{R})$ to increase and eventually exceed the available halo endpoint, rendering these recoils kinematically inaccessible.
Consequently, modest variations in $\delta$, the DM mass, or the high velocity tail of the halo can substantially modify the recoil support and event rate.

\subsection{Recoil rate and Galactic halo model}
\label{sec:recoil-halo}

The instantaneous recoil rate per unit detector mass is
\begin{equation}
 \frac{dR(E_{R},t)}{dE_{R}}
 =
 \frac{\rho_{\chi}}{m_{\chi}}
 \sum_{A}
 \frac{x_{A}}{\overline{m}_{\rm Xe}}
 \int_{v>v_{\min,A}}
 d^{3}v\,
 f_{\rm lab}(\mathbf v,t)\,
 v\,
 \frac{d\sigma_{12,A}}{dE_{R}},
 \qquad
 \overline{m}_{\rm Xe}=\sum_{A} x_{A} m_{A},
 \label{eq:rate}
\end{equation}
where $x_{A}$ is the number fraction of isotope $A$, $v=|\mathbf v|$,
and $f_{\rm lab}(\mathbf v,t)$ is the normalized DM velocity distribution in the laboratory frame. 
We approximate natural xenon by the seven isotopes with abundances above $\mathcal{O}(1\%)$~\cite{Raj:2019wpy}. Their natural number fractions $x_{A}$ are listed in Table~\ref{tab:xenon-abundances}.
\begin{table}[t]
\centering
\begin{tabular}{c|ccccccc}
\toprule
Isotope
& $^{128}\mathrm{Xe}$
& $^{129}\mathrm{Xe}$
& $^{130}\mathrm{Xe}$
& $^{131}\mathrm{Xe}$
& $^{132}\mathrm{Xe}$
& $^{134}\mathrm{Xe}$
& $^{136}\mathrm{Xe}$ \\
\midrule
$x_{A}$
& 0.019
& 0.264
& 0.041
& 0.212
& 0.269
& 0.104
& 0.089 \\
\bottomrule
\end{tabular}
\caption{Natural xenon isotope fractions. The rounded number fractions sum to $0.998$ and are normalized to unit sum in the seven isotope calculation.}
\label{tab:xenon-abundances}
\end{table}

Since the differential cross section in Eq.~\eqref{eq:dsigma} scales as $1/v^{2}$, the velocity dependence enters through the mean inverse speed
\begin{equation}
 \eta(v_{\min,A},t)
 =
 \int_{v>v_{\min,A}}d^{3}v\,
 \frac{f_{\rm lab}(\mathbf v,t)}{v}.
 \label{eq:eta}
\end{equation}

Using the same Standard Halo Model inputs as in Section~\ref{sec:kinematic-support}, we take the normalized Galactic
velocity distribution to be the Maxwellian form~\cite{Baxter:2021pqo}
\begin{equation}
 f_{\rm gal}(\mathbf u)
 =
 \frac{e^{-u^{2}/v_{0}^{2}}}{N_{\rm esc}(\pi v_{0}^{2})^{3/2}}
 \Theta(v_{\rm esc}-u),
 \qquad u=|\mathbf u|,
 \label{eq:galactic-distribution}
\end{equation}
where
\begin{equation}
 N_{\rm esc}
 =\operatorname{erf}(z)-\frac{2z}{\sqrt{\pi}}e^{-z^{2}},
 \qquad z=\frac{v_{\rm esc}}{v_{0}}.
 \label{eq:halo-normalization}
\end{equation}
The laboratory distribution is therefore
\begin{equation}
f_{\rm lab}(\mathbf v,t)
=
f_{\rm gal} \left(\mathbf u (\mathbf v,t)\right)
=
f_{\rm gal} \left(\mathbf v+\mathbf v_{E}(t)\right),
\label{eq:lab-distribution}
\end{equation}
with $\mathbf v_{E}(t)$ given by Eqs.~\eqref{eq:earth-velocity}--\eqref{eq:orbital-phase}.
The same velocity prescription therefore determines both the kinematic endpoint and the recoil rate.

For the time-averaged recoil spectrum, we approximate the temporal distribution of the LZ exposure as uniform over the data-taking interval,
\begin{equation}
 \overline{\frac{dR}{dE_{R}}}
 =
 \frac{1}{\Delta t}
 \int_{t_{i}}^{t_{f}}dt\,
 \frac{dR(E_{R},t)}{dE_{R}},
 \label{eq:time-average}
\end{equation}
where $\Delta t=t_{f}-t_{i}$,
$t_{i}=\text{27 March 2023}$, and
$t_{f}=\text{1 April 2024}$~\cite{LZ:2026axp}.
We account for the detector acceptance using the LZ NR signal efficiency $\epsilon_{\rm NR}(E_{R})$ reported in Ref.~\cite{LZ:2026axp}. The accepted instantaneous recoil spectrum is
\begin{equation}
 \frac{dR_{\rm acc}(E_{R},t)}{dE_{R}}
 =
 \epsilon_{\rm NR}(E_{R})
 \frac{dR(E_{R},t)}{dE_{R}},
 \label{eq:accepted}
\end{equation}
Using the total exposure $\mathcal{E}=2.84\,\mathrm{tonne\text{-}year}$~\cite{LZ:2026axp},
the corresponding time-averaged differential event count is
\begin{equation}
 \frac{dN_{\rm acc}}{dE_{R}}
 =
 \frac{\mathcal{E}}{\Delta t}
 \int_{t_{i}}^{t_{f}}dt\,
 \frac{dR_{\rm acc}(E_{R},t)}{dE_{R}}
 =
 \mathcal{E}\,\epsilon_{\rm NR}(E_{R})
 \overline{\frac{dR}{dE_{R}}}.
 \label{eq:accepted-count-spectrum}
\end{equation}
For diagnostics at the event date, we instead evaluate the instantaneous rate on 16 June 2023, the date of LZ230616~\cite{LZ:2026axp},
\begin{equation}
 t_{\rm event}=\text{16 June 2023},
 \qquad
 \left.\frac{dR}{dE_{R}}\right|_{\rm event}
 =
 \left.
 \frac{dR(E_{R},t)}{dE_{R}}
 \right|_{t=t_{\rm event}}.
 \label{eq:event-rate}
\end{equation}
The corresponding accepted spectrum follows from Eq.~\eqref{eq:accepted} evaluated at $t=t_{\rm event}$.


\subsection{LZ recoil rate and likelihood}
\label{sec:count-probe}

The endothermic splitting $\delta$ controls the recoil spectrum through the minimum incident speed in Eq.~\eqref{eq:vmin}. 
For fixed $m_{\chi}$ and $m_{r}$, increasing $\delta$ progressively removes low recoil phase space and shifts the surviving events toward higher recoil energies. 
The isotope dependent xenon response in Eq.~\eqref{eq:dsigma} and the published LZ NR signal efficiency then determine the accepted spectrum.
For the parameter space analysis, we use the expected number of accepted events in the recoil interval $\mathcal{B}$~\eqref{eq:diagnostic-band} associated with LZ230616,
\begin{equation}
 N_{\mathcal{B}}
 =
 \int_{\mathcal{B}}dE_{R}\,
 \frac{dN_{\rm acc}}{dE_{R}}
 =
 \mathcal{K}_{\mathcal{B}}(m_{\chi},\delta,m_{r})
 \left(
 \frac{c_{12}^{S}}{\Lambda_{r}^{2}}
 \right)^{2}.
 \label{eq:count-definition}
\end{equation}
The interval $\mathcal{B}$ is used as a diagnostic interval in true recoil energy, while detector acceptance is incorporated through $\epsilon_{\rm NR}(E_{R})$. The response coefficient $\mathcal{K}_{\mathcal{B}}(m_{\chi},\delta,m_{r})$ contains the isotope sum, radion propagator, halo velocity integral, nuclear response, detector efficiency, exposure, and the remaining mass-dependent factors. 

To assign a reproducible statistical interpretation to the expected count,
we introduce an auxiliary Poisson counting model in which the LZ230616 event is assigned to $\mathcal{B}$, the background contribution is neglected, and information outside $\mathcal{B}$ is not included.
For one assigned event, the corresponding likelihood is
\begin{equation}
 \mathcal{L}_{\mathcal{B}}(N_{\mathcal{B}})
 =
 N_{\mathcal{B}}e^{-N_{\mathcal{B}}},
 \label{eq:poisson-likelihood}
\end{equation}
which is maximized at $N_{\mathcal{B}}=1$. We therefore use $N_{\mathcal{B}}=1$ as the central expected signal count and define the likelihood ratio statistic
\begin{equation}
 q_{\mathcal{B}}
 =
 -2\ln
 \frac{\mathcal{L}_{\mathcal{B}}(N_{\mathcal{B}})}
 {\mathcal{L}_{\mathcal{B}}(1)}
 =
 2\left(
 N_{\mathcal{B}}-1-\ln N_{\mathcal{B}}
 \right).
 \label{eq:poisson-support}
\end{equation}
We adopt the relative likelihood support condition
\begin{equation}
 q_{\mathcal{B}}\leq1
 \quad\Longleftrightarrow\quad
 0.30\leq N_{\mathcal{B}}\leq2.36.
 \label{eq:count-band}
\end{equation}
Within this interval, the likelihood is at least $e^{-1/2}$ of its maximum.
This defines a likelihood support convention rather than a calibrated confidence interval. 
In particular, a single event does not justify identifying $q_{\mathcal{B}}\leq1$ with $68\%$ coverage through Wilks'
theorem~\cite{ParticleDataGroup:2024cfk}.
The corresponding constraint on the original radion parameters is
\begin{equation}
 \sqrt{\frac{0.30}{\mathcal{K}_{\mathcal{B}}}}
 \leq
 \frac{|c_{12}^{S}|}{\Lambda_{r}^{2}}
 \leq
 \sqrt{\frac{2.36}{\mathcal{K}_{\mathcal{B}}}}.
 \label{eq:lz-original-band}
\end{equation}
For fixed $m_{\chi}$, $\delta$, and $m_{r}$, the two boundaries are
parabolas in the $(\Lambda_{r},|c_{12}^{S}|)$ plane.
A more complete treatment would require the corresponding LZ likelihood, including detector response, backgrounds, and experimental nuisance parameters. 
Equation~\eqref{eq:lz-original-band} therefore provides an absolute event count constraint within the stated one-event approximation.

\subsection{Thermal abundance in the radion portal}
\label{sec:thermal-probe}

We assume that the dark sector is initially in thermal equilibrium with the visible bath, that the universe is radiation dominated during freeze-out, and that no later source repopulates the dark sector. 
Requiring the pseudo-Majorana pair to account for the full DM abundance~\cite{Planck:2018vyg}
gives the central target
\begin{equation}
 \Omega_{\chi}h^{2}=\Omega_{\chi_{1}+\chi_{2}}h^{2}=0.12.
 \label{eq:relic-target}
\end{equation}
For the explicit thermal calculation, we adopt the central configuration in Eq.~\eqref{eq:main-central-configuration} and the corresponding radion interaction in Eq.~\eqref{eq:main-central-interaction}, together with the visible trace coupling introduced in Section~\ref{sec:visible-radion}.
We work to leading order in $\delta/m_{\chi}$ during chemical freeze-out.

For the WIMP-scale masses considered here, freeze-out occurs at $T_{f} \sim 10\,\mathrm{GeV}$ and therefore $T_{f}\gg\delta$, with $\delta \sim 100\,\mathrm{keV}$. 
The two Majorana states have nearly equal equilibrium number densities during freeze-out.
If reactions interconverting $\chi_{1}$ and $\chi_{2}$ maintain chemical equilibrium, their total number density $n_{\chi}=n_{\chi_{1}}+n_{\chi_{2}}$ obeys~\cite{Griest:1990kh}
\begin{align}
 \dot n_{\chi}+3Hn_{\chi}
 =
 -\langle\sigma_{\rm eff}v_{\rm rel}\rangle
 \left[n_{\chi}^{2}-\left(n_{\chi}^{\rm eq}\right)^{2}\right],
 \label{eq:total-boltzmann}
\end{align}
with the thermally averaged effective annihilation rate
\begin{align}
 \langle\sigma_{\rm eff}v_{\rm rel}\rangle
 =
 \sum_{i,j=1}^{2}
 \frac{n_{\chi_{i}}^{\rm eq}n_{\chi_{j}}^{\rm eq}}{\left(n_{\chi}^{\rm eq}\right)^{2}}
 \langle\sigma_{ij}v_{\rm rel}\rangle
 \simeq
 \frac{1}{4}
 \left(\langle\sigma_{11}v_{\rm rel}\rangle
 +2\langle\sigma_{12}v_{\rm rel}\rangle
 +\langle\sigma_{22}v_{\rm rel}\rangle
 \right),
 \label{eq:coannihilation-weights}
\end{align}
where $n_{\chi}^{\rm eq}=n_{\chi_{1}}^{\rm eq}+n_{\chi_{2}}^{\rm eq}$, and the last
approximation uses
$n_{\chi_{1}}^{\rm eq}\simeq n_{\chi_{2}}^{\rm eq}\simeq n_{\chi}^{\rm eq}/2$.
For $m_{r}<m_{\chi}$, the leading annihilation channels are
\begin{equation}
 \chi_{1}\chi_{1}\to rr,\qquad
 \chi_{2}\chi_{2}\to rr,\qquad
 \chi_{1}\chi_{2}\to r^{*}\to{\rm SM}.
 \label{eq:annihilation-channels}
\end{equation}
The first two processes contain two transition vertices and therefore remain active even when the diagonal radion couplings are suppressed.
For the radion portal considered here, the leading short-distance annihilation rates are $p$-wave,
\begin{equation}
 \sigma v_{\rm rel}
 =
 b\,v_{\rm rel}^{2}
 \left[
 1+\mathcal{O}\left(v_{\rm rel}^{2},\frac{\delta}{m_{\chi}}\right)
 \right].
\end{equation}
The corresponding coefficients are~\cite{Benic:2014xho,DeSimone:2016fbz}
\begin{align}
 b_{rr}
 &=
 \left(\frac{c_{12}^{S}m_{\chi}}{\Lambda_{r}}\right)^{4}
 \frac{
 \sqrt{1-m_{r}^{2}/m_{\chi}^{2}}\,
 \left(
 9-8m_{r}^{2}/m_{\chi}^{2}
 +2m_{r}^{4}/m_{\chi}^{4}
 \right)}
 {24\pi m_{\chi}^{2}
 \left(2-m_{r}^{2}/m_{\chi}^{2}\right)^{4}},
 \label{eq:brr}\\
 b_{12}^{\rm SM}
 &=
 \left(\frac{c_{12}^{S}m_{\chi}}{\Lambda_{r}}\right)^{2}
 \frac{
 m_{\chi}\Gamma_{\rm SM}(2m_{\chi})
 }
 {2\left(4m_{\chi}^{2}-m_{r}^{2}\right)^{2}}.
 \label{eq:b12}
\end{align}
Using the equal equilibrium weights in Eq.~\eqref{eq:coannihilation-weights},
the effective annihilation coefficient is therefore
\begin{equation}
 b_{\rm eff}
 =
 \frac{b_{rr}+b_{12}^{\rm SM}}{2}.
 \label{eq:bsm}
\end{equation}
Here $\Gamma_{\rm SM}(M)$ denotes the total width into the retained SM final states, evaluated at invariant mass $M$.
For the thermal calculation, we retain the dominant tree-level SM final
states and approximate
\begin{equation}
 \Gamma_{\rm SM}(M)
 \simeq
 \sum_{f}\Gamma_{f\bar f}(M)
 +\Gamma_{WW}(M)
 +\Gamma_{ZZ}(M)
 +\Gamma_{hh}(M),
 \label{eq:gamma-sm}
\end{equation}
where the partial widths induced by the radion trace coupling in
Section~\ref{sec:visible-radion} are given by Ref.~\cite{Bae:2000pk}:
\begin{align}
 \Gamma_{f\bar f}(M)
 &=
 \frac{N_{c}^{f}m_{f}^{2}M}
 {8\pi\Lambda_{r}^{2}}
 \left(
 1-\frac{4m_{f}^{2}}{M^{2}}
 \right)^{3/2},
 \\
 \Gamma_{WW}(M)
 &=
 \frac{M^{3}}
 {16\pi\Lambda_{r}^{2}}
 \left(
 1-\frac{4m_{W}^{2}}{M^{2}}
 +\frac{12m_{W}^{4}}{M^{4}}
 \right)
 \sqrt{
 1-\frac{4m_{W}^{2}}{M^{2}}
 },
 \\
 \Gamma_{ZZ}(M)
 &=
 \frac{M^{3}}
 {32\pi\Lambda_{r}^{2}}
 \left(
 1-\frac{4m_{Z}^{2}}{M^{2}}
 +\frac{12m_{Z}^{4}}{M^{4}}
 \right)
 \sqrt{
 1-\frac{4m_{Z}^{2}}{M^{2}}
 },
 \\
 \Gamma_{hh}(M)
 &=
 \frac{M^{3}}
 {32\pi\Lambda_{r}^{2}}
 \left(
 1+\frac{2m_{h}^{2}}{M^{2}}
 \right)^{2}
 \sqrt{
 1-\frac{4m_{h}^{2}}{M^{2}}
 } .
 \label{eq:gamma-sm-partial}
\end{align}
Here $N_{c}^{f}=3$ for quarks and $N_{c}^{f}=1$ for leptons, and each channel is understood to vanish below its kinematic threshold. 
The anomaly-induced $gg$ and $\gamma\gamma$ channels are neglected in the thermal calculation, since at the
reference virtuality $M=2m_{\chi}\simeq 2\,\mathrm{TeV}$, the gluonic contribution satisfies
$\Gamma_{gg}/(\Gamma_{WW}+\Gamma_{ZZ}+\Gamma_{hh}) \simeq \alpha_{s}^{2}(M)b_{3}^{2}/(4\pi^{2})\simeq 0.8\%$, while the diphoton contribution is even smaller.
For $M=2m_{\chi}$ and $m_{\chi}\gg m_{W},m_{Z},m_{h},m_{t}$, the high mass limit gives~\cite{Giudice:2000av,Rizzo:2002pq}
\begin{equation}
 \Gamma_{\rm SM}(2m_{\chi})
 \simeq
 \frac{m_{\chi}^{3}}
 {\pi\Lambda_{r}^{2}} .
 \label{eq:gamma-sm-highmass}
\end{equation}
In the simultaneous limits $m_{r}\ll m_{\chi}$ and $m_{\chi}\gg m_{W},m_{Z},m_{h},m_{t}$,
the short-distance coefficient simplifies to
\begin{equation}
 b_{\rm eff}
 \simeq
 \frac{m_{\chi}^{2}}{\Lambda_{r}^{4}}
 \left[
 \frac{3|c_{12}^{S}|^{4}}{256\pi}
 +
 \frac{|c_{12}^{S}|^{2}}{64\pi}
 \right].
 \label{eq:thermal-scaling-original}
\end{equation}
The first term is the contribution from annihilation into radion pairs,
while the second arises from annihilation into the retained visible
states.
The numerical calculation uses the finite mass expressions in
Eqs.~\eqref{eq:gamma-sm}--\eqref{eq:gamma-sm-partial}, together with
the full $m_{r}$ dependence in Eqs.~\eqref{eq:brr} and~\eqref{eq:b12}.
Equations~\eqref{eq:gamma-sm-highmass} and
\eqref{eq:thermal-scaling-original} are used only to display the
asymptotic behavior.
Their different powers of $c_{12}^{S}$ allow the relic abundance to provide information independent of the LZ event count, which fixes $|c_{12}^{S}|/\Lambda_{r}^{2}$. 

We evaluate the thermal abundance at leading Born level as a first estimate, neglecting Sommerfeld enhancement~\cite{Cassel:2009wt,Slatyer:2009vg}.
At freeze-out, both Majorana states are thermally populated and their typical relative velocities are much higher than in the solar core.
The solar Sommerfeld factor therefore cannot be applied directly to the relic calculation; its consistent inclusion would require temperature-dependent coupled-channel annihilation rates and a new joint parameter fit.
The resulting joint solutions are consequently treated as Born reference solutions rather than precision determinations of the thermal parameters.
For a nonrelativistic Maxwell-Boltzmann distribution, the $p$-wave thermal average is~\cite{Griest:1990kh}
\begin{equation}
 \langle\sigma_{\rm eff}v_{\rm rel}\rangle
 =
 b_{\rm eff}\langle v_{\rm rel}^{2}\rangle
 =
 \frac{6b_{\rm eff}}{x},
 \qquad
 x=\frac{m_{\chi}}{T}.
 \label{eq:thermal-born}
\end{equation}
The nearly degenerate approximation is applied during chemical freeze-out
and does not assume equal late-time populations after conversion processes have decoupled.
Defining $Y_{\chi}=n_{\chi}/s$, the Boltzmann equation becomes
\begin{equation}
 \frac{dY_{\chi}}{dx}
 =
 -\frac{s}{Hx}
 \left(
 1+\frac{1}{3}\frac{d\ln g_{s}}{d\ln T}
 \right)
 \langle\sigma_{\rm eff}v_{\rm rel}\rangle
 \left[
 Y_{\chi}^{2}
 -
 \left(Y_{\chi}^{\rm eq}\right)^{2}
 \right],
 \label{eq:yield-equation}
\end{equation}
where
\begin{equation}
 s(T)=\frac{2\pi^{2}}{45}g_{s}(T)T^{3},
 \qquad
 H(T)=
 \sqrt{\frac{4\pi^{3}g_{\rho}(T)}{45}}
 \frac{T^{2}}{M_{\rm P}},
 \qquad
 Y_{\chi}^{\rm eq}
 =
 \frac{45}{\pi^{4}g_{s}(T)}
 x^{2}K_{2}(x).
 \label{eq:thermal-background}
\end{equation}
Here $M_{\rm P}=1.22\times 10^{19}~\mathrm{GeV}$ denotes the Planck mass.
We use the tabulated Standard Model thermodynamic functions $g_{s}(T)$ and $g_{\rho}(T)$ of Ref.~\cite{Saikawa:2020swg}.
The radion contribution to the background energy and entropy densities is neglected. 
We integrate Eq.~\eqref{eq:yield-equation} from equilibrium at $x=10$ to $x=10^{5}$, where the yield has approached its asymptotic value $Y_{\chi,\infty}$, which determines the present total relic abundance $\Omega_{\chi}h^{2}$~\eqref{eq:relic-target}.

The relic calculation above determines only the total abundance $\Omega_{\chi}h^{2}$, not the late-time partition between $\chi_{1}$ and $\chi_{2}$. 
In a nearly degenerate two-state dark sector, number-conserving down-scattering and other de-excitation processes
can continue after chemical freeze-out and reduce the excited-state population. 
The surviving fraction depends on the corresponding de-excitation rates and on the kinetic-decoupling history
\cite{Finkbeiner:2008gw}. 
We do not solve this separate late-time population evolution here, but assume in the direct-detection analysis
that the present halo is sufficiently dominated by the lighter state $\chi_{1}$, so that the endothermic process
$\chi_{1}A\to\chi_{2}A$ gives the leading transition signal.

\section{Joint analysis of the LZ event count and relic abundance}
\label{sec:results}

We now combine the direct detection and cosmological information in the radion-portal parameter space
$(m_{\chi},\delta,m_{r},\Lambda_{r},c_{12}^{S})$.
A necessary prerequisite for the direct detection analysis is that the diagnostic interval $\mathcal{B}$ defined in
Eq.~\eqref{eq:diagnostic-band} has kinematic recoil support.
It requires that at least one retained xenon isotope must admit a recoil energy $E_{R}\in\mathcal{B}$ satisfying
\begin{equation}
v_{\mathcal{B}}^{\star} \equiv
 \min_{\substack{A\\ E_{R}\in\mathcal{B}}}
 v_{\min,A}(E_{R})
 \leq
 \max_{t} \ v_{\max}(t),
 \label{eq:B-support-condition}
\end{equation}
where $v_{\min,A}(E_{R})$ and $v_{\max}(t)$ are given by
Eqs.~\eqref{eq:vmin} and~\eqref{eq:halo-endpoint}, respectively.
For the adopted recoil response and detector efficiency, this corresponds to $\mathcal{K}_{\mathcal{B}}(m_{\chi},\delta,m_{r})>0$.
Within the kinematically supported region, the joint solution is determined by the LZ event count and the thermal relic abundance,
\begin{equation}
\left\{
\begin{aligned}
 |c_{12}^{S}|
 &=
 \Lambda_{r}^{2}
 \sqrt{
 \frac{N_{\mathcal{B}}}
 {\mathcal{K}_{\mathcal{B}}(m_{\chi},\delta,m_{r})}
 },
 \\
 \Omega_{\chi}h^{2}
 &=0.12 .
\end{aligned}
\right.
\label{eq:joint_con_1}
\end{equation}
The central solution uses $N_{\mathcal{B}}=1$, while varying $N_{\mathcal{B}}$ within the likelihood support of
Eq.~\eqref{eq:count-band} traces the corresponding family of solutions.

For $m_{\chi}=1\,\mathrm{TeV}$, the LZ spin-independent (SI) $\mathcal{O}_{1}^{s}$ result provides the additional requirement on the cross section defined in Eq.~\eqref{eq:sigma0}
\begin{equation}
 \sigma_{\rm SI,-}^{N}(\delta)
 \leq
 \sigma_{N}^{0}
 \leq
 \sigma_{\rm SI,+}^{N}(\delta),
 \label{eq:lz-si-constraint}
\end{equation}
where
$\sigma_{N}^{0}=\sigma_{\rm SI}^{N}$ under the matching derived in Appendix~\ref{sec:lz-normalization}.
Equation~\eqref{eq:lz-si-constraint} therefore selects the subset of the joint solution set compatible with the two-sided $90\%$ confidence interval reported by LZ.

\subsection{Dark matter parameter space
\texorpdfstring{$(m_{\chi},\delta)$}{(mchi, delta)}}
\label{sec:mass-plane}

We examine the DM mass $m_{\chi}$ and excitation energy $\delta$ for three representative radion masses,
$m_{r}=10$, $30$, $50\,\mathrm{GeV}$, to characterize the radion dependence of the joint solution region.
For each $m_{r}$, we identify the region in the $(m_{\chi},\delta)$ plane where kinematic recoil support exists and the central LZ event count, $N_{\mathcal{B}}=1$, can be satisfied simultaneously with $\Omega_{\chi}h^{2}=0.12$.
The joint solution region is bounded by two distinct conditions.
One boundary marks the loss of kinematic recoil support in $\mathcal{B}$ and is determined by Eq.~\eqref{eq:B-support-condition}.
The other marks where the central conditions $N_{\mathcal{B}}=1$ and $\Omega_{\chi}h^{2}=0.12$
cease to admit a positive finite solution for $(\Lambda_{r},|c_{12}^{S}|)$, as derived below.
The LZ SI consistency condition is additionally indicated on the published $m_{\chi}=1\,\mathrm{TeV}$ slice.
The numerical scan covers $200\leq m_{\chi}\leq1500\,\mathrm{GeV}$ and $100\leq\delta\leq450\,\mathrm{keV}$.

\begin{figure}[!htbp]
\centering
\includegraphics[height=5.8cm]{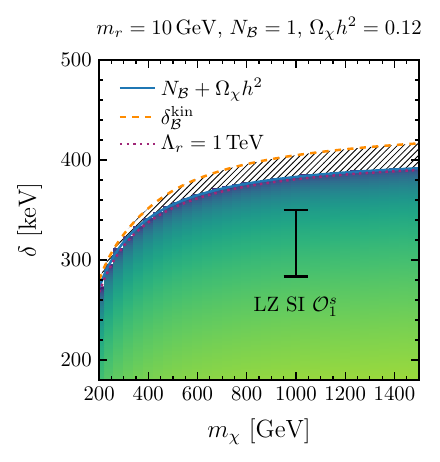} 
\hspace{-6mm}
\includegraphics[height=5.8cm]{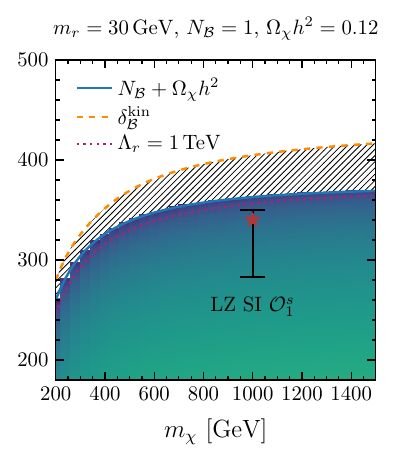}
\hspace{-6mm}
\includegraphics[height=5.8cm]{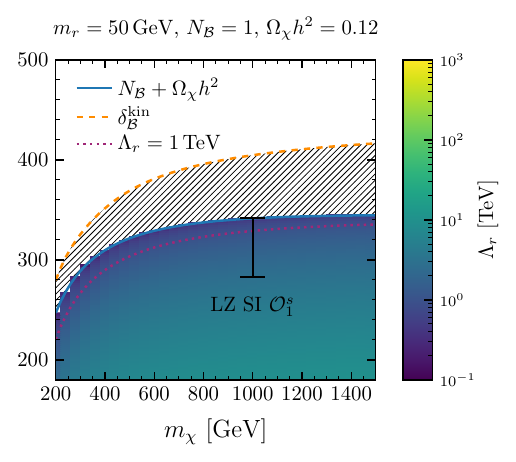}
\caption{DM mass and splitting scan for $m_{r}=10$, $30$, $50\,\mathrm{GeV}$.
Colored regions satisfy $N_{\mathcal{B}}=1$ and $\Omega_{\chi}h^{2}=0.12$, with color indicating the derived $\Lambda_{r}$.
The hatched region has kinematic recoil support but no positive joint solution, while the dashed boundary marks the kinematic ceiling.
The dotted contour marks the reference scale $\Lambda_{r}=1\,\mathrm{TeV}$.
The thick segment at $m_{\chi}=1\,\mathrm{TeV}$ marks the portion selected by the published LZ SI interval.
The red star in the $m_{r}=30\,\mathrm{GeV}$ panel marks the reference point at $m_{\chi}=1\,\mathrm{TeV}$ and
$\delta=340\,\mathrm{keV}$, for which the central joint solution gives $\Lambda_{r}\simeq2.268\,\mathrm{TeV}$.}
\label{fig:mass-plane}
\end{figure}

The thermal relic condition is governed by the effective annihilation coefficient $b_{\rm eff}$. 
Using Eqs.~\eqref{eq:brr}--\eqref{eq:bsm}, it now takes the form
\begin{align}
 b_{\rm eff}
 =
 \frac{m_{\chi}^{2}\Lambda_{r}^{4}}
 {48\pi\mathcal{K}_{\mathcal{B}}^{2}}
 \frac{
 \sqrt{1-\kappa_{r\chi}^{2}}
 \left(9-8\kappa_{r\chi}^{2}
 +2\kappa_{r\chi}^{4}\right)}
 {\left(2-\kappa_{r\chi}^{2}\right)^{4}}
 +
 \frac{\Lambda_{r}^{2}\Gamma_{\rm SM}(2m_{\chi})}
 {4m_{\chi}\mathcal{K}_{\mathcal{B}}
 \left(4-\kappa_{r\chi}^{2}\right)^{2}}, 
 \quad
 \kappa_{r \chi} = \frac{m_{r}}{m_{\chi}}.
 \label{eq:scan-beff}
\end{align}
For each $m_{\chi}$, let $b_{\star}(m_{\chi})$ denote the effective annihilation coefficient that yields
$\Omega_{\chi}h^{2}=0.12$ after numerical integration of Eq.~\eqref{eq:yield-equation}.
Note that each retained partial width scales as $\Lambda_{r}^{-2}$,
the combination $\Lambda_{r}^{2}\Gamma_{\rm SM}(2m_{\chi})$ in the second term is thus independent of $\Lambda_{r}$.
Since the first term is positive, a joint solution exists if and only if
\begin{equation}
 \frac{\Lambda_{r}^{2}\Gamma_{\rm SM}(2m_{\chi})}
 {4m_{\chi}\mathcal{K}_{\mathcal{B}}
 \left(4-\kappa_{r\chi}^{2}\right)^{2}}
 <
 b_{\star}(m_{\chi}),
 \label{eq:joint-existence}
\end{equation}
which defines the termination of the joint solution region.
Within this region, the corresponding solution in the $(\Lambda_{r},|c_{12}^{S}|)$ plane is
\begin{equation}
\begin{aligned}
 \Lambda_{r}
 &=
 \left[
 \frac{
 48\pi\mathcal{K}_{\mathcal{B}}^{2}
 \left(2-\kappa_{r\chi}^{2}\right)^{4}}
 {m_{\chi}^{2}
 \sqrt{1-\kappa_{r\chi}^{2}}
 \left(9-8\kappa_{r\chi}^{2}
 +2\kappa_{r\chi}^{4}\right)}
 \left(
 b_{\star}(m_{\chi})
 -
 \frac{\Lambda_{r}^{2}\Gamma_{\rm SM}(2m_{\chi})}
 {4m_{\chi}\mathcal{K}_{\mathcal{B}}
 \left(4-\kappa_{r\chi}^{2}\right)^{2}}
 \right)
 \right]^{1/4},
 \\[4pt]
 |c_{12}^{S}|
 &=
 \frac{\Lambda_{r}^{2}}
 {\sqrt{\mathcal{K}_{\mathcal{B}}}}.
\end{aligned}
\label{eq:joint-solution}
\end{equation}
Fig.~\ref{fig:mass-plane} separates the kinematic and joint-solution boundaries.
For a given $m_{\chi}$, the kinematic boundary is
\begin{equation}
 \delta_{\mathcal{B}}^{\rm kin}(m_{\chi})
 =
 \max_{\substack{A\\ E_{R}\in\mathcal{B}}}
 \left[
 \sqrt{2m_{A}E_{R}}\,\max_{t} \ v_{\max}(t)
 -
 \frac{m_{A}E_{R}}{\mu_{\chi A}}
 \right],
 \label{eq:delta-B-kinematic}
\end{equation}
which is actually the boundary form of the kinematic-support condition in Eq.~\eqref{eq:B-support-condition}.
It depends on $m_{\chi}$ and the halo endpoint, but not on $m_{r}$.
At $m_{\chi}=1\,\mathrm{TeV}$,
Eq.~\eqref{eq:delta-B-kinematic} gives $\delta_{\mathcal{B}}^{\rm kin}\simeq404.7\,\mathrm{keV}$.
The joint solution~\eqref{eq:joint-solution} terminates at a smaller splitting when Eq.~\eqref{eq:joint-existence} is saturated.
For $m_{\chi}=1\,\mathrm{TeV}$, the corresponding endpoints are
approximately $\delta = 383.0$, $362.8$, $341.5\,\mathrm{keV}$ for
$m_{r}=10$, $30$, $50\,\mathrm{GeV}$, respectively.
The relevant xenon momentum transfers satisfy $Q^{2}\ll m_{r}^{2}$, so the propagator factor
approaches its contact limit, which implies $\mathcal{K}_{\mathcal{B}}\propto m_{r}^{-4}$.
The second term in Eq.~\eqref{eq:scan-beff} therefore scales approximately as $m_{r}^{4}$, 
apart from the mild dependence through $\left(4-\kappa_{r\chi}^{2}\right)^{-2}$.
$\mathcal{K}_{\mathcal{B}}$ also decreases as $\delta$ approaches the kinematic boundary because the available recoil support is progressively reduced.
Since $b_{\star}(m_{\chi})$ is fixed at a given $m_{\chi}$, the joint solution region contracts as $m_{r}$ increases.
To provide a reference for the radion interaction scale within the joint solution, Fig.~\ref{fig:mass-plane} also shows the dotted contour $\Lambda_{r}=1\,\mathrm{TeV}$.
For a given $m_{\chi}$, the region between this contour and the joint-solution boundary has $\Lambda_{r}<1\,\mathrm{TeV}$, while
smaller $\delta$ corresponds to $\Lambda_{r}>1\,\mathrm{TeV}$.
The former region has a stronger visible radion coupling $1/\Lambda_{r}$ and is therefore less conservative for the weakly
coupled effective description adopted here.
This contour is consequently used only as a reference scale rather than as a sharp boundary.

\vspace{1cm}

\vspace{0.5cm}

\begin{figure}[!htbp]
\centering
\includegraphics[width=0.60\textwidth]{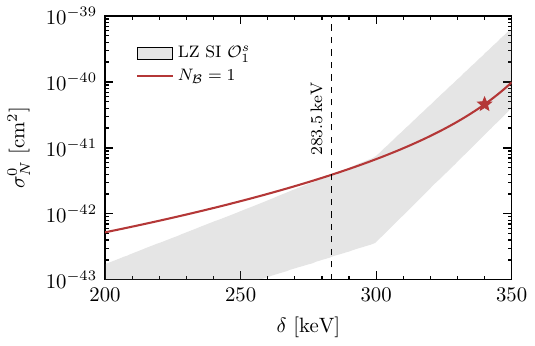}
\caption{Reference nucleon cross section $\sigma_{N}^{0}(\delta)$ implied by $N_{\mathcal{B}}=1$ at $m_{\chi}=1\,\mathrm{TeV}$.
The gray band denotes the LZ SI $\mathcal{O}_{1}^{s}$ two-sided $90\%$ confidence interval.
The results for $m_{r}=10$, $30$, $50\,\mathrm{GeV}$ are indistinguishable on this scale.
The star marks the reference point at $\delta=340\,\mathrm{keV}$, for which the joint solution gives
$\sigma_{N}^{0}\simeq4.58\times10^{-41}\,\mathrm{cm^{2}}$.
The dashed vertical line marks the lower limit, $\delta\simeq283.5\,\mathrm{keV}$.
}
\label{fig:lz-si-slice}
\end{figure}

In addition, $\sigma_{N}^{0}\propto (m_{r}^{4}\mathcal{K}_{\mathcal{B}})^{-1}$, so the scaling
$\mathcal{K}_{\mathcal{B}}\propto m_{r}^{-4}$ approximately cancels the $m_{r}$ dependence of the corresponding LZ SI $\mathcal{O}_{1}^{s}$ ranges in $\delta$.
Fig.~\ref{fig:lz-si-slice} shows this independence directly on the published $m_{\chi}=1\,\mathrm{TeV}$ slice.
The curves for $m_{r}=10$, $30$, $50\,\mathrm{GeV}$ are indistinguishable on this scale.
The lower limit, near $\delta\simeq283.5\,\mathrm{keV}$, is determined by the upper cross-section boundary $\sigma_{\rm SI,+}^{N}(\delta)$ and is therefore not an artifact of the scanned $\delta$ range.
This determines the nearly common lower endpoint of the LZ SI segments in Fig.~\ref{fig:mass-plane}.
The apparent upper endpoint at $350\,\mathrm{keV}$, however, is set only by the largest mass splitting included in the published LZ scan and should not be interpreted as a physical upper limit.
For $m_{r}=10$ and $30\,\mathrm{GeV}$, the joint solutions extend to $\delta\simeq383.0$ and $362.8\,\mathrm{keV}$, respectively, beyond the available LZ SI range ending at $350\,\mathrm{keV}$.
The current LZ SI constraint therefore cannot assess the remaining joint solution region above this endpoint.
By contrast, for $m_{r}=50\,\mathrm{GeV}$, the joint solution terminates near $\delta\simeq341.5\,\mathrm{keV}$ within the LZ range. The LZ SI constraint therefore allows the entire surviving joint-solution interval above the lower limit.

\vspace{1cm}

\subsection{Radion parameter space
\texorpdfstring{$(m_{r},\Lambda_{r}^{-1})$}{(mr, Lambda r)}}
\label{sec:rad_plane}

\begin{figure}[!htbp]
\centering
\includegraphics[width=0.60\textwidth]{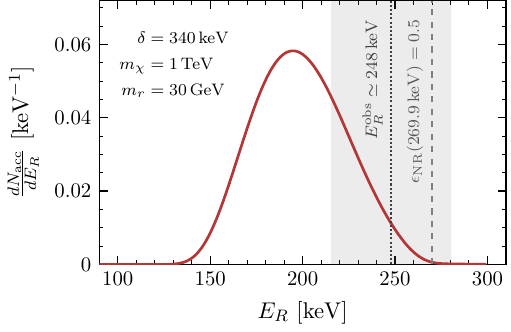}
\caption{Time-averaged accepted recoil spectrum for the reference point in Eq.~\eqref{eq:ref_point_0}, which peaks at $E_{R}\simeq194.74\,\mathrm{keV}$. 
The shaded region denotes the diagnostic interval $\mathcal{B}$, the dotted line marks the LZ observed recoil energy $E_{R}^{\rm obs} \simeq 248\,\mathrm{keV}$, and the dashed line marks the larger recoil energy at which $\epsilon_{\rm NR}(E_{R})=0.5$.}
\label{fig:reference-absolute}
\end{figure}

Motivated by the solutions in Figs.~\ref{fig:mass-plane} and
\ref{fig:lz-si-slice}, we fix $m_{\chi}=1\,\mathrm{TeV}$ and
$\delta=340\,\mathrm{keV}$ and scan the
$(m_{r},\Lambda_{r}^{-1})$ plane. For the subsequent analysis, we
select the reference point
\begin{equation}
\boxed{
\begin{alignedat}{3}
 m_{\chi} &= 1\,\mathrm{TeV},
 &\qquad \delta &= 340\,\mathrm{keV},
 &\qquad m_{r} &= 30\,\mathrm{GeV},
 \\
 \Lambda_{r} &\simeq 2.268\,\mathrm{TeV},
 & c_{12}^{S} &\simeq 3.195,
 & c_{D} &= 1 .
\end{alignedat}}
\label{eq:ref_point_0}
\end{equation}
It satisfies the central joint conditions in Eq.~\eqref{eq:joint_con_1}.
The corresponding accepted recoil spectrum, shown in Fig.~\ref{fig:reference-absolute}, peaks at
$E_{R}\simeq194.74\,\mathrm{keV}$ and predicts $3.87$ accepted events over the tabulated recoil range,
with $1.00$ within the diagnostic interval $\mathcal{B}$ and $2.87$ below it.
The reference nucleon cross section $\sigma_{N}^{0}\simeq4.58\times10^{-41}\, \mathrm{cm}^{2}$ lies within the published two-sided
LZ SI $\mathcal{O}_{1}^{s}$ interval.
Since this experimental interval incorporates the inelastic signal template over the extended analysis
region, the comparison provides a broader consistency check than the auxiliary $N_{\mathcal{B}}=1$ condition,
subject to the LZ signal-template assumptions.

\vspace{1cm}

\vspace{1cm}

\begin{figure}[!htbp]
\centering
\includegraphics[width=0.60\textwidth]{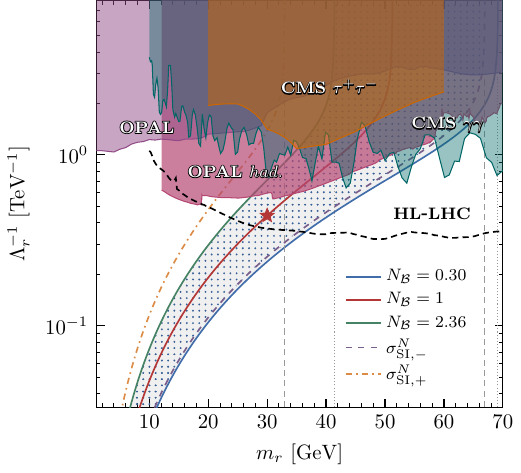}
\caption{
Radion parameter space for $m_{\chi}=1\,\mathrm{TeV}$ and $\delta=340\,\mathrm{keV}$.
Gray shading denotes the LZ event-count support, and the dotted region marks the additional LZ SI selection.
Solid contours indicate fixed event counts, while colored dashed contours show the LZ SI $\mathcal{O}_{1}^{s}$ boundaries.
Vertical dotted and dashed lines mark the corresponding endpoints.
The shaded regions indicate the OPAL and CMS exclusions, while the black dashed curve shows the projected HL-LHC diphoton sensitivity.
The star marks the reference point in Eq.~\eqref{eq:ref_point_0}.
}
\label{fig:rad_limit_collider}
\end{figure}

We now keep $(m_{\chi},\delta)$ fixed and examine how these conditions map onto the radion parameter space as $m_{r}$ is varied.
Combining Eq.~\eqref{eq:joint_con_1} with
Eqs.~\eqref{eq:brr}--\eqref{eq:bsm}, the relic-abundance condition along a fixed $N_{\mathcal{B}}$ contour becomes
\begin{equation}
 \frac{m_{\chi}^{2}\Lambda_{r}^{4}}
 {48\pi\mathcal{K}_{\mathcal{B}}^{2}}
 \frac{
 \sqrt{1-\kappa_{r\chi}^{2}}
 \left(9-8\kappa_{r\chi}^{2}
 +2\kappa_{r\chi}^{4}\right)}
 {\left(2-\kappa_{r\chi}^{2}\right)^{4}}
 N_{\mathcal{B}}^{2}
 +
 \frac{\Lambda_{r}^{2}\Gamma_{\rm SM}(2m_{\chi})}
 {4m_{\chi}\mathcal{K}_{\mathcal{B}}
 \left(4-\kappa_{r\chi}^{2}\right)^{2}}
 N_{\mathcal{B}}
 =
 b_{\star}(m_{\chi}).
 \label{eq:relic-surface-count}
\end{equation}
For each fixed $N_{\mathcal{B}}$,
Eq.~\eqref{eq:relic-surface-count} defines a solution branch in the $(m_{r},\Lambda_{r}^{-1})$ plane.
In Fig.~\ref{fig:rad_limit_collider}, we show the central branch $N_{\mathcal{B}}=1$ together with the region spanned by the likelihood
support in Eq.~\eqref{eq:count-band}.
For a given $\Lambda_{r}$, larger $N_{\mathcal{B}}$ requires smaller $m_{r}$.
Conversely, for a given $m_{r}$, increasing $N_{\mathcal{B}}$ shifts the solution toward larger $\Lambda_{r}^{-1}$.
The $N_{\mathcal{B}}=2.36$, $1$, $0.30$ branches terminate at $m_{r}\simeq41.38$, $51.28$, $69.17\,\mathrm{GeV}$,
respectively.
Each $N_{\mathcal{B}}$ branch has its own domain of existence, so the termination of the upper count branch does not remove the central or lower count solutions.
The reference point lies well within the event count support.

The two LZ SI $\mathcal{O}_{1}^{s}$ boundaries can be mapped onto the same relic abundance surface.
Using Eq.~\eqref{eq:sigma0}, the lower and upper cross section
boundaries in Eq.~\eqref{eq:lz-si-constraint} correspond to
\begin{equation}
 \mathcal{S}_{\pm}(m_{r})
 \equiv
 \left.
 \left(
 \frac{c_{12}^{S}}{\Lambda_{r}^{2}}
 \right)^{2}
 \right|_{\sigma_{N}^{0}
 =\sigma_{\rm SI,\pm}^{N}}
 =
 \frac{\pi m_{r}^{4}}
 {\mu_{\chi N}^{2}m_{\chi}^{2}m_{N}^{2}f_{N}^{2}(0)}
 \sigma_{\rm SI,\pm}^{N}(\delta).
 \label{eq:si-coupling-boundaries}
\end{equation}
Through Eq.~\eqref{eq:count-definition}, the same boundaries can also
be expressed as the corresponding event counts,
\begin{equation}
 N_{\mathcal{B},\pm}^{\rm SI}(m_{r})
 =
 \mathcal{K}_{\mathcal{B}}\,
 \mathcal{S}_{\pm}(m_{r}).
 \label{eq:si-count-boundaries}
\end{equation}
Substituting Eq.~\eqref{eq:si-coupling-boundaries} into the relic condition gives
\begin{equation}
 \frac{m_{\chi}^{2}\Lambda_{r}^{4}}
 {48\pi}
 \frac{
 \sqrt{1-\kappa_{r\chi}^{2}}
 \left(9-8\kappa_{r\chi}^{2}
 +2\kappa_{r\chi}^{4}\right)}
 {\left(2-\kappa_{r\chi}^{2}\right)^{4}}
 \mathcal{S}_{\pm}^{2}
 +
 \frac{\Lambda_{r}^{2}\Gamma_{\rm SM}(2m_{\chi})}
 {4m_{\chi}
 \left(4-\kappa_{r\chi}^{2}\right)^{2}}
 \mathcal{S}_{\pm}
 =
 b_{\star}(m_{\chi}),
 \label{eq:relic-surface-si}
\end{equation}
which therefore defines the two dashed curves in Fig.~\ref{fig:rad_limit_collider}.
$\sigma_{\rm SI,+}^{N}$ therefore gives the upper dashed curve in $\Lambda_{r}^{-1}$, while $\sigma_{\rm SI,-}^{N}$ gives the lower one.
Note that, unlike the solid curves, the dashed curves do not correspond to fixed values of $N_{\mathcal{B}}$.
Their event counts in Eq.~\eqref{eq:si-count-boundaries} vary with $m_{r}$.

In particular, the two branches also have finite domains of existence.
Since the first term in Eq.~\eqref{eq:relic-surface-si} is positive,
a positive finite solution requires
\begin{equation}
 \frac{\Lambda_{r}^{2}\Gamma_{\rm SM}(2m_{\chi})}
 {4m_{\chi}
 \left(4-\kappa_{r\chi}^{2}\right)^{2}}
 \mathcal{S}_{\pm}(m_{r})
 <
 b_{\star}(m_{\chi}).
 \label{eq:si-branch-existence}
\end{equation}
Since $\mathcal{S}_{+}>\mathcal{S}_{-}$, the $\sigma_{\rm SI,+}^{N}$ branch has the smaller domain of existence in
$m_{r}$ and terminates first, at $m_{r}\simeq32.95\,\mathrm{GeV}$.
Beyond this point, the upper LZ SI boundary no longer intersects the relic solution branch, while
$\sigma_{\rm SI,-}^{N}$ continues to delimit the region.
The termination of the $\sigma_{\rm SI,-}^{N}$ branch therefore sets the upper limit
\begin{equation}
m_{r} \lesssim 66.93\,\mathrm{GeV},
\end{equation}
for solutions compatible with the published LZ SI $\mathcal{O}_{1}^{s}$ band on this fixed
$m_{\chi}=1\,\mathrm{TeV}$, $\delta=340\,\mathrm{keV}$ slice.

\vspace{1cm}

\vspace{1cm}

Fig.~\ref{fig:rad_limit_collider} also compares the joint LZ event-count and relic-abundance solutions with existing collider constraints.
The decay mode independent and hadronic OPAL Higgsstrahlung searches, together with the CMS $\tau^{+}\tau^{-}$ and $\gamma\gamma$ searches, exclude the regions above their respective 95\% CL boundaries.
The hadronic OPAL constraint strengthens the LEP exclusion, while the CMS diphoton search provides the strongest constraint over part of the mass range.
The two CMS reinterpretations include the phenomenological gluonic QCD correction $K_{g}$ in production and decay, as detailed below.
These exclusions remove part of the joint solution region, while the reference point in Eq.~\eqref{eq:ref_point_0} remains allowed.
We also show the projected median expected 95\% CL sensitivity of the HL-LHC diphoton search at $\sqrt{s}=14\,\mathrm{TeV}$
with $3\,\mathrm{ab}^{-1}$.
At $m_{r}=30.3\,\mathrm{GeV}$, the corrected projected sensitivity reaches $\left(\Lambda_{r}^{-1}\right)_{\rm HL}^{95,\rm exp} \simeq0.370\,\mathrm{TeV}^{-1}$,
below the reference value $\Lambda_{r}^{-1}\simeq0.441\,\mathrm{TeV}^{-1}$ at $m_{r}=30\,\mathrm{GeV}$.
This suggests that the HL-LHC could probe the reference region under the stated projection assumptions.
The collider constraints and HL-LHC projection are detailed in Appendices~\ref{app:opal-higgsstrahlung}--\ref{app:hllhc-diphoton}.

\subsection{Solar capture and IceCube constraints}
\label{sec:solar_icecube}

Solar capture provides a complementary probe of the radion portal through secondary neutrinos from DM annihilation in the Sun.
The thermal Higgsino study of Ref.~\cite{Pospelov:2026ewn} obtains a model-dependent splitting bound of $\delta\gtrsim566\,\mathrm{keV}$ from IceCube data. 
This bound cannot be directly applied to the radion portal because of the different capture,
annihilation, and neutrino production mechanisms.

\begin{figure}[t]
\centering
\includegraphics[width=0.4\textwidth]{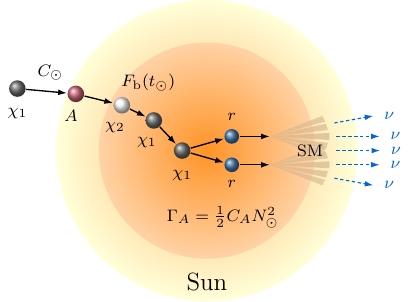}
\caption{
Solar capture, conversion, and annihilation within the thermal-core prescription.
Endothermic scattering $\chi_{1}A\to\chi_{2}A$ captures halo DM, while subsequent exothermic down-scattering can
re-eject the outgoing $\chi_{1}$ or leave it gravitationally bound.
The retained population is assumed to thermalize in the solar core, where Sommerfeld-enhanced $\chi_{1}\chi_{1}\to rr$
annihilation produces radions that decay into SM particles and yield secondary neutrinos.
}
\label{fig:solar-capture-schematic}
\end{figure}

We first calculate capture, orbital re-ejection, and Sommerfeld-enhanced annihilation at the reference point~\eqref{eq:ref_point_0}.
We then scan $(m_{r},\Lambda_{r}^{-1})$ at fixed $(m_{\chi},\delta)=(1\,\mathrm{TeV},340\,\mathrm{keV})$,
imposing the Born-level relic abundance condition.
The resulting non-equilibrium annihilation rates are compared with rate limits reconstructed from the observed
ten-year IceCube $b\bar{b}$ and $\tau^{+}\tau^{-}$ results~\cite{IceCube:2025fcu}, without assuming
capture-annihilation equilibrium in our model.
Fig.~\ref{fig:solar-capture-schematic} illustrates the corresponding physical processes in the Sun.

\vspace{1cm}

\vspace{1cm}

An incident halo particle with asymptotic speed $u$ is accelerated by the solar gravitational potential to
\begin{equation}
 w^{2}=u^{2}+v_{{\rm esc},\odot}^{2}(R)
 \label{eq:solar_esc}
\end{equation}
at solar radius $R$, where $v_{{\rm esc},\odot}(R)$ is the local solar escape speed. 
Capture requires scattering to remove sufficient orbital energy for the outgoing DM particle to become gravitationally bound.
For the ground state $\chi_{1}$-dominated halo assumed above, the relevant process is the radion-mediated transition
$\chi_{1}A\to\chi_{2}A$. 
Its inelastic threshold and recoil kinematics are given by Eqs.~\eqref{eq:star} and \eqref{eq:recoiledges}, evaluated at the local incident speed $w$.

For a fixed recoil energy $E_{R}$, the inelastic scattering is
kinematically allowed only when
\begin{equation}
 w
 \geq
 v_{\min,A}(E_{R}),
 \label{eq:solar-scattering-condition}
\end{equation}
where $v_{\min,A}(E_{R})$ is given in Eq.~\eqref{eq:vmin}. In
addition, energy conservation implies that the scattering removes
$E_{R}+\delta$ from the incident DM kinetic energy. Gravitational
capture therefore requires
\begin{equation}
E_{R}+\delta \geq \frac{1}{2}m_{\chi}u^{2}.
\label{eq:solar_bind_1}
\end{equation}
The solar capture calculation uses the radion-mediated nuclear differential cross section in Eq.~\eqref{eq:dsigma}, but replaces the LZ velocity convolution in Eq.~\eqref{eq:rate} by the solar velocity distribution
and nuclear density profile, together with the gravitational acceleration and binding condition above. 
The corresponding optically thin capture rate for stationary solar nuclei is
\begin{equation}
C_{\odot}^{(0)}
=
\frac{\rho_{\chi}}{m_{\chi}}
\sum_{A} 
\int_{0}^{R_{\odot}}  4\pi R^{2}dR\,n_{A}(R)
\int_{0}^{\infty}dE_{R}
\int_{u_{\min,A}(E_{R},R)}^{u_{\max}(E_{R})} du\, \frac{f_{\odot}(u)}{u} w^{2} \frac{d\sigma_{12,A}(w)}{dE_{R}},
\label{eq:solar_cap_0}
\end{equation}
where
\begin{align}
u_{\min,A}(E_{R},R)
&= \sqrt{\max \left[v_{\min,A}^{2}(E_{R}) - v_{{\rm esc},\odot}^{2}(R), 0\right]},
\\
u_{\max}(E_{R})
&= \min \left[u_{\rm halo}^{\max}, \sqrt{\frac{2(E_{R}+\delta)}{m_{\chi}}} \right].
\label{eq:solar-speed-window}
\end{align}
The lower limit follows from the inelastic condition $w\geq v_{\min,A}(E_{R})$, while the upper limit combines the gravitational binding condition~\eqref{eq:solar_bind_1} with the finite support $u\leq u_{\rm halo}^{\max}$ of the solar frame halo distribution.
A recoil energy contributes only when $u_{\min,A}(E_{R},R)\leq u_{\max}(E_{R})$.
The normalized speed distribution $f_{\odot}(u)$ follows from the halo model of Section~\ref{sec:recoil-halo} transformed to the solar frame using
\begin{equation}
\v_{\odot} = \v_{\rm LSR} + \v_{\rm pec},
\end{equation}
so that, for the truncated halo adopted above,
\begin{equation}
u_{\rm halo}^{\max} = v_{\rm esc,gal}+v_{\odot},
\end{equation}
with $v_{\rm esc,gal}=544\,\mathrm{km/s}$ denoting the Galactic halo escape speed.
Because the LZ recoil rate and solar capture rate depend on the same transition coupling $c_{12}^{S}/\Lambda_{r}^{2}$, the general solar capture rate can be written as
\begin{equation}
C_{\odot} = \mathcal{K}_{\odot} \left(\frac{c_{12}^{S}}{\Lambda_{r}^{2}}\right)^{2},
\label{eq:solar-count-matching}
\end{equation}
where $\mathcal{K}_{\odot}$ contains the solar composition, velocity
distribution, gravitational capture, and nuclear response as in Eq.~\eqref{eq:solar_cap_0}. 
No free coupling is introduced in the solar capture calculation.
Using the BS05(AGS,OP) solar profile~\cite{Bahcall:2004pz}, the AGS05 abundances~\cite{Asplund:2004eu}, and the same nuclear response and finite radion propagator as in the LZ calculation, the reference point gives
\begin{equation}
C_{\odot}^{(0)} \simeq 4.71\times10^{21}\,\mathrm{s}^{-1},
\qquad
C_{\odot}^{(1)} \simeq 4.75\times10^{21}\,\mathrm{s}^{-1}.
\label{eq:solar_cap_ref}
\end{equation}
$C_{\odot}^{(0)}$ is obtained for stationary nuclei, while $C_{\odot}^{(1)}$ includes thermal nuclear motion explicitly formulated in Appendix~\ref{app:solar-thermal}. 
Thermal nuclear motion therefore changes the total capture rate by only $0.9\%$.
Iron dominates the capture rate, contributing about $89.4\%$. 
The central optical depth is $0.030$, supporting the single-scattering approximation.

The initial endothermic scattering $\chi_{1}A\to\chi_{2}A$ captures the excited state $\chi_{2}$ on a gravitationally bound,
nonthermal orbit in the Sun.
During subsequent passages through the solar medium, $\chi_{2}$ can undergo exothermic down-scattering
$\chi_{2}A\to\chi_{1}A$, releasing the excitation energy $\delta$.
For a freshly captured particle, this energy release may overcome its gravitational binding energy and re-eject the outgoing
$\chi_{1}$ from the Sun~\cite{Nussinov:2009ft,Menon:2009qj,Blennow:2018xwu}.
Otherwise, $\chi_{1}$ remains bound and contributes to the retained population.
We calculate the re-ejection probability and conversion delays by following the orbital evolution of captured particles until
their first down-scattering, as detailed in Appendix~\ref{app:solar-thermal}.

Let $\tau_{21}$ denote the time from initial capture to the first down-scattering, and $\epsilon_{\rm orb}'$ the specific
orbital energy of the outgoing $\chi_{1}$.
We define the cumulative fractions
\begin{align}
F_{\rm b}(t)
&\equiv
P\!\left(
\tau_{21}\leq t,\,
\epsilon_{\rm orb}'<0
\mid \text{initial capture}
\right),
\nonumber\\
P_{\text{re-ej}}(t)
&\equiv
P\!\left(
\tau_{21}\leq t,\,
\epsilon_{\rm orb}'\geq 0
\mid \text{initial capture}
\right),
\label{eq:solar-native-retention}
\end{align}
where $F_{\rm b}(t)$ counts conversions into bound $\chi_{1}$, while $P_{\text{re-ej}}(t)$ counts re-ejections, both occurring
by time $t$.
At the reference point, the orbital calculation gives $F_{\rm b}(t_{\odot})\simeq0.904$ and $P_{\text{re-ej}}(t_{\odot})\simeq0.096$.
For the retained population, we adopt the thermal-core prescription.
Assuming a constant initial capture rate, the effective source is
\begin{equation}
C_{\rm core}(t) = C_{\odot}^{(1)}F_{\rm b}(t),
\label{eq:solar-core-source}
\end{equation}
which accounts for both re-ejection and conversion delays.

We approximate the retained DM as thermalized in the solar core, with
\begin{equation}
T_{\chi}\simeq T_{\odot,c}=1.3\,\mathrm{keV},
\qquad
V_{\rm eff}\simeq2.07\times10^{26}\,\mathrm{cm}^{3},
\label{eq:solar-core-reference}
\end{equation}
where $V_{\rm eff}$ is the effective volume of the retained DM distribution at $m_{\chi}=1\,\mathrm{TeV}$\footnote{
Incomplete thermalization can produce a more extended DM distribution, increasing $V_{\rm eff}$ and suppressing the annihilation rate~\cite{Peter:2009mk,Widmark:2017yvd,Blennow:2018xwu}. Thus, the thermal-core prescription can overestimate the signal relative to incomplete relaxation.}.
Let $N_{1,2}(t)$ denote the numbers of retained $\chi_{1,2}$ particles, and define
\begin{equation}
N_{\odot}(t)=N_{1}(t)+N_{2}(t),
\qquad
f_{2}(t)=\frac{N_{2}(t)}{N_{\odot}(t)},
\qquad
N_{1}(t)=[1-f_{2}(t)]N_{\odot}(t).
\label{eq:solar-populations}
\end{equation}
Since the retained particles enter the thermal core as $\chi_{1}$ and thermal excitation is strongly suppressed for $\delta\gg T_{\odot,c}$, we adopt $f_{2}\simeq0$ in the annihilation calculation, as discussed in Appendix~\ref{app:solar-thermalization}.

For the retained thermal-core population, the total annihilation event rate is
\begin{equation}
\Gamma_{A}
=\frac{1}{2}C_{11}N_{1}^{2}+C_{12}N_{1}N_{2}+\frac{1}{2}C_{22}N_{2}^{2},
\qquad
C_{ij}=\frac{\langle\sigma_{ij}v_{\rm rel}\rangle_{\odot}}{V_{\rm eff}}.
\label{eq:solar-two-state-annihilation}
\end{equation}
At the Born-level, the coefficients in Eqs.~\eqref{eq:brr} and \eqref{eq:b12} give
\begin{equation}
\begin{aligned}
b_{\odot}(f_{2})&=[(1-f_{2})^{2}+f_{2}^{2}]b_{rr}
+2f_{2}(1-f_{2})b_{12}^{\rm SM},\\
\langle\sigma v_{\rm rel}\rangle_{\odot}^{\rm Born}
&=\frac{6T_{\odot,c}}{m_{\chi}}b_{\odot}(f_{2}),
\qquad
C_{A}^{\rm Born}=\frac{\langle\sigma v_{\rm rel}\rangle_{\odot}^{\rm Born}}{V_{\rm eff}}.
\end{aligned}
\label{eq:solar-born-coefficient}
\end{equation}
At the reference point,
\begin{equation}
b_{rr}=2.94\times10^{-8}\,\mathrm{GeV}^{-2},
\qquad
b_{12}^{\rm SM}=3.90\times10^{-9}\,\mathrm{GeV}^{-2}.
\end{equation}
For $f_{2}\simeq0$, the corresponding Born values are
\begin{equation}
\langle\sigma v_{\rm rel}\rangle_{\odot}^{\rm Born}
\simeq2.68\times10^{-33}\,\mathrm{cm}^{3}\mathrm{s}^{-1},
\qquad
C_{A}^{\rm Born}\simeq1.29\times10^{-59}\,\mathrm{s}^{-1}.
\label{eq:solar-reference-coefficient}
\end{equation}
The low relative velocities of DM in the solar core suppress the annihilation rate through the $p$-wave factor
$6T_{\odot,c}/m_{\chi}=7.8\times10^{-9}$, but can also make repeated radion exchange important, giving rise to the Sommerfeld enhancement.
In the normalized $(\chi_{1}\chi_{1},\chi_{2}\chi_{2})$ two-particle basis, the dominant scalar transition induces
the coupled-channel potential
\begin{equation}
V_{S}(r)=
\begin{bmatrix}
0 & -\dfrac{\alpha_{\chi}e^{-m_{r}r}}{r}\\[6pt]
-\dfrac{\alpha_{\chi}e^{-m_{r}r}}{r} & 2\delta
\end{bmatrix},
\qquad
\alpha_{\chi}=\frac{(g_{12}^{S})^{2}}{4\pi},
\label{eq:solar-sommerfeld-potential}
\end{equation}
with $\alpha_{\chi}\simeq0.158$ at the reference point $m_{r}=30\,\mathrm{GeV}$.
The off-diagonal interaction mixes the incoming $\chi_{1}\chi_{1}$ pair with the heavier $\chi_{2}\chi_{2}$
channel, which lies $2\delta$ above the lighter threshold.
Although this channel is kinematically closed at typical solar thermal velocities, its virtual contribution can
enhance the short-distance annihilation amplitude.
We obtain the $p$-wave Sommerfeld factor $S_{1}(v_{\rm rel})$ by solving the coupled Schr\"odinger equation with
the centrifugal barrier~\cite{Cassel:2009wt,Slatyer:2009vg}.
The radial solutions, annihilation projection, and thermal averaging are detailed in Appendix~\ref{app:solar-thermalization}.
Since the Born annihilation rate is proportional to $v_{\rm rel}^{2}$, the thermally averaged enhancement is
\begin{equation}
\overline{S}_{1}
\equiv
\frac{\langle v_{\rm rel}^{2}S_{1}(v_{\rm rel})\rangle_{\odot}}
{\langle v_{\rm rel}^{2}\rangle_{\odot}}.
\label{eq:solar-sommerfeld-coefficient}
\end{equation}
We obtain $\overline{S}_{1}\simeq71.0$ at the reference point, and it runs along the relic-constrained radion scan.
The enhancement partially compensates for the $p$-wave suppression, yielding
\begin{equation}
\langle\sigma v_{\rm rel}\rangle_{\odot}
\simeq1.90\times10^{-31}\,\mathrm{cm}^{3}\mathrm{s}^{-1},
\qquad
C_{A}=\overline{S}_{1}C_{A}^{\rm Born}
\simeq9.17\times10^{-58}\,\mathrm{s}^{-1}.
\end{equation}
The Sommerfeld enhancement can also affect thermal freeze-out, although the substantially higher relative velocities generally
reduce its importance compared with the solar core.

With the time-dependent source in Eq.~\eqref{eq:solar-core-source},
the retained population evolves according to
\begin{equation}
\frac{dN_{\odot}}{dt}
=C_{\rm core}(t)-2\Gamma_{A}(t)
=C_{\rm core}(t)-C_{A}N_{\odot}^{2}(t),
\qquad
\Gamma_{A}(t)=\frac{C_{A}}{2}N_{\odot}^{2}(t).
\label{eq:solar_evo_1}
\end{equation}
Solving this equation with the source obtained from the orbital Monte Carlo, we find $N_{\odot}(t_{\odot})\simeq6.03\times10^{38}$
and $\Gamma_{A}(t_{\odot})\simeq 1.67\times10^{20}\,\mathrm{s}^{-1}$ at the reference point.
The ratio of the annihilation depletion rate to the effective core source is
\begin{equation}
\frac{2\Gamma_{A}(t_{\odot})}
{C_{\rm core}(t_{\odot})}
\simeq0.0777,
\label{eq:solar_ann_cap_ratio_1}
\end{equation}
indicating that capture-annihilation equilibrium has not been reached.
For comparison, in the thermal Higgsino scenario of Ref.~\cite{Pospelov:2026ewn}, electroweak interactions
provide sufficient orbital cooling and an unsuppressed low-velocity annihilation rate, allowing capture-annihilation equilibrium.

At the reference point, annihilation is dominated by $\chi_{1}\chi_{1}\to rr$.
Since $\Gamma_{r}/m_{r}\ll1$, radion production and decay factorize in the narrow width approximation.
Including the gluon trace anomaly, the top loop~\cite{Cheung:2000rw}, and the $K_{g}$ correction specified in
Appendix~\ref{app:opal-higgsstrahlung},
the radion branching fractions at $m_{r}=30\,\mathrm{GeV}$ are
\begin{equation}
\mathcal{B}(r\to b\bar{b},gg,\tau^{+}\tau^{-},c\bar{c})
\simeq(0.373,0.565,0.0366,0.0250).
\label{eq:rad_br_1}
\end{equation}

\begin{figure}[!htbp]
\centering
\includegraphics[width=0.60\textwidth]{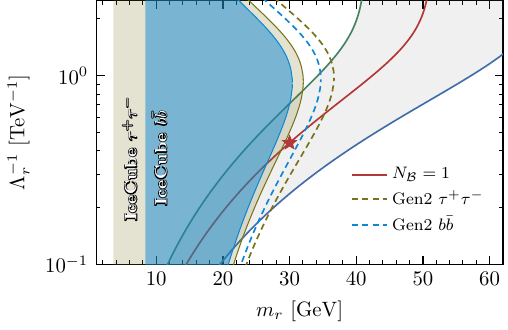}
\caption{
Energetic-daughter proxy comparison with reconstructed IceCube direct-channel rate scales for
$m_{\chi}=1\,\mathrm{TeV}$ and $\delta=340\,\mathrm{keV}$.
Gray shading denotes the LZ event-count support.
The red curve represents the central $N_{\mathcal{B}}=1$ solution, while the star marks the reference point.
Blue (olive) shading marks where the $b\bar{b}$ ($\tau^{+}\tau^{-}$) proxy exceeds the corresponding
reconstructed IceCube direct-channel rate scale.
Dashed curves show the projected IceCube-Gen2 sensitivities.
These comparisons are not model-specific exclusions or calibrated Gen2 sensitivities.
}
\label{fig:rad_limit_IC}
\end{figure}

IceCube derives upper limits on the SI DM-proton scattering cross section from solar neutrino searches for assumed direct annihilation channels $X$, under the capture-annihilation equilibrium assumption~\cite{IceCube:2025fcu}.
Reversing this conversion gives the corresponding annihilation-rate template limits,
\begin{equation}
\Gamma_{X}^{\mathrm{lim,IC}}(m_{\chi})
=\frac{1}{2}C_{\odot}^{\mathrm{IC,SI}}
\!\left(m_{\chi},\sigma_{\chi p}^{\mathrm{SI,lim},X}(m_{\chi})\right),
\qquad X=b\bar{b},\tau^{+}\tau^{-},
\label{eq:icecube_rate_conversion}
\end{equation}
where $C_{\odot}^{\mathrm{IC,SI}}$ denotes the conventional elastic capture rate used in the experimental conversion.
Using the observed $90\%$ CL IceCube SI limits at $m_{\chi}=1\,\mathrm{TeV}$ and the $W^{+}W^{-}$ annihilation-rate normalization inferred in Ref.~\cite{Pospelov:2026ewn}, we estimate
\begin{equation}
\left(\Gamma_{b\bar{b}}^{\mathrm{lim,IC}}, \Gamma_{\tau\tau}^{\mathrm{lim,IC}} \right)
\simeq \left(1.68\times10^{20}, 8.63\times10^{18} \right)\,\mathrm{s}^{-1}.
\label{eq:icecube_template_limits}
\end{equation}
These reconstructed limits can be compared with our non-equilibrium annihilation prediction without imposing
capture-annihilation equilibrium on the radion scenario.
For an approximate comparison with the IceCube templates for direct DM annihilation of $m_{\chi}=1\,\mathrm{TeV}$,
we focus on the energetic radion decay products.
At the reference point, each radion carries $E_{r}\simeq m_{\chi}=1\,\mathrm{TeV}$.
For $r\to X$, with $X=b\bar{b},\tau^{+}\tau^{-}$, only one daughter has energy above $500\,\mathrm{GeV}$.
The rate of selected daughters is thus $2\mathcal{B}(r\to X)\Gamma_{A}(t_{\odot})$.
Expressing this rate in equivalent pair units, we define
\begin{equation}
\Gamma_{X}(t_{\odot})
\equiv\mathcal{B}(r\to X)\Gamma_{A}(t_{\odot}),
\qquad
R_{X}^{\mathrm{IC}}
\equiv\frac{\Gamma_{X}(t_{\odot})}
{\Gamma_{X}^{\mathrm{lim,IC}}},
\label{eq:solar_onepair_proxy}
\end{equation}
At the reference point, the annihilation rates and ratios are
\begin{equation}
\begin{aligned}
\left(\Gamma_{b\bar{b}}(t_{\odot}), \Gamma_{\tau\tau}(t_{\odot}) \right)
&\simeq
\left(6.23\times10^{19}, 6.11\times10^{18}\right)\,\mathrm{s}^{-1},
\\
\left(R_{b\bar{b}}^{\mathrm{IC}}, R_{\tau\tau}^{\mathrm{IC}}\right)
&\simeq(0.371,0.707).
\end{aligned}
\label{eq:ice_ratio}
\end{equation}
At the reference point, the energetic-daughter proxy rates in both channels are comparable in magnitude to the
corresponding rate scales reconstructed from IceCube's direct-annihilation templates.
We extend this comparison across the $(m_{r},\Lambda_{r}^{-1})$ plane along the Born-level relic-abundance solution.
The corresponding IceCube template exclusion regions are shown in Fig.~\ref{fig:rad_limit_IC}.
These template-based comparisons are not model-specific exclusions, which require the radion-cascade neutrino
spectra and detector response.

\vspace{0.5cm}

\vspace{0.5cm}

The sizable solar annihilation rate motivates future neutrino telescopes, including the IceCube Upgrade~\cite{IceCube:2026rbh},
IceCube-Gen2~\cite{IceCube-Gen2:2020qha}, and TRIDENT~\cite{Wang:2026kka}.
In particular, IceCube-Gen2 aims to expand the instrumented optical volume to about $8\,\mathrm{km}^{3}$, with high-energy point-source projections indicating a sensitivity improvement of approximately a factor of $5$~\cite{IceCube-Gen2:2020qha,Kowalski:2025lr}, which gives the projected limits and corresponding signal-to-limit ratios at the reference point
\begin{equation}
\begin{aligned}
\left(\Gamma_{b\bar{b}}^{\mathrm{lim,Gen2}},
\Gamma_{\tau\tau}^{\mathrm{lim,Gen2}}\right)
&\simeq
\left(3.36\times10^{19},1.73\times10^{18}\right)\,\mathrm{s}^{-1},
\\
\left(R_{b\bar{b}}^{\mathrm{Gen2}}, R_{\tau\tau}^{\mathrm{Gen2}}\right)
&\simeq(1.85,3.53).
\end{aligned}
\label{eq:gen2_future_limit}
\end{equation}
The IceCube-Gen2 dashed curves in Fig.~\ref{fig:rad_limit_IC} show the corresponding projected boundaries,
while a model-specific sensitivity requires the radion-cascade neutrino spectra and detector response.

\section{Conclusions}
\label{sec:conclusions}

We have investigated a radion portal for pseudo-Majorana DM as a particle interpretation of LZ230616.
In the pseudo-Majorana scenario considered here, the radion generates an unsuppressed scalar transition of order $m_{\chi}/\Lambda_{r}$, which is absent in the well-known pseudo-Dirac regime.
The diagonal scalar couplings are instead suppressed relative to the transition by the small ratio $\delta/m_{\chi}$.
This interaction structure therefore realizes a scalar endothermic interaction for the high recoil signal without introducing an independent transition operator or an unsuppressed elastic channel.

We combined the accepted LZ event count with the thermal relic abundance directly in the parameter space
$(m_{\chi},\delta,m_{r},\Lambda_{r},c_{12}^{S})$. 
These two observables probe different combinations of the same radion interactions and therefore determine the coupling and interaction scale rather than introducing an independent direct detection normalization. 
The joint analysis admits a reference solution with $m_{\chi}=1\,\mathrm{TeV}$, $\delta=340\,\mathrm{keV}$, and
$m_{r}=30\,\mathrm{GeV}$, for which $\Lambda_{r}\simeq2.268\,\mathrm{TeV}$ and $c_{12}^{S}\simeq3.195$.
It satisfies the central event count and relic abundance conditions, while lying within the published LZ SI $\mathcal{O}_{1}^{s}$ band.
The reference point also remains compatible with the current OPAL and CMS collider constraints, while the CMS diphoton search
provides the strongest exclusion over much of the radion mass range.
Our projected HL-LHC diphoton sensitivity suggests that future measurements could probe the reference region under the adopted
statistical assumptions.
Solar capture provides a complementary probe through orbital re-ejection and Sommerfeld-enhanced annihilation of the
retained population.
Our reference solution remains out of capture-annihilation equilibrium,
while its energetic-daughter proxy rates are comparable in magnitude to the reconstructed IceCube $b\bar{b}$ and
$\tau^{+}\tau^{-}$ direct-channel rate scales.
The radion parameter scan and an illustrative IceCube-Gen2 rescaling suggest potential future sensitivity,
although model-specific constraints require the radion-cascade neutrino spectra and detector response.

Our results connect the inelastic interpretation of LZ230616 to the thermal relic abundance and solar neutrino signals
through a common pseudo-Majorana mass and radion interaction structure.
At low energies, the scalar transition governs endothermic nuclear scattering, producing kinematically filtered
high-recoil events and enabling solar capture.
In the early universe, radion-mediated annihilation determines the thermal relic abundance, while the retained
solar population can produce secondary neutrinos.
These complementary observables therefore probe the same dark sector across distinct physical regimes.
The radion portal also suggests a possible geometric origin in warped extra dimensions.
Appendix~\ref{app:orbifold-fixed-metric} presents a stabilized $\mathcal{S}^{1}/\mathbb{Z}_{2}$ construction in which
two bulk Dirac fermions with opposite parities supply complementary dark-sector light modes.
Metric and GW scalar fluctuations contribute to the effective radion couplings through the bulk fermion profiles and interactions. The interplay of metric and GW scalar fluctuations can generate opposite-sign radion couplings to two nearly degenerate Majorana states.
An illustrative 5D benchmark realizes the required scalar transition.
The radiative stability of this alignment, effects of heavier KK modes, and complete matching to the
single-radion phenomenological analysis remain to be established.

\appendix

\section{Takagi rotation and radion couplings}
\label{sec:takagi}

We compare the pseudo-Dirac and pseudo-Majorana limits of the same real Weyl-field Lagrangian. For two left-handed fields
$\psi=(\eta,\xi)^{\mathrm{T}}$, the mass and radion interactions are
\begin{equation}
 \mathcal{L}_{\rm mass}+\mathcal{L}_{r\chi}
 =-\frac12\psi^{\text{T}} M\psi-\frac r2\psi^{\text{T}} G\psi+\mathrm{h.c.},
\label{eq:L_M_D_rad_0}
\end{equation}
with
\begin{equation}
 M=\left[\begin{matrix}\mu_{\eta}&M_{D}\\M_{D}&\mu_{\xi}\end{matrix}\right],
 \qquad
 G=\frac1{\Lambda_{r}}
 \left[\begin{matrix}
  c_{\eta}\mu_{\eta}&c_{D} M_{D}\\
  c_{D} M_{D}&c_{\xi}\mu_{\xi}
 \end{matrix}\right].
 \label{eq:MG}
\end{equation}
We take all parameters to be real, with $M_{D},\Lambda_{r}>0$, and use
\begin{equation}
 \Sigma=\mu_{\eta}+\mu_{\xi},\qquad
 \Delta=\mu_{\eta}-\mu_{\xi},\qquad
 D=\sqrt{4M_{D}^{2}+\Delta^{2}}.
 \label{eq:takagi-abbreviations}
\end{equation}
The orthogonal matrix
\begin{equation}
 O=\begin{bmatrix}s_{\theta}&c_{\theta}\\-c_{\theta}&s_{\theta}\end{bmatrix},
 \qquad
 s_{\theta}=\sqrt{\frac{D-\Delta}{2D}},\qquad
 c_{\theta}=\sqrt{\frac{D+\Delta}{2D}}
 \label{eq:takagi-U}
\end{equation}
gives
\begin{equation}
 O^{\mathrm{T}}MO=\operatorname{diag}(\lambda_{-},\lambda_{+}),
 \qquad
 \lambda_{\mp}=\frac{\Sigma\mp D}{2},\qquad
 \lambda_{-}\lambda_{+}=\det M=\mu_{\eta}\mu_{\xi}-M_{D}^{2}.
 \label{eq:takagi-positive-check}
\end{equation}
For $\Sigma>0$, $\det M<0$ gives one negative signed eigenvalue,
whereas $\det M>0$ gives two positive eigenvalues. 
The boundary $\det M=0$ has one massless eigenstate. 
These signs distinguish the Takagi branches in the chosen real basis. 
The pseudo-Dirac and pseudo-Majorana regimes additionally require the respective mass
hierarchies specified below. In particular, the determinant sign alone is not a substitute for these hierarchies.

For $\psi=U\zeta$, where $U$ makes both masses positive, define
$\widehat{G}=U^{\mathrm{T}}GU$ and the four-component Majorana fields
$\chi_{i}=(\zeta_{i\alpha},\zeta_{i}^{\dagger\dot\alpha})^{\mathrm{T}}$.
The two-component to four-component identity~\cite{Dreiner:2008tw}
then gives
\begin{equation}
 \mathcal{L}_{r\chi}
 =-\frac{r}{2}\sum_{i,j=1}^{2}
 \bar\chi_{i}\left(g^{S}_{ij}+ig^{A}_{ij}\gamma^{5}\right)\chi_{j},
 \qquad
 g^{S}=\operatorname{Re}\widehat{G},\quad
 g^{A}=-\operatorname{Im}\widehat{G}.
 \label{eq:takagi-four-component}
\end{equation}

\subsection{Pseudo-Dirac regime and pseudoscalar transition}
\label{sec:takagi-pseudodirac}

In the pseudo-Dirac regime,
$\mu_{0}\equiv\max(|\mu_{\eta}|,|\mu_{\xi}|)\ll M_{D}$ and
$\Sigma>0$, so $\det M<0$ and $\lambda_{-}<0<\lambda_{+}$.
The positive physical masses, their average, and their splitting are
\begin{equation}
 m_{\chi_{1}}=\frac{D-\Sigma}{2},\qquad
 m_{\chi_{2}}=\frac{D+\Sigma}{2},\qquad
 m_{\chi}=\frac{D}{2},\qquad
 \delta=\Sigma.
 \label{eq:takagi-physical-masses}
\end{equation}
The Takagi matrix is
$U=O\operatorname{diag}(-i,1)$, giving
\begin{equation}
 \widehat{G}
 =\begin{bmatrix}-H_{11}&-iH_{12}\\-iH_{12}&H_{22}\end{bmatrix},
 \qquad H=O^{\mathrm{T}}GO.
 \label{eq:takagiresponse}
\end{equation}
Thus the diagonal couplings are scalar and the transition is
pseudoscalar, with exact entries
\begin{equation}
 \begin{aligned}
 g^{S}_{11}
 &=\frac{1}{2\Lambda_{r}}
 \left[\frac{4c_{D}M_{D}^{2}+\Delta(c_{\eta}\mu_{\eta}-c_{\xi}\mu_{\xi})}{D}
 -(c_{\eta}\mu_{\eta}+c_{\xi}\mu_{\xi})\right],\\[3pt]
 g^{S}_{22}
 &=\frac{1}{2\Lambda_{r}}
 \left[\frac{4c_{D}M_{D}^{2}+\Delta(c_{\eta}\mu_{\eta}-c_{\xi}\mu_{\xi})}{D}
 +(c_{\eta}\mu_{\eta}+c_{\xi}\mu_{\xi})\right],\\[3pt]
 g^{A}_{12}=g^{A}_{21}
 &=\frac{M_{D}}{\Lambda_{r}D}
 \left[(c_{\eta}-c_{D})\mu_{\eta}
 -(c_{\xi}-c_{D})\mu_{\xi}\right],
 \end{aligned}
 \label{eq:takagiresponseentries}
\end{equation}
while $g^{S}_{12}=g^{S}_{21}=g^{A}_{11}=g^{A}_{22}=0$. The complete
interaction therefore reads
\begin{equation}
 \mathcal{L}_{r\chi}
 =-\frac{r}{2}\left(g^{S}_{11}\bar\chi_{1}\chi_{1}
 +g^{S}_{22}\bar\chi_{2}\chi_{2}\right)
 -r\,g^{A}_{12}\bar\chi_{2}i\gamma^{5}\chi_{1}.
 \label{eq:takagi-full-interaction}
\end{equation}
The off-diagonal terms combine as written because the pseudoscalar Majorana bilinear is symmetric under interchange of the two fields.

For fixed coefficients $c_{D},c_{\eta},c_{\xi}$ in the pseudo-Dirac limit,
\begin{equation}
 \begin{aligned}
 g^{S}_{11},\ g^{S}_{22}
 &=\frac{c_{D}M_{D}}{\Lambda_{r}}
 +\mathcal{O}\!\left(\frac{\mu_{0}}{\Lambda_{r}}\right),\\
 g^{A}_{12}
 &=\frac{(c_{\eta}-c_{D})\mu_{\eta}
 -(c_{\xi}-c_{D})\mu_{\xi}}{2\Lambda_{r}}
 +\mathcal{O}\!\left(\frac{\mu_{0}^{3}}{M_{D}^{2}\Lambda_{r}}\right).
 \end{aligned}
 \label{eq:gP12expand}
\end{equation}
The transition is therefore smaller than the diagonal couplings for $\mathcal{O}(1)$ coefficients and unsuppressed $c_{D}$. The estimate $g^{A}_{12}=\mathcal{O}(\delta/\Lambda_{r})$ additionally requires
$\mu_{\eta},\mu_{\xi}=\mathcal{O}(\delta)$ without cancellations.
In the aligned case $c_{\eta}=c_{\xi}=c_{D}=c$, the matrix $G=cM/\Lambda_{r}$ is diagonal in the positive-mass basis, and
$g^{A}=0$.

The pseudoscalar nature of the transition provides a further nonrelativistic suppression. 
For the on-shell transition $\chi_{1}(p_{1})\to\chi_{2}(p_{2})$,
the spin-averaged ratio of the pseudoscalar and scalar bilinears is
\begin{equation}
 \frac{\overline{|\bar u_{2}i\gamma^{5}u_{1}|^{2}}}
 {\overline{|\bar u_{2}u_{1}|^{2}}}
 =\frac{\delta^{2}-t}{4m_{\chi}^{2}-t}
 \simeq\frac{Q^{2}+\delta^{2}}{4m_{\chi}^{2}},
 \qquad t=(p_{2}-p_{1})^{2}\simeq-Q^{2}.
 \label{eq:takagi-exact-bilinear-ratio}
\end{equation}
The first equality is exact for on-shell spinors. At $Q\simeq246\,\mathrm{MeV}$, $m_{\chi}=1\,\mathrm{TeV}$, and
$\delta=340\,\mathrm{keV}$, this factor is approximately $1.5\times10^{-8}$. For $\mu_{\eta},\mu_{\xi}=\mathcal{O}(\delta)$,
the additional squared-coupling suppression is of order
$(\delta/m_{\chi})^{2}\simeq1.2\times10^{-13}$, yielding a combined suppression of order $10^{-21}$. 
By contrast, the diagonal scalar channel has neither the pseudoscalar bilinear suppression nor an
inelastic threshold. The pseudo-Dirac hierarchy therefore does not realize the dominant scalar transition of the main analysis.

\subsection{Pseudo-Majorana regime and scalar transition}
\label{sec:takagi-majorana}

For the pseudo-Majorana hierarchy, we take
\begin{equation}
 \mu_{\eta}\simeq\mu_{\xi}\simeq\mu_{\chi}\gg M_{D},
 \qquad \mu_{\chi}=\frac{\mu_{\eta}+\mu_{\xi}}{2}>0.
 \label{eq:pseudo-majorana-hierarchy}
\end{equation}
In the positive mass branch,
\begin{equation}
 \det M=\mu_{\eta}\mu_{\xi}-M_{D}^{2}>0,
 \qquad \Sigma>0,
 \label{eq:majorana-branch}
\end{equation}
we have
\begin{equation}
 m_{\chi_{1}}=\frac{\Sigma-D}{2},\qquad
 m_{\chi_{2}}=\frac{\Sigma+D}{2},\qquad
 m_{\chi}=\frac{\Sigma}{2}=\mu_{\chi},\qquad
 \delta=D.
 \label{eq:majorana-masses}
\end{equation}
Both real eigenvalues are positive, so $U=O$, with no relative Takagi rephasing. Consequently,
\begin{equation}
 \widehat{G}=O^{\mathrm{T}}GO,\qquad
 g^{S}=\widehat{G},\qquad g^{A}=0.
 \label{eq:majorana-dictionary}
\end{equation}
The exact scalar entries are
\begin{equation}
 \begin{aligned}
 g^{S}_{11}
 &=\frac{1}{\Lambda_{r}}\left(
 c_{\eta}\mu_{\eta}s_{\theta}^{2}
 +c_{\xi}\mu_{\xi}c_{\theta}^{2}
 -2c_{D}M_{D}s_{\theta}c_{\theta}\right),\\[3pt]
 g^{S}_{22}
 &=\frac{1}{\Lambda_{r}}\left(
 c_{\eta}\mu_{\eta}c_{\theta}^{2}
 +c_{\xi}\mu_{\xi}s_{\theta}^{2}
 +2c_{D}M_{D}s_{\theta}c_{\theta}\right),\\[3pt]
 g^{S}_{12}=g^{S}_{21}
 &=\frac{1}{\Lambda_{r}}\left[
 (c_{\eta}\mu_{\eta}-c_{\xi}\mu_{\xi})s_{\theta}c_{\theta}
 +c_{D}M_{D}(s_{\theta}^{2}-c_{\theta}^{2})\right].
 \end{aligned}
 \label{eq:majorana-rotated-entries}
\end{equation}
In particular,
\begin{equation}
 g^{S}_{12}=\frac{M_{D}}{\Lambda_{r}D}
 \left[(c_{\eta}-c_{D})\mu_{\eta}
 -(c_{\xi}-c_{D})\mu_{\xi}\right],
 \label{eq:maj_trans-exact}
\end{equation}
which has the same algebraic numerator as the pseudo-Dirac
transition but is now scalar. A scalar transition of order
$m_{\chi}/\Lambda_{r}$ is possible because $M_{D}/D$ need not be
small. This requires nonuniversal radion couplings: for
$c_{\eta}=c_{\xi}=c_{D}=c$, the relation $G=cM/\Lambda_{r}$ instead
gives $g^{S}=c\operatorname{diag}(m_{\chi_{1}},m_{\chi_{2}})/\Lambda_{r}$.

The physical average mass and splitting do not fix the three real mass-matrix parameters. Define
\begin{equation}
 x=\frac{\Delta}{\delta},\qquad |x|<1,
 \label{eq:majorana-x}
\end{equation}
so that
\begin{equation}
 \mu_{\eta}=m_{\chi}+\frac{x\delta}{2},\qquad
 \mu_{\xi}=m_{\chi}-\frac{x\delta}{2},\qquad
 M_{D}=\frac{\delta}{2}\sqrt{1-x^{2}}.
 \label{eq:majorana-parametrization}
\end{equation}
These relations give $m_{\chi_{1,2}}=m_{\chi}\mp\delta/2$ and require $m_{\chi}>\delta/2$. The parameter $x$ controls the
Majorana-mass difference at fixed spectrum, with $x=0$ giving
$\mu_{\eta}=\mu_{\xi}$.
Introduce
\begin{equation}
 c_{+}=\frac{c_{\eta}+c_{\xi}}{2},\qquad
 c_{-}=\frac{c_{\eta}-c_{\xi}}{2}.
 \label{eq:majorana-response-combinations}
\end{equation}
Using $s_{\theta}^{2}=(1-x)/2$,
$c_{\theta}^{2}=(1+x)/2$, and
$2s_{\theta}c_{\theta}=\sqrt{1-x^{2}}$, the exact entries become
\begin{align}
 g^{S}_{11}
 &=\frac{1}{\Lambda_{r}}
 \left\{m_{\chi}(c_{+}-c_{-}x)
 +\frac{\delta}{2}\left[c_{-}x-c_{+}x^{2}-c_{D}(1-x^{2})\right]\right\},
 \label{eq:majorana-g11}\\
 g^{S}_{22}
 &=\frac{1}{\Lambda_{r}}
 \left\{m_{\chi}(c_{+}+c_{-}x)
 +\frac{\delta}{2}\left[c_{-}x+c_{+}x^{2}+c_{D}(1-x^{2})\right]\right\},
 \label{eq:majorana-g22}\\
 g^{S}_{12}
 &=\frac{\sqrt{1-x^{2}}}{\Lambda_{r}}
 \left[c_{-}m_{\chi}+\frac{\delta x}{2}(c_{+}-c_{D})\right].
 \label{eq:majorana-g12}
\end{align}
The sum and difference of the diagonal entries are
\begin{align}
 g^{S}_{11}+g^{S}_{22}
 &=\frac{1}{\Lambda_{r}}
 \left(2m_{\chi}c_{+}+\delta c_{-}x\right),\\
 g^{S}_{22}-g^{S}_{11}
 &=\frac{1}{\Lambda_{r}}
 \left\{2m_{\chi}c_{-}x
 +\delta\left[c_{+}x^{2}+c_{D}(1-x^{2})\right]\right\}.
 \label{eq:majorana-diagonal-combinations}
\end{align}
For $\delta\ll m_{\chi}$, both diagonal entries are suppressed relative to the transition if
$|c_{+}|\ll|c_{-}|$ and $|x|\ll1$, with $c_{-}\neq0$
and $c_{D}/c_{-}$ not parametrically enhanced. In this region,
\begin{equation}
 g^{S}_{12}\simeq\frac{m_{\chi}}{\Lambda_{r}}
 c_{-}\sqrt{1-x^{2}},
 \label{eq:majorana-transition-leading}
\end{equation}
while the leading terms of the diagonal combinations are
$2m_{\chi}c_{+}/\Lambda_{r}$ and
$2m_{\chi}c_{-}x/\Lambda_{r}$, respectively.
Thus the required alignment amounts to
$|\mu_{\eta}-\mu_{\xi}|\ll\delta$ and
$c_{\eta}\simeq-c_{\xi}$, stronger than approximate equality of
the Majorana masses at the scale $m_{\chi}$.
These conditions are realized by the conventional radion in the stabilized orbifold of Appendix~\ref{app:orbifold-fixed-metric}, where the bulk profiles and GW interactions generate opposite effective responses to positive Majorana masses~\cite{Azatov:2008vm}.

At the central configuration,
\begin{equation}
 \mu_{\eta}=\mu_{\xi}=\mu_{\chi}=m_{\chi},\qquad
 M_{D}=\frac{\delta}{2},\qquad
 c_{\eta}=-c_{\xi}\equiv c_{M}\neq0,
 \label{eq:majorana-central-configuration}
\end{equation}
we have $x=c_{+}=0$ and $c_{-}=c_{M}$, giving
\begin{equation}
 g^{S}=\frac{1}{\Lambda_{r}}
 \begin{bmatrix}
 -c_{D}\delta/2&c_{M}m_{\chi}\\
 c_{M}m_{\chi}&c_{D}\delta/2
 \end{bmatrix},\qquad g^{A}=0.
 \label{eq:majorana-central-couplings}
\end{equation}
The resulting interaction is
\begin{equation}
 \mathcal{L}_{r\chi}
 =-r\,\frac{c_{M}m_{\chi}}{\Lambda_{r}}\bar\chi_{2}\chi_{1}
 +\frac{c_{D}\delta}{4\Lambda_{r}}r
 \left(\bar\chi_{1}\chi_{1}-\bar\chi_{2}\chi_{2}\right),
 \label{eq:majorana-central-interaction}
\end{equation}
where the symmetric Majorana transition bilinear cancels the factor
$1/2$ in the off-diagonal sum, without an additional Hermitian
conjugate~\cite{Dreiner:2008tw}. The relative scalar couplings satisfy
\begin{equation}
 \left|\frac{g^{S}_{11}}{g^{S}_{12}}\right|
 =\left|\frac{g^{S}_{22}}{g^{S}_{12}}\right|
 =\left|\frac{c_{D}}{c_{M}}\right|\frac{\delta}{2m_{\chi}}.
 \label{eq:majorana-central-ratio}
\end{equation}
For the reference point in Eq.~\eqref{eq:ref_point_0}, this ratio is $5.3\times10^{-8}$, establishing the suppression of the elastic
scalar vertices while retaining the dominant transition.

\section{Opposite radion responses from the \texorpdfstring{$\mathcal{S}^{1}/\mathbb{Z}_{2}$}{S1/Z2} orbifold}
\label{app:orbifold-fixed-metric}

The conventional radion coupling to the energy-momentum trace suggests mass-proportional responses with a common sign. 
For bulk fermions, however, the effective radion couplings also depend on the KK profiles and their interactions with the stabilizing GW scalar. 
We show that these contributions can generate 4D opposite-sign radion responses in
Eq.~\eqref{eq:main-central-configuration} after dimensional reduction.

We consider a stabilized $\mathcal{S}^{1}/\mathbb{Z}_{2}$ geometry, with the covering circle $-\pi R<y\leq \pi R$ and the orbifold identification $y\sim-y$.
The physical interval is $0\leq y\leq \pi R$, with a UV brane at $y=0$ and an IR brane at $y=\pi R\equiv-\pi R$, where the SM is localized.
In the scalar metric gauge, the metric and stabilizing field are written as~\cite{Csaki:2000zn,Csaki:2007ns,Chacko:2014pqa}
\begin{equation}
 \begin{aligned}
 ds^{2}&= g_{MN} dx^{M} dx^{N}
 =e^{-2[A(y)+F(x,y)]}
 \eta_{\mu\nu}dx^{\mu}dx^{\nu}
 -[1+2F(x,y)]^{2}dy^{2},\\
 F(x,y)&=\frac{r(x)}{\Lambda_{r}}\mathcal{F}(y),
 \qquad
 \phi(x,y)=\phi_{0}(y)
 +\frac{r(x)}{\Lambda_{r}}\mathcal{H}(y).
 \end{aligned}
 \label{eq:fixed-orbifold-radion-metric}
\end{equation}
Here $A(y)=A(|y|)$ is the background warp factor and $r(x)$ is the canonically normalized 4D radion.
The functions $\mathcal{F}(y)$ and $\mathcal{H}(y)$ describe its orbifold-even metric and stabilizer profiles, respectively.
We adopt $\mathcal{F}(\pi R)=-1$, consistently with the radion convention in Eq.~\eqref{eq:rsm}.
The corresponding curved 5D Dirac matrices are given by the Clifford algebra
\begin{equation}
\left\{\Gamma^{M}, \Gamma^{N} \right\}=2g^{MN},
\qquad
M,N=0,1,2,3,5.
\end{equation}
Choosing a diagonal vielbein, we define
\begin{equation}
\begin{aligned}
\Gamma^{\mu}
&=e^{A(y)+F(x,y)}\gamma^{\mu}
= e^{A(y)} \left[1+F(x,y) \right]\gamma^{\mu} + \mathcal{O}(F^{2}),
\\
\Gamma^{5}
&=\frac{i\gamma^{5}}{1+2F(x,y)}
= i\left[1-2F(x,y) \right]\gamma^{5} + \mathcal{O}(F^{2}),
\end{aligned}
\label{eq:fixed-orbifold-gamma}
\end{equation}
where $\gamma^{\mu}$ are the 4D Dirac matrices and
$\gamma^{5}=i\gamma^{0}\gamma^{1}\gamma^{2}\gamma^{3}$.
Setting $r=0$ recovers the background configuration.
In the weak-backreaction limit, $A(|y|)\simeq k|y|$, where $k$ is the AdS curvature scale, and the background metric reduces to the Randall-Sundrum form.

For stabilization, we introduce an orbifold-even GW scalar
\begin{equation}
\phi(x,-y)=\phi(x,y),
\end{equation}
with the following superpotential, bulk potential, and brane potentials~\cite{Csaki:2000zn}:
\begin{equation}
\begin{aligned}
 W(\phi)&=\frac{6k}{\kappa_{5}^{2}}-u\phi^{2},
 &
 V_{5}(\phi)&=
 \frac{1}{8}\left(\frac{\partial W}{\partial\phi}\right)^{2}
 -\frac{\kappa_{5}^{2}}{6}W(\phi)^{2},
 \nonumber\\
 \lambda_{\mathrm{UV}}(\phi)&=W(\phi)
 +\frac{\gamma_{\mathrm{UV}}}{2}(\phi-U)^{2},
 &
 \lambda_{\mathrm{IR}}(\phi)&=-W(\phi)
 +\frac{\gamma_{\mathrm{IR}}}{2}(\phi-V)^{2}.
\end{aligned}
\label{eq:fixed-orbifold-stabilization}
\end{equation}
$\kappa_{5}$ is the 5D gravitational coupling,
while $\gamma_{\mathrm{UV}}$ and $\gamma_{\mathrm{IR}}$ characterize the stiffness of the brane potentials.
The first-order background equations,
\begin{equation}
 \phi_{0}'(y)=\frac{1}{2}
 \left.\frac{\partial W}{\partial\phi}\right|_{\phi=\phi_{0}}
 =-u\phi_{0},
 \qquad
 A'(y)=\frac{\kappa_{5}^{2}}{6}W(\phi_{0}),
\end{equation}
together with the stiff-boundary conditions
$\phi_{0}(0)=U$ and $\phi_{0}(\pi R)=V$, yield
\begin{equation}
\phi_{0}(y)=Ue^{-u|y|},
\qquad
A(y)=k|y|+\frac{\kappa_{5}^{2}U^{2}}{12}
\left(e^{-2u|y|}-1\right),
\qquad
u \pi R =\ln\frac{U}{V}.
\label{eq:fixed-orbifold-background}
\end{equation}

The radion describes fluctuations of the proper distance between the UV and IR branes.
The SM is localized on the IR brane, whose induced metric is
$\gamma_{\mu\nu}(x)=e^{-2[A(\pi R)+F(x,\pi R)]}\eta_{\mu\nu}$.
Varying the SM action with respect to this metric gives
\begin{equation}
\delta S_{\mathrm{SM}}
=-\frac12\int d^{4}x\,\sqrt{-\gamma}\,
T_{\mathrm{SM}}^{\mu\nu}\delta\gamma_{\mu\nu},
\end{equation}
where $\delta\gamma_{\mu\nu}=-2F(x,\pi R)\gamma_{\mu\nu}$ at linear order.
$\mathcal{F}(\pi R)=-1$ reproduces the radion coupling in Eq.~\eqref{eq:rsm}, with $\Lambda_{r}$ fixed by canonical normalization.
To determine the radion couplings to bulk fermions, we obtain its metric and stabilizer profiles from the linearized
Einstein--scalar equations~\cite{Csaki:2000zn,Csaki:2007ns}.
The two profiles are related by
\begin{equation}
\phi_{0}'\mathcal{H}
=\frac{3}{\kappa_{5}^{2}}\left(\mathcal{F}'-2A'\mathcal{F}\right),
\label{eq:fixed-orbifold-scalar-constraint}
\end{equation}
and the bulk equation for $\mathcal{F}(y)$ reads
\begin{equation}
\mathcal{F}''-2\left(A'+\frac{\phi_{0}''}{\phi_{0}'}\right)\mathcal{F}'
+\left(-4A''+4A'\frac{\phi_{0}''}{\phi_{0}'}
+m_{r}^{2}e^{2A}\right)\mathcal{F}=0.
\label{eq:fixed-orbifold-scalar-equations}
\end{equation}
In the stiff-boundary limit, $\mathcal{H}(0)=\mathcal{H}(\pi R)=0$, giving
$(\mathcal{F}'-2A'\mathcal{F})|_{0,\pi R}=0$.
In the weak-backreaction regime, $\kappa_{5}^{2}\phi_{0}^{2}(y)\ll1$,
the leading radion metric profile is
\begin{equation}
\mathcal{F}(y)\simeq-e^{2k(|y|-\pi R)}.
\label{eq:fixed-orbifold-radion-profile}
\end{equation}
To obtain the leading nonvanishing stabilizer fluctuation,
we retain the first backreaction correction to the metric profile. Defining $z=k|y|$, $\ell=k \pi R$, $\epsilon=u/k$, and
$\mathcal{F}(y)=-e^{2[A(y)-A(\pi R)]}P(z)$,
Eq.~\eqref{eq:fixed-orbifold-scalar-constraint} gives
\begin{equation}
\mathcal{H}(y)=
\frac{3e^{2[A(y)-A(\pi R)]}}{\kappa_{5}^{2}\epsilon\phi_{0}(y)}
\frac{dP}{dz},
\qquad
\frac{dP}{dz}\bigg|_{z=0,\ell}=0,
\quad P(\ell)=1.
\label{eq:fixed-orbifold-stabilizer-exact}
\end{equation}
At the first nonvanishing order in the weak-backreaction expansion, the stabilizer profile is~\cite{Csaki:2000zn}
\begin{equation}
\mathcal{H}(y)\simeq
\epsilon e^{-2\ell}\phi_{0}(y)\Bigg[
e^{2z}-1
-\frac{e^{2\ell}-1}{e^{(4+2\epsilon)\ell}-1}
\left(e^{(4+2\epsilon)z}-1\right)
\Bigg],
\label{eq:fixed-orbifold-stabilizer-profile}
\end{equation}
which vanishes at both branes.

We introduce two independent 5D Dirac spinors $\Psi_{\xi}$ and $\Psi_{\eta}$ with opposite orbifold parities,
\begin{equation}
\Psi_{\xi}(x,-y)=-\gamma^{5}\Psi_{\xi}(x,y),\qquad
\Psi_{\eta}(x,-y)=+\gamma^{5}\Psi_{\eta}(x,y).
\label{eq:fixed-orbifold-fermion-parities}
\end{equation}
Thus $\Psi_{\xi L}$ and $\Psi_{\eta R}$ are even, while $\Psi_{\xi R}$ and $\Psi_{\eta L}$ are odd and vanish on both branes, but may be nonzero in the bulk~\cite{Leng:2020ofk}. This geometric parity is distinct from the exact dark $\mathbb{Z}_{2}$ of Section~\ref{sec:mass-sector}. The GW scalar contributes linearly to the bulk mass functions~\cite{Chacko:2014pqa,vonHarling:2016vhf}. The fermion action is
\begin{equation}
\begin{aligned}
S_{\Psi}=\int d^{5}x\,\sqrt{|g|}\bigg\{&
\sum_{j=\xi,\eta}\left[\frac{i}{2}\overline{\Psi}_{j}\Gamma^{M}\overleftrightarrow{D}_{M}\Psi_{j}
-\frac{1}{2}\left(\mathcal{M}_{M,j}(\phi)\overline{\Psi}_{j}\Psi_{j}^{c}+\mathrm{h.c.}\right)\right]\\
&-\left(\mathcal{M}_{D}(\phi)\overline{\Psi}_{\eta}\Psi_{\xi}+\mathrm{h.c.}\right)\bigg\}.
\end{aligned}
\label{eq:fixed-orbifold-fermion-action}
\end{equation}
For the bulk Majorana operators, we adopt the charge-conjugation prescription of Huber and Shafi~\cite{Huber:2003sf}, also employed by Iyer and Vempati~\cite{Iyer:2013hca},
with $\Psi_{j}^{c}\equiv C_{5}\gamma^{0}\Psi_{j}^{*}$ and $C_{5}=C_{4}=-i\gamma^{2}\gamma^{0}$. 
For simplicity, we omit the conventional diagonal bulk Dirac mass terms proportional to
$\bar{\Psi}_{j}\Psi_{j}$, and take
\begin{equation}
\mathcal{M}_{M,j}(\phi)=m_{M,j}+\lambda_{j}\phi,
\qquad
\mathcal{M}_{D}(\phi)=m_{D}+\lambda_{D}\phi.
\label{eq:fixed-orbifold-bulk-masses}
\end{equation}
Varying the fermion action gives the coupled 5D equations of motion,
\begin{equation}
\begin{aligned}
i\Gamma^{M}D_{M}\Psi_{\xi}
-\mathcal{M}_{M,\xi}(\phi)\Psi_{\xi}^{c}
-\mathcal{M}_{D}(\phi)\Psi_{\eta}&=0,\\
i\Gamma^{M}D_{M}\Psi_{\eta}
-\mathcal{M}_{M,\eta}(\phi)\Psi_{\eta}^{c}
-\mathcal{M}_{D}(\phi)\Psi_{\xi}&=0.
\end{aligned}
\label{eq:fixed-orbifold-fermion-5D-eom}
\end{equation}
At $r(x)=0$, the covariant Dirac operator reduces to
\begin{equation}
i\Gamma^{M}D_{M}
=i e^{A(y)}\gamma^{\mu}\partial_{\mu}
-\gamma^{5}\left(\partial_{y}-2A'(y)\right).
\label{eq:fixed-orbifold-fermion-Dirac-operator}
\end{equation}
The 5D Dirac spinor field is expanded through the KK decomposition as
\begin{equation}
\Psi_{j}(x,y)=\frac{e^{2A(y)}}{\sqrt{2\pi R}}\sum_{n}\left[f_{j L}^{(n)}(y)\chi_{j L}^{(n)}(x)+f_{j R}^{(n)}(y)\chi_{j R}^{(n)}(x)\right],
\label{eq:fixed-orbifold-fermion-kk-decomposition}
\end{equation}
where $\chi_{j}^{(n)c}=\chi_{j}^{(n)}$ are the 4D KK modes satisfying the Dirac equation $(i\gamma^{\mu}\partial_{\mu}-m_{j,n})\chi_{j}^{(n)}=0$. 
A bulk Majorana mass makes the KK spectrum non-vector-like, with 4D Majorana mass eigenstates~\cite{Huber:2003sf}. 
The equations of motion for the profile functions become~\cite{Huber:2003sf,Iyer:2013hca}
\begin{equation}
\begin{aligned}
\partial_{y}f_{jL}^{(n)}
&=-\left[m_{j,n}e^{A(y)}-m_{M,j}
-\lambda_{j}\phi_{0}(y)\right]f_{jR}^{(n)},\\
\partial_{y}f_{jR}^{(n)}
&=\left[m_{j,n}e^{A(y)}-m_{M,j}
-\lambda_{j}\phi_{0}(y)\right]f_{jL}^{(n)}.
\end{aligned}
\label{eq:fixed-orbifold-fermion-eom}
\end{equation}
The general profile solutions are
\begin{equation}
\begin{aligned}
f_{\xi L}^{(n)}(y)&=N_{\Psi}\cos\theta_{\xi,n}(y),&
f_{\xi R}^{(n)}(y)&=N_{\Psi}\sin\theta_{\xi,n}(y),\\
f_{\eta L}^{(n)}(y)&=-N_{\Psi}\sin\theta_{\eta,n}(y),&
f_{\eta R}^{(n)}(y)&=N_{\Psi}\cos\theta_{\eta,n}(y),
\end{aligned}
\label{eq:fixed-orbifold-fermion-profiles}
\end{equation}
with the phase function
\begin{equation}
\theta_{j,n}(y)
=m_{j,n}\int_{0}^{y}ds\,e^{A(s)}
-m_{M,j}y-\frac{\lambda_{j}U}{u}(1-e^{-uy}),
\qquad
y \in [0, \pi R],
\label{eq:fixed-orbifold-fermion-phase}
\end{equation}
extended to the full covering circle by $\theta_{j,n}(-y)=-\theta_{j,n}(y)$.
The orbifold parity conditions therefore impose
\begin{equation}
\begin{aligned}
f_{\xi L}^{(n)}(-y)&=f_{\xi L}^{(n)}(y),&f_{\xi R}^{(n)}(-y)&=-f_{\xi R}^{(n)}(y),\\
f_{\eta L}^{(n)}(-y)&=-f_{\eta L}^{(n)}(y),&f_{\eta R}^{(n)}(-y)&=f_{\eta R}^{(n)}(y),
\end{aligned}
\label{eq:fixed-orbifold-profile-parities}
\end{equation}
which force the odd profiles to vanish at $y=0,\pi R$, leading to
\begin{equation}
\theta_{j,n}(0)=0,
\qquad 
\theta_{j,n}(\pi R)=q_{j,n}\pi,\quad q_{j,n}\in\mathbb{Z}.
\end{equation}
The orthonormal condition takes
\begin{equation}
\frac{1}{2\pi R}\int_{-\pi R}^{\pi R}dy\,e^{A(y)}
\left[f_{jL}^{(n)*}f_{jL}^{(m)}+f_{jR}^{(n)*}f_{jR}^{(m)}\right]=\delta_{nm},
\label{eq:fixed-orbifold-fermion-normalization}
\end{equation}
which fixes
\begin{equation}
N_{\Psi}=\left[\frac{1}{2\pi R}\int_{-\pi R}^{\pi R}dy\,e^{A(y)}\right]^{-1/2}
\simeq\sqrt{\frac{k\pi R}{e^{k\pi R}-1}}.
\label{eq:fixed-orbifold-fermion-N}
\end{equation}
The opposite parities also ensure the cross-chiral orthogonality,
\begin{equation}
\frac{1}{2\pi R}\int_{-\pi R}^{\pi R}dy\,e^{A(y)}f_{jL}^{(n)*}(y)f_{jR}^{(m)}(y)=0.
\label{eq:fixed-orbifold-chiral-orthogonality}
\end{equation}
The 4D KK mass spectrum is given by
\begin{equation}
\begin{aligned}
m_{j,n}&= \frac{N_{\Psi}^{2}}{\pi R} 
\left|q_{j,n}\pi+m_{M,j}\pi R +\lambda_{j}\frac{U-V}{u}\right|
\\
&\simeq\frac{k}{e^{k \pi R}-1} \left|q_{j,n}\pi+m_{M,j}\pi R +\lambda_{j}\frac{U-V}{u}\right|.
\end{aligned}
\label{eq:fixed-orbifold-kk-spectrum}
\end{equation}

Taking $\zeta_{j}^{(n)}$ as the independent left-handed component of the KK mode $\chi_{j}^{(n)}$, dimensional reduction gives
\begin{equation}
\mathcal{L}_{\mathrm{mass}}^{(4)}
=-\sum_{n,m}\left[
\frac{\mu_{\eta,nm}}{2}\zeta_{\eta}^{(n)}\zeta_{\eta}^{(m)}
+\frac{\mu_{\xi,nm}}{2}\zeta_{\xi}^{(n)}\zeta_{\xi}^{(m)}
+M_{D,nm}\zeta_{\eta}^{(n)}\zeta_{\xi}^{(m)}
+\mathrm{h.c.}\right],
\label{eq:fixed-orbifold-reduced-mass}
\end{equation}
where $\mu_{\xi,nm}$ and $\mu_{\eta,nm}$ are the symmetric Majorana mass matrices, while $M_{D,nm}$ denotes the
Dirac mixing matrix:
\begin{align}
\mu_{j,nm}&=m_{j,n}\delta_{nm},
\label{eq:4D_m_Majo_0}\\
M_{D,nm}&=\frac{N_{\Psi}^{2}}{2\pi R}
\int_{-\pi R}^{\pi R}dy\,
(m_{D}+\lambda_{D}\phi_{0})
\cos(\theta_{\eta,n}+\theta_{\xi,m}).
\label{eq:4D_m_Dirac_0}
\end{align}

The radion interactions with the bulk fermions arise from the metric fluctuation and the direct coupling to the GW
scalar. We define the fermionic energy-momentum tensor by the conventional metric variation,
\begin{equation}
T_{MN}=\frac{2}{\sqrt{|g|}}\frac{\delta S_{\Psi}}{\delta g^{MN}},
\end{equation}
where the GW scalar and fermion fields are held fixed. 
For the bulk fermion action in Eq.~\eqref{eq:fixed-orbifold-fermion-action},
the energy-momentum tensor takes the form
\begin{equation}
T_{MN}
=\frac{i}{4}\sum_{j=\xi,\eta}
\overline{\Psi}_{j}
\left(
\Gamma_{M}\overleftrightarrow{D}_{N}
+\Gamma_{N}\overleftrightarrow{D}_{M}
\right)\Psi_{j}
-g_{MN}\mathcal{L}_{\Psi},
\end{equation}
where $\mathcal{L}_{\Psi}$ denotes the complete 5D fermion Lagrangian. Using the bulk fermion equations
of motion~\eqref{eq:fixed-orbifold-fermion-5D-eom}, the energy-momentum tensor reduces to
\begin{equation}
T_{MN}=\frac{i}{4}\sum_{j=\xi,\eta}
\overline{\Psi}_{j}
\left(
\Gamma_{M}\overleftrightarrow{D}_{N}
+\Gamma_{N}\overleftrightarrow{D}_{M}
\right)\Psi_{j}.
\end{equation}
The metric perturbation in Eq.~\eqref{eq:fixed-orbifold-radion-metric} gives
\begin{equation}
\delta g^{\mu\nu}=2F g^{\mu\nu}\big|_{r=0},
\qquad
\delta g^{55}=-4F g^{55}\big|_{r=0}.
\end{equation}
Consequently, the metric contribution to the linear radion interaction is~\cite{Csaki:2007ns}
\begin{equation}
\mathcal{L}_{r\zeta}^{(4)} \supset
\frac{r}{\Lambda_{r}}
\int_{-\pi R}^{\pi R}dy\,e^{-4A(y)}
\mathcal{F}(y)
\left(T_{\mu}{}^{\mu}-2T_{5}{}^{5}\right)_{r=0},
\label{eq:fixed-orbifold-radion-stress}
\end{equation}
with
\begin{equation}
\begin{aligned}
\left(T_{\mu}{}^{\mu}-2T_{5}{}^{5}\right)_{r=0}
=&\frac{3i}{2}\sum_{j=\xi,\eta} e^{A(y)}
\overline{\Psi}_{j}\gamma^{\mu}
\overleftrightarrow{\partial}_{\mu}\Psi_{j}
\\
&-\left[\sum_{j=\xi,\eta}
\mathcal{M}_{M,j}(\phi_{0})
\overline{\Psi}_{j}\Psi_{j}^{c}
+2\mathcal{M}_{D}(\phi_{0})
\overline{\Psi}_{\eta}\Psi_{\xi}
+\mathrm{h.c.}
\right].
\end{aligned}
\label{eq:fixed-orbifold-stress-explicit}
\end{equation}
In addition, the radion-induced GW fluctuation generates a direct contribution through the scalar-dependent bulk
fermion masses~\cite{Chacko:2014pqa},
\begin{equation}
\mathcal{L}_{r\zeta}^{(4)} \supset
-\frac{r}{\Lambda_{r}}
\int_{-\pi R}^{\pi R}dy\,e^{-4A(y)}
\mathcal{H}(y)
\left[
\sum_{j=\xi,\eta}\frac{\lambda_{j}}{2} \overline{\Psi}_{j}\Psi_{j}^{c}
+\lambda_{D} \overline{\Psi}_{\eta}\Psi_{\xi}
+\mathrm{h.c.}
\right]_{r=0}.
\label{eq:fixed-orbifold-radion-GW}
\end{equation}

Substituting the KK decomposition into Eqs.~\eqref{eq:fixed-orbifold-radion-stress} and \eqref{eq:fixed-orbifold-radion-GW}, the linear radion interactions take the form
\begin{equation}
\begin{aligned}
\mathcal{L}_{r\zeta}^{(4)}
&=\frac{r}{\Lambda_{r}}
\sum_{j=\xi,\eta}\sum_{n,m}
\frac{i}{2}K_{j,nm}
\zeta_{j}^{(n)\dagger}
\bar{\sigma}^{\mu}
\overleftrightarrow{\partial}_{\mu}
\zeta_{j}^{(m)}
\\
&\quad-\frac{r}{\Lambda_{r}}\sum_{n,m}
\bigg[
\frac{\mu_{\eta,nm}^{\prime}}{2}
\zeta_{\eta}^{(n)}\zeta_{\eta}^{(m)}
+\frac{\mu_{\xi,nm}^{\prime}}{2}
\zeta_{\xi}^{(n)}\zeta_{\xi}^{(m)}
+M_{D,nm}^{\prime}
\zeta_{\eta}^{(n)}\zeta_{\xi}^{(m)}
+\mathrm{h.c.}
\bigg].
\end{aligned}
\label{eq:fixed-orbifold-radion-interactions}
\end{equation}
$K_{j,nm}$ denotes the radion-induced kinetic coupling, while $\mu_{j,nm}^{\prime}$ and $M_{D,nm}^{\prime}$
denote the Majorana and Dirac couplings, respectively.
They are given by
\begin{align}
K_{j,nm}
&=\frac{3N_{\Psi}^{2}}{2\pi R}
\int_{-\pi R}^{\pi R}dy\,e^{A}\mathcal{F}
\cos(\theta_{j,n}-\theta_{j,m}),
\label{eq:fixed-orbifold-kinetic-response}\\
\mu_{j,nm}^{\prime}
&=\frac{N_{\Psi}^{2}}{2\pi R}
\int_{-\pi R}^{\pi R}dy\,
\left[
2\mathcal{F}\mathcal{M}_{M,j}(\phi_{0})
+\lambda_{j}\mathcal{H}
\right]
\cos(\theta_{j,n}-\theta_{j,m}),
\label{eq:fixed-orbifold-mirror-response}\\
M_{D,nm}^{\prime}
&=\frac{N_{\Psi}^{2}}{2\pi R}
\int_{-\pi R}^{\pi R}dy\,
\left[
2\mathcal{F}\mathcal{M}_{D}(\phi_{0})
+\lambda_{D}\mathcal{H}
\right]
\cos(\theta_{\eta,n}+\theta_{\xi,m}).
\label{eq:fixed-orbifold-dirac-response}
\end{align}
The radion-induced kinetic operators can be eliminated using the coupled 4D equations of motion derived from Eq.~\eqref{eq:fixed-orbifold-reduced-mass},
\begin{equation}
\begin{aligned}
i\bar{\sigma}^{\mu}\partial_{\mu}\zeta_{\eta}^{(n)}
&=m_{\eta,n}\zeta_{\eta}^{(n)\dagger}
+\sum_{m}M_{D,nm}\zeta_{\xi}^{(m)\dagger},\\
i\bar{\sigma}^{\mu}\partial_{\mu}\zeta_{\xi}^{(n)}
&=m_{\xi,n}\zeta_{\xi}^{(n)\dagger}
+\sum_{m}M_{D,mn}\zeta_{\eta}^{(m)\dagger}.
\end{aligned}
\end{equation}
Together with their Hermitian conjugates, these equations convert the kinetic radion interactions into mass bilinears. The resulting radion Lagrangian becomes
\begin{equation}
\mathcal{L}_{r\zeta}^{(4)}
=-\frac{r}{\Lambda_{r}}\sum_{n,m}\bigg[
\frac{\mathcal{G}_{\eta,nm}}{2}
\zeta_{\eta}^{(n)}\zeta_{\eta}^{(m)}
+\frac{\mathcal{G}_{\xi,nm}}{2}
\zeta_{\xi}^{(n)}\zeta_{\xi}^{(m)}
+\mathcal{G}_{D, nm}
\zeta_{\eta}^{(n)}\zeta_{\xi}^{(m)}
+\mathrm{h.c.}\bigg],
\end{equation}
where the effective scalar responses are
\begin{equation}
\begin{aligned}
\mathcal{G}_{j,nm}
&=\mu_{j,nm}^{\prime}
-\frac{m_{j,n}+m_{j,m}}{2}K_{j,nm},\\
\mathcal{G}_{D, nm}
&=M_{D,nm}^{\prime}
-\frac12\sum_{p}\left[
K_{\eta,np}M_{D,pm}
+M_{D,np}K_{\xi,pm}
\right].
\end{aligned}
\label{eq:fixed-orbifold-mass-response}
\end{equation}

We identify the two zero KK modes with the Weyl fields of our 4D effective model,
\begin{equation}
\xi\equiv\zeta_{\xi}^{(0)},
\qquad
\eta\equiv\zeta_{\eta}^{(0)}.
\label{eq:fixed-orbifold-weyl-matching}
\end{equation}
The mass and radion interactions in the two-state basis $\psi=(\eta,\xi)^{\mathrm{T}}$ take the form of
Eq.~\eqref{eq:L_M_D_rad_0}, with
\begin{equation}
\begin{aligned}
\mu_{\eta}&=m_{\eta,0},&
\mu_{\xi}&=m_{\xi,0},&
M_{D}&=M_{D,00},\\
c_{\eta}\mu_{\eta}&=\mathcal{G}_{\eta,00},&
c_{\xi}\mu_{\xi}&=\mathcal{G}_{\xi,00},&
c_{D}M_{D}&=\mathcal{G}_{D,00}.
\end{aligned}
\label{eq:fixed-orbifold-four-dimensional-matching}
\end{equation}
Since the lightest KK mode minimizes the absolute value in Eq.~\eqref{eq:fixed-orbifold-kk-spectrum}, the Majorana
mass difference satisfies
\begin{equation}
|\mu_{\eta}-\mu_{\xi}|
\leq\frac{N_{\Psi}^{2}}{2R}
\simeq\frac{\pi k}{2}e^{-k\pi R},
\label{eq:fixed-orbifold-mass-difference-bound}
\end{equation}
where the approximation assumes weak backreaction and $k\pi R\gg1$. This provides a general bound at the warped
IR scale, independently of the bulk fermion mass parameters.
Equations~\eqref{eq:fixed-orbifold-mirror-response}--\eqref{eq:fixed-orbifold-mass-response} show that the effective radion responses receive both geometric and GW-dependent contributions, whose competition can produce either sign even for positive Majorana masses. 
The sum of the Majorana radion responses is
\begin{equation}
\begin{aligned}
c_{\eta}\mu_{\eta}+c_{\xi}\mu_{\xi}
&=\frac{N_{\Psi}^{2}}{2\pi R}
\int_{-\pi R}^{\pi R}dy\,
\bigg[
2(m_{M,\eta}+m_{M,\xi})\mathcal{F}-3(\mu_{\eta}+\mu_{\xi})e^{A}\mathcal{F}\\
&\quad\quad\quad\quad\quad\quad\quad\ \ +(\lambda_{\eta}+\lambda_{\xi})
(2\mathcal{F}\phi_{0}+\mathcal{H})
\bigg]\\
&\simeq\mu_{\eta}+\mu_{\xi}
-\bigg[
m_{M,\eta}+m_{M,\xi}
+\frac{2(2+\epsilon)}{4+\epsilon}
(\lambda_{\eta}+\lambda_{\xi})V
\bigg]e^{-k\pi R}.
\end{aligned}
\label{eq:fixed-orbifold-general-response-sum}
\end{equation}
Using the lightest-mode mass bound, the cancellation of this sum requires
\begin{equation}
0\lesssim
\frac{m_{M,\eta}+m_{M,\xi}}{k}
+\frac{2(2+\epsilon)}{4+\epsilon}
\frac{(\lambda_{\eta}+\lambda_{\xi})V}{k}
\lesssim\pi.
\label{eq:fixed-orbifold-response-cancellation-bound}
\end{equation}

The target 4D configuration is a stronger Majorana degeneracy, requiring
\begin{equation}
\mu_{\eta}\simeq\mu_{\xi}\simeq m_{\chi},
\qquad
M_{D}\simeq\frac{\delta}{2},
\qquad
c_{\eta}\simeq-c_{\xi}\simeq c_{M},
\qquad
\delta=m_{\chi_{2}}-m_{\chi_{1}}>0.
\label{eq:fixed-orbifold-mirror-alignment}
\end{equation}
As an illustrative 5D realization, we consider 
\begin{equation}
m_{M,\eta}=m_{M,\xi}\equiv m_{M},
\qquad
\lambda_{\eta}=-\lambda_{\xi}\equiv\lambda.
\label{eq:fixed-orbifold-mirror-masses}
\end{equation}
The Majorana mass and the effective 4D coefficients therefore become
\begin{equation}
m_{\chi}
=\frac{N_{\Psi}^{2}}{\pi R}
\left[
m_{M}\pi R
+\frac{\pi}{2}(q_{\eta,0}+q_{\xi,0})
\right],
\qquad
c_{M}
=\frac{\lambda N_{\Psi}^{2}}{2\pi R\,m_{\chi}}
\int_{-\pi R}^{\pi R}dy\,(2\mathcal{F}\phi_{0}+\mathcal{H}),
\label{eq:fixed-orbifold-mirror-coefficients}
\end{equation}
which directly give the Majorana mass difference and the sum of their radion responses
\begin{equation}
\begin{aligned}
\mu_{\eta}-\mu_{\xi}
&=\frac{N_{\Psi}^{2}}{\pi R}
\left[
(q_{\eta,0}-q_{\xi,0})\pi + \frac{2\lambda(U-V)}{u}
\right],\\
c_{\eta}\mu_{\eta}+c_{\xi}\mu_{\xi}
&=\frac{N_{\Psi}^{2}}{2\pi R}
\bigg[
4m_{M}\int_{-\pi R}^{\pi R}dy\,\mathcal{F}
-3(m_{\eta,0}+m_{\xi,0})
\int_{-\pi R}^{\pi R}dy\,e^{A}\mathcal{F}
\bigg].
\end{aligned}
\end{equation}
In the weak backreaction limit, the degeneracy condition in Eq.~\eqref{eq:fixed-orbifold-mirror-alignment} therefore reduces to
\begin{equation}
\begin{aligned}
q_{\eta,0}-q_{\xi,0} &=-\frac{2\lambda(U-V)}{\pi u},\\
m_{M}e^{-k\pi R}&= m_{\chi}.
\end{aligned}
\end{equation}
For illustration, we take the radion coupling scale $\Lambda_{r}=2.268~\mathrm{TeV}$ and choose the following 5D parameters, with $k\simeq2.4\times10^{18}~\mathrm{GeV}$:
\begin{equation}
\begin{aligned}
\frac{u}{k}&\simeq-0.195,\quad
k\pi R\simeq35.5,\quad
\kappa_{5}^{2}V^{2}\simeq0.0235, \quad 
\frac{m_{M}}{k}\simeq1.093, \\
\frac{\lambda V}{k}&\simeq-3.678,\quad
\left(\frac{m_{D}}{k},
\frac{\lambda_{D}V}{k}\right)
\simeq(0.974,-0.613)\times10^{-6}.
\end{aligned}
\label{eq:fixed-orbifold-benchmark-input}
\end{equation}
The resulting radion mass and effective 4D parameters are
\begin{equation}
\begin{aligned}
m_{r}&\simeq30~\mathrm{GeV},\qquad
m_{\chi_{1,2}}\simeq1~\mathrm{TeV}\mp0.17~\mathrm{MeV},\\
M_{D}&\simeq0.17~\mathrm{MeV},\qquad
c_{\eta}\simeq-c_{\xi}\simeq3.19,\qquad
c_{D}\simeq1,\\
|g_{12}^{S}|&\simeq1.41,\qquad
\left|\frac{g_{11}^{S}}{g_{12}^{S}}\right|
\simeq\left|\frac{g_{22}^{S}}{g_{12}^{S}}\right|
\sim6\times10^{-8}.
\end{aligned}
\label{eq:fixed-orbifold-benchmark-result}
\end{equation}
This benchmark demonstrates the existence of opposite-sign radion responses in a stabilized warped construction.
The zero modes are separated from the higher KK modes by masses of order a TeV. 
The scalar transition remains dominant, with small corrections to the suppressed elastic couplings from higher-mode mixing.
The negative value of $u$ is compatible with bulk scalar stability, since $m_{\phi}^{2}=u(u+4k)\simeq-0.74k^{2} > -4k^{2}$ satisfies
the AdS$_{5}$ Breitenlohner-Freedman bound~\cite{Breitenlohner:1982bm}.
Related GW stabilization mechanisms with relevant scalar deformations have also been investigated in Refs.~\cite{Eroncel:2023uqf,Csaki:2023pwy}.
For our benchmark, the scalar backreaction remains small, $\kappa_{5}^{2}\phi_{0}^{2}\lesssim0.024$,
and the scalar-radion fluctuation spectrum is positive under the stiff-brane boundary conditions.
The background is therefore locally stable against linear scalar-metric perturbations within this approximation, while global stability and control of the 5D gravitational effective theory require further assessment.

\section{Derivation of the nuclear differential cross section}
\label{sec:cross-section-derivation}

We derive the differential cross section used in Eq.~\eqref{eq:dsigma} for the inelastic process
\begin{equation}
\chi_{1}(p_{{1}})+A(k) \to \chi_{2}(p_{{2}})+A(k').
\label{eq:appendix-scattering-process}
\end{equation}

For the nearly degenerate pseudo-Majorana pair, we neglect relative corrections of $\mathcal{O}(\delta/m_{\chi})$ in
the mass and flux normalization factors, identifying $m_{\chi_{1}},m_{\chi_{2}}\simeq m_{\chi}$ and
$\mu_{\chi_{1}A}\simeq\mu_{\chi A}$ where appropriate.
The splitting $\delta=m_{\chi_{2}}-m_{\chi_{1}}>0$ is retained in the inelastic kinematics.
We define the four-momentum transfer by
\begin{equation}
 q^{\mu}\equiv k'^{\mu}-k^{\mu}
 =p_{{1}}^{\mu}-p_{{2}}^{\mu},
 \qquad
 t\equiv q^{2},
 \label{eq:appendix-momentum-transfer}
\end{equation}
and denote the magnitude of its spatial component in the target rest frame by
\begin{equation}
 Q\equiv|\mathbf q|.
 \label{eq:appendix-Q-definition}
\end{equation}
For a target nucleus initially at rest,
\begin{equation}
 k^{\mu}=(m_{A},\mathbf 0),
 \qquad
 k'^{\mu}=(m_{A}+E_{R},\mathbf q),
\end{equation}
where $E_{R}$ is the kinetic energy of the recoiling nucleus. The on-shell condition $k'^{2}=m_{A}^{2}$ gives
\begin{equation}
 E_{R}
 =\sqrt{m_{A}^{2}+Q^{2}}-m_{A},
 \qquad
 Q^{2}=2m_{A}E_{R}+E_{R}^{2}.
 \label{eq:appendix-recoil-energy}
\end{equation}
The invariant momentum transfer therefore satisfies the relation
\begin{equation}
 t=(k'-k)^{2}=-2m_{A}E_{R}.
 \label{eq:appendix-t-recoil}
\end{equation}
In the nonrelativistic limit $E_{R}\ll m_{A}$,
\begin{equation}
 E_{R}\simeq\frac{Q^{2}}{2m_{A}},
 \qquad
 Q^{2}\simeq-t.
 \label{eq:appendix-Q-recoil}
\end{equation}
For a general $2\to2$ process, let $p_{i}^{*}$ and $p_{f}^{*}$ denote the magnitudes of the initial and final three-momenta in the center-of-mass frame, respectively. 
The scattering angle $\theta_{*}$ is defined between the incoming and outgoing DM momenta.
The recoil-energy Jacobian is
\begin{equation}
 \left|\frac{d\cos\theta_{*}}{dE_{R}}\right|
 =\frac{m_{A}}{p_{i}^{*}p_{f}^{*}}.
 \label{eq:appendix-ER-jacobian}
\end{equation}
For scalar exchange, the unpolarized cross section is independent of the azimuthal angle. 
The standard two-body phase-space expression therefore gives
\begin{equation}
 \frac{d\sigma_{12,A}}{dE_{R}}
 =
 \frac{m_{A}}{32\pi s(p_{i}^{*})^{2}}
 \overline{|\mathcal{M}_{A}|^{2}}.
 \label{eq:appendix-dsdER-general}
\end{equation}
In the nonrelativistic limit, with $\v_{\mathrm{rel}}\equiv \v_{\chi_{1}}-\v_{A}$,
we have $p_{i}^{*}\simeq\mu_{\chi_{1}A}v_{\mathrm{rel}}$
and $s(p_{i}^{*})^{2}\simeq m_{\chi_{1}}^{2}m_{A}^{2}v_{\mathrm{rel}}^{2}$.

For the scalar radion interaction in Eq.~\eqref{eq:nuclearamplitude}, we write
\begin{equation}
 \mathcal{M}_{A}
 =
 C_{A}(Q)(\bar u_{2}u_{1})(\bar u_{A}u_{A}),
 \qquad
 C_{A}(Q)
 =
 \frac{c_{12}^{S}m_{\chi}m_{N}f_{N}(t)}
 {\Lambda_{r}^{2}(m_{r}^{2}+Q^{2})}
 AF_{A}(Q).
 \label{eq:appendix-CA}
\end{equation}
The nonrelativistic spin-averaged bilinears satisfy
\begin{align}
\overline{|\bar u_{2}u_{1}|^{2}}
=
2(p_{1}\cdot p_{2}+m_{\chi_{1}}m_{\chi_{2}}) 
\simeq4m_{\chi_{1}}m_{\chi_{2}},
\qquad
\overline{|\bar u_{A}u_{A}|^{2}}
\simeq4m_{A}^{2}.
\end{align}
Consequently, 
\begin{equation}
\frac{d\sigma_{12,A}}{dE_{R}}
\simeq
\frac{m_{A}}{2\pi v_{\mathrm{rel}}^{2}}
|C_{A}(Q)|^{2},
\label{eq:appendix-dsdER-CA}
\end{equation}
Using $f_{N}(t)\simeq f_{N}(0)=1$ and the reference nucleon cross section in Eq.~\eqref{eq:sigma0},
we obtain
\begin{equation}
 \frac{d\sigma_{12,A}}{dE_{R}}
 =
 \frac{m_{A}A^{2}}
 {2\mu_{\chi N}^{2}v_{\mathrm{rel}}^{2}}
 \sigma_{N}^{0}F_{A}^{2}(Q)
 \left(
 \frac{m_{r}^{2}}{m_{r}^{2}+Q^{2}}
 \right)^{2}.
 \label{eq:appendix-dsdER-final}
\end{equation}
In the target rest frame, $v_{\mathrm{rel}}=v$, reproducing Eq.~\eqref{eq:dsigma}.

In addition, we note that the differential cross section~\eqref{eq:appendix-dsdER-final} applies only within the
kinematically allowed recoil range. At leading nonrelativistic order, energy conservation requires
\begin{equation}
\v_{\mathrm{rel}}\cdot\q
= \frac{Q^{2}}{2\mu_{\chi A}}+\delta,
\qquad
\mu_{\chi A} = \frac{m_{\chi}m_{A}}{m_{\chi}+m_{A}}.
\label{eq:appendix-energy-delta}
\end{equation}
This implies
\begin{equation}
v_{\mathrm{rel}} \geq \frac{Q}{2\mu_{\chi A}}+\frac{\delta}{Q}.
\label{eq:appendix-vmin-Q}
\end{equation}
Using $Q^{2}\simeq2m_{A}E_{R}$, we recover
\begin{equation}
v_{\min,A}(E_{R}) = \frac{1}{\sqrt{2m_{A}E_{R}}}
\left( \frac{m_{A}E_{R}}{\mu_{\chi A}}+\delta \right),
\label{eq:appendix-vmin}
\end{equation}
which is the endothermic support condition used in Section~\ref{sec:rate}.

\section{Matching to the LZ \texorpdfstring{$\mathcal{O}_{1}^{s}$}{O1s} normalization}
\label{sec:lz-normalization}

The supplementary LZ analysis~\cite{LZ:2026axp} gives a two-sided $90\%$ confidence
interval for the inelastic isoscalar operator $\mathcal{O}_{1}^{s}$.
Assuming scalar coupling, the LZ reference cross section can be written as\footnote{
LZ denotes the DM nucleon cross section by $\sigma_{\rm SI}^{N}$ in
its normalization equation, while the vertical axis of the supplementary exclusion plot is labeled by $\sigma_{\rm SI}$. 
The two denote the same per-nucleon cross section normalization for the isoscalar $\mathcal{O}_{1}^{s}$ interaction.}
\begin{equation}
 \sigma_{\rm SI}^{N}
 =\frac{\mu_{\chi N}^{2}}{\pi}\left|c_{1}^{s}\right|^{2},
 \label{eq:lz-si-definition}
\end{equation}
where $c_{1}^{s}$ is the contact coefficient with mass dimension $-2$.
The Higgs vacuum expectation value introduced in the original
normalization of Ref.~\cite{LZ:2026axp} cancels in
Eq.~\eqref{eq:lz-si-definition}. Thus $\sigma_{\rm SI}^{N}$ is a
reference cross section per nucleon, rather than a xenon cross section
or an event rate.

In the approximation $f_{p}(t)=f_{n}(t)\simeq1$ and
$m_{p}\simeq m_{n}$ used in Section~\ref{sec:matching}, the radion
mediated scalar transition is isoscalar. Its leading nonrelativistic matching gives
\begin{equation}
 c_{1,r}^{s}(Q)
 =\frac{c_{12}^{S}m_{\chi}m_{N}f_{N}(t)}
 {\Lambda_{r}^{2}(m_{r}^{2}+Q^{2})},
 \qquad Q^{2}\simeq-t.
 \label{eq:lz-radion-matching}
\end{equation}
At zero momentum transfer, $c_{1}^{s}=c_{1,r}^{s}(0)$ therefore gives
\begin{equation}
 \sigma_{\rm SI}^{N}
 =\frac{\mu_{\chi N}^{2}}{\pi}
 \left[
 \frac{c_{12}^{S}m_{\chi}m_{N}f_{N}(0)}
 {\Lambda_{r}^{2}m_{r}^{2}}
 \right]^{2}
 =\sigma_{N}^{0},
 \label{eq:lz-normalization-identity}
\end{equation}
which identifies the radion-mediated reference nucleon cross section with
the normalization constrained by the published LZ $\mathcal{O}_{1}^{s}$ interval.
The finite momentum recoil spectrum retains the radion propagator in Eq.~\eqref{eq:lz-radion-matching}.
We therefore impose the two-sided constraint in Eq.~\eqref{eq:lz-si-constraint} at the same
$m_{\chi}$ and $\delta$ as an additional consistency condition on our parameter space.
Unlike the auxiliary event-count condition $N_{\mathcal{B}}=1$, the published LZ interval
constrains the signal template over the extended analysis region, providing a broader experimental
consistency check subject to the adopted signal template assumptions.

\section{Collider probes}

\subsection{OPAL Higgsstrahlung constraint on the radion scale}
\label{app:opal-higgsstrahlung}

The OPAL Collaboration performed a decay mode independent search for a neutral scalar $S^{0}$ produced through the Higgsstrahlung process $e^{+}e^{-}\to ZS^{0}$~\cite{OPAL:2002ifx}.
For the narrow scalar interpretation, the production rate is parameterized by
\begin{equation}
k(m_{S^{0}})
\equiv
\frac{
\sigma(e^{+}e^{-}\to ZS^{0})
}{
\left.
\sigma_{\rm SM}(e^{+}e^{-}\to ZH)
\right|_{m_{H}=m_{S^{0}}}
},
\label{eq:opal_k_definition}
\end{equation}
and the corresponding limit is independent of the specific decay branching fractions of $S^{0}$.

For the radion interaction defined in Eqs.~\eqref{eq:rsm} and~\eqref{eq:smtrace}, the $ZZr$ vertex has the same
Lorentz structure as the SM $ZZH$ vertex, with the coupling ratio
\begin{equation}
\frac{g_{rZZ}}{g_{HZZ}^{\rm SM}}
=
\frac{v}{\Lambda_{r}},
\end{equation}
where $v\simeq246\,\mathrm{GeV}$.
Identifying $m_{S^{0}}=m_{r}$ therefore gives
\begin{equation}
k(m_{r})
=
\frac{v^{2}}{\Lambda_{r}^{2}}.
\label{eq:radion_opal_k}
\end{equation}
Let $k_{95}^{\rm obs}(m_{r})$ denote the observed 95\% CL upper limit from OPAL. The allowed radion parameter space satisfies
\begin{equation}
\frac{v^{2}}{\Lambda_{r}^{2}}
\leq
k_{95}^{\rm obs}(m_{r}),
\end{equation}
or equivalently
\begin{equation}
\Lambda_{r}^{-1}
\leq
\frac{\sqrt{k_{95}^{\rm obs}(m_{r})}}{v}
=
4.065\,
\sqrt{k_{95}^{\rm obs}(m_{r})}\,
\mathrm{TeV}^{-1}.
\label{eq:opal_radion_limit}
\end{equation}
The region above this curve is therefore excluded at 95\% CL in the $(m_{r},\Lambda_{r}^{-1})$ plane.

For the numerical implementation, we use the public HBDataSet representation of the decay mode independent OPAL limit~\cite{OPAL:2002ifx}, distributed with HiggsTools~\cite{Bahl:2022igd}.
At $m_{r}=30\,\mathrm{GeV}$, the observed limit $k_{95}^{\rm obs}=0.1887$ gives
\begin{equation}
\Lambda_{r}^{-1}\lesssim1.766\,\mathrm{TeV}^{-1}.
\end{equation}
The reference point with $\Lambda_{r}^{-1}\simeq0.441\,\mathrm{TeV}^{-1}$ therefore remains allowed.

The OPAL Collaboration also performed a flavour-independent Higgsstrahlung search for a scalar decaying into hadronic final states~\cite{OPAL:2003ujp}.
The observed 95\% CL upper limit from this hadronic search constrains
\begin{equation}
k_{\rm had}(m_{r})
\equiv
\frac{\sigma(e^{+}e^{-}\to Zr)\,\mathcal{B}(r\to\mathrm{hadrons})}
{\left.\sigma_{\rm SM}(e^{+}e^{-}\to ZH)\right|_{m_{H}=m_{r}}}
=
\frac{v^{2}}{\Lambda_{r}^{2}}\,\mathcal{B}_{\rm had}(m_{r}),
\label{eq:opal_hadron_k}
\end{equation}
where $\mathcal{B}_{\rm had}=\mathcal{B}_{b\bar b}+\mathcal{B}_{c\bar c}+\mathcal{B}_{s\bar s}+\mathcal{B}_{gg}$ in our retained decay channels.
For the on-shell radion widths in the range relevant to Fig.~\ref{fig:rad_limit_collider}, we use running quark masses and one-loop $\alpha_{s}(m_{r})$, with $\alpha_{s}(m_{Z})=0.1181$, $m_{b}(m_{b})=4.18\,\mathrm{GeV}$, $m_{c}(m_{c})=1.27\,\mathrm{GeV}$, and $m_{s}(2\,\mathrm{GeV})=0.093\,\mathrm{GeV}$.
The gluonic width contains the QCD trace anomaly and finite top-loop contribution~\cite{Bae:2000pk,Cheung:2000rw},
\begin{equation}
\Gamma_{gg}^{\rm LO}(m_{r})
=
\frac{\alpha_{s}^{2}(m_{r})m_{r}^{3}}{32\pi^{3}\Lambda_{r}^{2}}
\left|7+F_{t}(m_{r})\right|^{2},
\label{eq:opal_gluon_width_LO}
\end{equation}
with $F_{t}$ defined in Appendix~\ref{app:cms-diphoton}.
We apply a gluonic QCD correction to this width~\cite{Spira:1995rr},
\begin{equation}
\Gamma_{gg}^{K_{g}}(m_{r})=K_{g}(m_{r})\Gamma_{gg}^{\rm LO}(m_{r}),
\qquad
K_{g}(m_{r})=1+\frac{\alpha_{s}(m_{r})}{\pi}
\left(\frac{95}{4}-\frac{7n_{f}}{6}\right),\quad n_{f}=5.
\label{eq:opal_gluon_Kg}
\end{equation}
We retain the other partial widths at their stated leading-order accuracy and recompute the total width and all branching fractions after applying $K_{g}$.
The hadronic search thus gives
\begin{equation}
\Lambda_{r}^{-1}
\leq
\frac{1}{v}
\left[\frac{k_{95,\rm had}^{\rm obs}(m_{r})}{\mathcal{B}_{\rm had}(m_{r})}\right]^{1/2}.
\label{eq:opal_hadron_limit}
\end{equation}
For the numerical implementation, we digitize the observed exclusion curve published in Ref.~\cite{OPAL:2003ujp} and use its original sampled mass points in the interval $12\lesssim m_{r}\lesssim70\,\mathrm{GeV}$.
Near the reference mass, at the digitized point $m_{r}\simeq29.97\,\mathrm{GeV}$, we obtain $k_{95,\rm had}^{\rm obs}\simeq0.0208$ and $\mathcal{B}_{\rm had}\simeq0.963$, giving
\begin{equation}
\Lambda_{r}^{-1}\lesssim0.597\,\mathrm{TeV}^{-1}.
\label{eq:opal_hadron_reference_limit}
\end{equation}
This constraint is stronger than the decay mode independent bound at the same mass, while the reference point remains allowed.

\subsection{CMS \texorpdfstring{$\tau^{+}\tau^{-}$}{tau tau} constraint on the radion coupling scale}
\label{app:cms-tautau}

The CMS Collaboration performed a search for a spin-$0$ resonance $\phi$ produced through gluon fusion and decaying into
$\tau^{+}\tau^{-}$~\cite{CMS:2026mwx}.
The analysis uses $61.9\,\mathrm{fb}^{-1}$ of proton-proton collision data at $\sqrt{s}=13.6\,\mathrm{TeV}$ and covers the mass range $20\leq m_{\phi}\leq60\,\mathrm{GeV}$.
The observed 95\% CL upper limits are reported on
\begin{equation}
\sigma(pp\to\phi)\,
\mathcal{B}(\phi\to\tau^{+}\tau^{-}).
\label{eq:cms_sigma_limit}
\end{equation}
We recast these limits for the radion $r$, which is also dominantly produced through gluon fusion in the mass range of interest.

For the numerical implementation, we calculate the leading-order radion production cross section with \textsc{MadGraph5\_aMC@NLO}~\cite{Alwall:2014hca}
using the CT18NNLO parton distribution functions (PDFs)~\cite{Hou:2019efy}.
For the on-shell decays, we use the partial widths from the visible trace coupling, including the anomaly, top loop, and $K_{g}$ correction to $\Gamma_{gg}$ specified in Appendix~\ref{app:opal-higgsstrahlung}.
For $X=\tau^{+}\tau^{-}$ or $\gamma\gamma$, the resulting branching fraction and the reference gluon-fusion signal rate obey
\begin{equation}
\mathcal{B}_{X}^{K_{g}}(m_{r}) =\frac{\mathcal{B}_{X}^{\rm LO}(m_{r})}{F_{g}(m_{r})},
\qquad
\sigma_{X}^{\rm ref}(m_{r})=\sigma_{X}^{\rm ref,LO}(m_{r})\frac{K_{ggF}^{\rm prod}(m_{r})}{F_{g}(m_{r})},
\label{eq:cms_gluon_Kg_reweight}
\end{equation}
with the reweighting factor
\begin{equation}
F_{g}(m_{r}) \equiv1+[K_{g}(m_{r})-1]\mathcal{B}_{gg}^{\rm LO}(m_{r}).
\end{equation}
In the present numerical recast, we take $K_{ggF}^{\rm prod}(m_{r})=K_{g}(m_{r})$ as a phenomenological production correction.
Taking $\Lambda_{r}^{\rm ref}=1\,\mathrm{TeV}$, we define the corrected reference signal rate as
\begin{equation}
\sigma_{\tau\tau}^{\rm ref}(m_{r})
\equiv
\sigma^{\rm ref}(pp\to r)\,
\mathcal{B}(r\to\tau^{+}\tau^{-}).
\label{eq:cms_radion_reference}
\end{equation}
Since the radion coupling to the SM energy-momentum trace is
proportional to $\Lambda_{r}^{-1}$, its gluon fusion production cross
section scales as $\Lambda_{r}^{-2}$.  All SM partial widths carry the
same overall dependence on $\Lambda_{r}$ in the absence of
Higgs-radion mixing, so the branching fractions are independent of
this scale.  The predicted signal rate can therefore be rescaled as
\begin{equation}
\sigma_{\tau\tau}(m_{r},\Lambda_{r})
=
\sigma_{\tau\tau}^{\rm ref}(m_{r})
\left(
\frac{\Lambda_{r}^{\rm ref}}{\Lambda_{r}}
\right)^{2}.
\label{eq:cms_radion_scaling}
\end{equation}
Let $\sigma_{\tau\tau}^{95,\rm obs}(m_{r})$ denote the observed CMS upper limit in Eq.~\eqref{eq:cms_sigma_limit}.  
Defining the 95\% CL boundary by
\begin{equation}
\sigma_{\tau\tau}(m_{r},\Lambda_{r}^{95})
=
\sigma_{\tau\tau}^{95,\rm obs}(m_{r}),
\end{equation}
Eq.~\eqref{eq:cms_radion_scaling} gives
\begin{equation}
\Lambda_{r}^{95}(m_{r})
=
\Lambda_{r}^{\rm ref}
\left[
\frac{
\sigma_{\tau\tau}^{\rm ref}(m_{r})
}{
\sigma_{\tau\tau}^{95,\rm obs}(m_{r})
}
\right]^{1/2}.
\label{eq:cms_radion_limit}
\end{equation}
The allowed region therefore satisfies
\begin{equation}
\Lambda_{r}^{-1}
\leq
\left(\Lambda_{r}^{\rm ref}\right)^{-1}
\left[
\frac{
\sigma_{\tau\tau}^{95,\rm obs}(m_{r})
}{
\sigma_{\tau\tau}^{\rm ref}(m_{r})
}
\right]^{1/2}.
\label{eq:cms_radion_inverse_limit}
\end{equation}
The region above this curve is excluded at 95\% CL in the $(m_{r},\Lambda_{r}^{-1})$ plane.

For the numerical limits, we use the observed values provided in the public HEPData record associated with the CMS analysis.
The strongest observed constraint occurs at
$m_{r}=35\,\mathrm{GeV}$, where
\begin{equation}
\Lambda_{r}^{-1} < 1.087\,\mathrm{TeV}^{-1}.
\label{eq:cms_radion_strongest}
\end{equation}
At the reference mass $m_{r}=30\,\mathrm{GeV}$, the corresponding constraint is
\begin{equation}
\Lambda_{r}^{-1} < 1.413\,\mathrm{TeV}^{-1}.
\end{equation}
The reference point with $\Lambda_{r}^{-1}\simeq0.441\,\mathrm{TeV}^{-1}$ therefore remains allowed.

\subsection{CMS diphoton constraint on the radion coupling scale}
\label{app:cms-diphoton}

The CMS Collaboration performed a search for a narrow spin-$0$ resonance $\phi$ produced through gluon fusion and decaying into
a photon pair~\cite{CMS:2026zsp}.
The analysis uses $54.4\,\mathrm{fb}^{-1}$ of proton-proton collision data at $\sqrt{s}=13\,\mathrm{TeV}$ and covers the mass range
$10\leq m_{\phi}\leq70\,\mathrm{GeV}$.
The observed 95\% CL upper limits are reported on
\begin{equation}
\sigma(gg\to\phi)\, \mathcal{B}(\phi\to\gamma\gamma).
\label{eq:cms_gammagamma_sigma_limit}
\end{equation}
We recast these limits for the radion $r$, whose production in this mass range is dominated by gluon fusion.

For the numerical implementation, we calculate the leading-order radion production cross section with
\textsc{MadGraph5\_aMC@NLO}~\cite{Alwall:2014hca} using the CT18NNLO PDFs~\cite{Hou:2019efy},
and apply the same phenomenological gluon-fusion production factor $K_{ggF}^{\rm prod}=K_{g}$ as in Appendix~\ref{app:cms-tautau}.
The renormalization and factorization scales are fixed to $\mu_{R}=\mu_{F}=m_{r}$.
The $\gamma\gamma$ partial width is unchanged by this gluonic correction, while its branching fraction is recomputed with the $K_{g}$-corrected total radion width through Eq.~\eqref{eq:cms_gluon_Kg_reweight}.
For the trace coupling in Eq.~\eqref{eq:rsm}, the diphoton width receives contributions from the electromagnetic trace anomaly
together with the $W$ boson and top quark loops~\cite{Bae:2000pk,Cheung:2000rw},
\begin{equation}
\Gamma_{\gamma\gamma}(m_{r})
=
\frac{\alpha^{2}m_{r}^{3}}
{256\pi^{3}\Lambda_{r}^{2}}
\left[
-\frac{11}{3}
-F_{W}(m_{r})
+\frac{8}{3}F_{t}(m_{r})
\right]^{2},
\label{eq:radion_gammagamma_width}
\end{equation}
where
\begin{align}
F_{t}(m_{r})
&=
\frac{4m_{t}^{2}}{m_{r}^{2}}
\left[
1+
\left(
1-\frac{4m_{t}^{2}}{m_{r}^{2}}
\right)
\arcsin^{2}\!\left(
\frac{m_{r}}{2m_{t}}
\right)
\right],
\\
F_{W}(m_{r})
&=
2+
\frac{12m_{W}^{2}}{m_{r}^{2}}
+
\frac{12m_{W}^{2}}{m_{r}^{2}}
\left(
2-\frac{4m_{W}^{2}}{m_{r}^{2}}
\right)
\arcsin^{2}\!\left(
\frac{m_{r}}{2m_{W}}
\right).
\end{align}
Throughout the mass range considered here, $m_{r}<2m_{W}<2m_{t}$, so both loop functions are real. In the limit $m_{r}\ll 2m_{W},2m_{t}$, one has $F_{W}\to 7$ and $\frac{8}{3}F_{t}\to\frac{16}{9}$, yielding an amplitude coefficient of $-\frac{80}{9}$ corresponding to the effective electromagnetic trace anomaly of the light fields. $\mathcal{B}(r\to\gamma\gamma)$ is therefore insensitive to these loop functions.

Taking $\Lambda_{r}^{\rm ref}=1\,\mathrm{TeV}$, we define the reference diphoton signal rate as
\begin{equation}
\sigma_{\gamma\gamma}^{\rm ref}(m_{r})
\equiv
\sigma^{\rm ref}(gg\to r)\,
\mathcal{B}(r\to\gamma\gamma).
\label{eq:cms_gammagamma_reference}
\end{equation}
Since the gluon fusion cross section scales as $\Lambda_{r}^{-2}$ and all radion partial widths have the same overall
$\Lambda_{r}^{-2}$ dependence in the absence of Higgs-radion mixing, the branching fractions are independent of $\Lambda_{r}$.
The signal rate at an arbitrary radion scale is therefore
\begin{equation}
\sigma_{\gamma\gamma}(m_{r},\Lambda_{r})
=
\sigma_{\gamma\gamma}^{\rm ref}(m_{r})
\left(
\frac{\Lambda_{r}^{\rm ref}}{\Lambda_{r}}
\right)^{2}.
\label{eq:cms_gammagamma_scaling}
\end{equation}
Let $\sigma_{\gamma\gamma}^{95,\rm obs}(m_{r})$ denote the observed CMS upper limit in Eq.~\eqref{eq:cms_gammagamma_sigma_limit}.
The allowed radion parameter space consequently satisfies
\begin{equation}
\Lambda_{r}^{-1}
\leq
\left(\Lambda_{r}^{\rm ref}\right)^{-1}
\left[
\frac{
\sigma_{\gamma\gamma}^{95,\rm obs}(m_{r})
}{
\sigma_{\gamma\gamma}^{\rm ref}(m_{r})
}
\right]^{1/2}.
\label{eq:cms_gammagamma_radion_limit}
\end{equation}
The region above this curve is excluded at 95\% CL in the $(m_{r},\Lambda_{r}^{-1})$ plane.
For the numerical limits, we use the observed values provided in the public HEPData record associated with the CMS analysis.
The strongest radion constraint occurs at $m_{r}=43.4\,\mathrm{GeV}$, where
\begin{equation}
\sigma_{\gamma\gamma}^{95,\rm obs} = 0.08196\,\mathrm{pb},
\qquad
\sigma_{\gamma\gamma}^{\rm ref} = 0.19975\,\mathrm{pb},
\end{equation}
which gives
\begin{equation}
\Lambda_{r}^{-1} < 0.641\,\mathrm{TeV}^{-1}.
\label{eq:cms_gammagamma_strongest}
\end{equation}
Near the reference mass, the CMS point at
$m_{r}=30.3\,\mathrm{GeV}$ gives
\begin{equation}
\Lambda_{r}^{-1} < 0.740\,\mathrm{TeV}^{-1}.
\end{equation}
The reference point with $\Lambda_{r}^{-1}\simeq0.441\,\mathrm{TeV}^{-1}$ therefore remains allowed.

\subsection{Projected HL-LHC sensitivity to the radion diphoton signal}
\label{app:hllhc-diphoton}

To estimate the prospective HL-LHC sensitivity to the radion diphoton signal, we extrapolate the median expected 95\% CL upper limits from the CMS search discussed in Appendix~\ref{app:cms-diphoton}~\cite{CMS:2026zsp}.
We adopt the HL-LHC benchmark of $\sqrt{s}=14\,\mathrm{TeV}$ with an integrated luminosity of $3\,\mathrm{ab}^{-1}$~\cite{ATLAS:2025eii}.
The CMS median expected limits are obtained under the background-only hypothesis and provide the experimental baseline for our projection.

In the background-dominated statistical regime, the upper limit on the signal event count scales approximately with the square
root of the background count.
Assuming unchanged selected background cross sections and effective signal acceptance, the expected upper limit on the signal cross section therefore scales as $\mathcal{L}^{-1/2}$.
Taking $\mathcal{L}_{\rm CMS}=54.4\,\mathrm{fb}^{-1}$ and $\mathcal{L}_{\rm HL}=3000\,\mathrm{fb}^{-1}$, we obtain
\begin{equation}
\sigma_{\gamma\gamma,\rm HL}^{95,\rm exp}(m_{r})
=
\sigma_{\gamma\gamma,\rm CMS}^{95,\rm exp}(m_{r})
\left(
\frac{\mathcal{L}_{\rm CMS}}
{\mathcal{L}_{\rm HL}}
\right)^{1/2}.
\label{eq:hllhc_gammagamma_projection}
\end{equation}
In this simplified extrapolation, the selected background cross sections are held fixed when increasing
$\sqrt{s}$ from $13$ to $14\,\mathrm{TeV}$, while the radion production cross section is recalculated at $14\,\mathrm{TeV}$. 
Potential limitations from background modeling, photon identification, and other systematic uncertainties at high luminosity are not
explicitly modeled, so the projected sensitivity should be regarded as potentially optimistic.
For the radion prediction, we recalculate the leading-order gluon fusion production cross section at $\sqrt{s}=14\,\mathrm{TeV}$ using the same numerical setup as in Appendix~\ref{app:cms-diphoton}, and apply the same phenomenological production factor $K_{ggF}^{\rm prod}=K_{g}$.
The renormalization and factorization scales remain fixed to $\mu_{R}=\mu_{F}=m_{r}$, and the $K_{g}$-corrected radion branching fractions are the same at $13$ and $14\,\mathrm{TeV}$ for fixed $m_{r}$.
Taking $\Lambda_{r}^{\rm ref}=1\,\mathrm{TeV}$, we define
\begin{equation}
\sigma_{\gamma\gamma}^{\rm ref,14}(m_{r})
\equiv
\sigma_{14}^{\rm ref}(gg\to r)\,
\mathcal{B}(r\to\gamma\gamma).
\label{eq:hllhc_gammagamma_reference}
\end{equation}
The projected median expected 95\% CL sensitivity to the inverse radion scale is then
\begin{equation}
\left(\Lambda_{r}^{-1}\right)_{\rm HL}^{95,\rm exp}
=
\left(\Lambda_{r}^{\rm ref}\right)^{-1}
\left[\frac{\sigma_{\gamma\gamma,\rm HL}^{95,\rm exp}(m_{r})}{\sigma_{\gamma\gamma}^{\rm ref,14}(m_{r})}\right]^{1/2}.
\label{eq:hllhc_gammagamma_radion_limit}
\end{equation}

For the numerical implementation, we evaluate the radion production cross section at $14\,\mathrm{TeV}$ and the diphoton branching fraction from $m_{r}=10$ to $70\,\mathrm{GeV}$.
The strongest projected sensitivity occurs at $m_{r}=49.3\,\mathrm{GeV}$, where
\begin{equation}
\Lambda_{r}^{-1} \lesssim 0.321\,\mathrm{TeV}^{-1}.
\label{eq:hllhc_gammagamma_strongest}
\end{equation}
At $m_{r}=30.3\,\mathrm{GeV}$, we obtain
\begin{equation}
\sigma_{\gamma\gamma,\rm HL}^{95,\rm exp}
=
0.03261\,\mathrm{pb},
\qquad
\sigma_{\gamma\gamma}^{\rm ref,14}
=
0.23883\,\mathrm{pb},
\end{equation}
which gives the projected sensitivity
\begin{equation}
\Lambda_{r}^{-1} \lesssim 0.370\,\mathrm{TeV}^{-1}.
\label{eq:hllhc_gammagamma_reference_limit}
\end{equation}
This projected reach lies below the reference value $\Lambda_{r}^{-1}\simeq0.441\,\mathrm{TeV}^{-1}$ at
$m_{r}=30\,\mathrm{GeV}$, suggesting that the HL-LHC could probe this region of the radion parameter space under the stated assumptions.

\section{Solar capture and re-ejection}
\label{app:solar-thermal}

Endothermic scattering $\chi_{1}A\to\chi_{2}A$ can gravitationally capture halo DM in the excited state
$\chi_{2}$. We determine the initial orbits of these captured particles, including the thermal motion of solar
nuclei, and follow their subsequent evolution until the first exothermic down-scattering $\chi_{2}A\to\chi_{1}A$,
which can re-eject the particle or leave a bound $\chi_{1}$~\cite{DiMauro:2026dqp,Blennow:2018xwu}.

At solar radius $R$, a halo particle with asymptotic speed $u$ has speed $w$, where $w^{2}=u^{2}+v_{{\rm esc},\odot}^{2}(R)$. Solar nuclei follow the normalized local Maxwell-Boltzmann distribution $f_{A}(\v_{A};R)$ at temperature $T_{\odot}(R)$. Choosing the incident DM direction as the longitudinal axis, we write $v_{A,\parallel}=\sqrt{2}\,\sigma_{A}\xi_{H}$ and $v_{A,\perp}^{2}=2\sigma_{A}^{2}\xi_{L}$, where $\sigma_{A}^{2}=T_{\odot}(R)/m_{A}$, $-\infty<\xi_{H}<\infty$, and $0\leq\xi_{L}<\infty$. The nuclear thermal measure is then $\pi^{-1/2}e^{-\xi_{H}^{2}-\xi_{L}}d\xi_{H}d\xi_{L}$. The relative speeds before and after endothermic scattering satisfy
\begin{equation}
 v_{\rm rel}^{2}=(w-\sqrt{2}\,\sigma_{A}\xi_{H})^{2}+2\sigma_{A}^{2}\xi_{L},
 \qquad
 v_{\rm rel}'=\sqrt{v_{\rm rel}^{2}-\frac{2\delta}{\mu_{\chi A}}}.
 \label{eq:solar-thermal-compact-kinematics}
\end{equation}
The allowed thermal domain $\mathcal{D}_{A}(R,u)$ consists of $(\xi_{H},\xi_{L})$ for which $v_{\rm rel}^{2}>2\delta/\mu_{\chi A}$.
With $\theta_{*}$ the angle between incoming and outgoing relative velocities in the two-body center of mass frame, the capture rate is
\begin{align}
C_{\odot}^{(1)}
&=
\frac{\rho_{\chi}}{m_{\chi}}
\sum_{A}
\int_{0}^{R_{\odot}}
4\pi R^{2}dR\,n_{A}(R)
\int_{0}^{u_{\rm halo}^{\max}}
du\,
\frac{f_{\odot}(u)}{u}\,w
\frac{1}{\sqrt{\pi}}
\int_{\mathcal{D}_{A}(R,u)}
d\xi_{H}\,d\xi_{L}\,
e^{-\xi_{H}^{2}-\xi_{L}}\,
v_{\rm rel}
\nonumber\\
&\quad\times
\int_{-1}^{1}
d\cos\theta_{*}\,
P_{A}^{\rm cap}(\theta_{*};R,u,\xi_{H},\xi_{L})
\frac{d\sigma_{12,A}}{d\cos\theta_{*}},
\label{eq:solar_cap_1}
\end{align}
where $P_{A}^{\rm cap}$ is the fraction of scattering azimuths for which the outgoing $\chi_{2}$ is gravitationally bound immediately after the initial endothermic scattering. 
For the azimuthally symmetric scalar cross section, the fraction is
\begin{equation}
P_{A}^{\rm cap}(\theta_{*};R,u,\xi_{H},\xi_{L})
=
\begin{cases}
1, & \kappa_{A}\geq 1,\\[2mm]
1-\dfrac{\arccos\kappa_{A}}{\pi}, & -1<\kappa_{A}<1,\\[3mm]
0, & \kappa_{A}\leq -1,
\end{cases}
\label{eq:solar-thermal-pcap}
\end{equation}
where
\begin{equation}
\begin{aligned}
\kappa_{A}
&=\frac{v_{{\rm esc},\odot}^{2}(R)-v_{\rm cm}^{2}-b_{A}^{2}
-2b_{A}v_{{\rm cm},\parallel}\cos\theta_{*}}
{2b_{A}v_{{\rm cm},\perp}\sin\theta_{*}},
\qquad
b_{A}=\beta_{A}v_{\rm rel}',
\\
\beta_{A}&=\frac{m_{A}}{m_{\chi}+m_{A}},
\qquad
v_{{\rm cm},\parallel}=w\cos\alpha_{A}-\beta_{A}v_{\rm rel},
\qquad
v_{{\rm cm},\perp}=w\sin\alpha_{A},
\\
\cos\alpha_{A}
&=\frac{w-\sqrt{2}\,\sigma_{A}\xi_{H}}{v_{\rm rel}},
\qquad
v_{\rm cm}^{2}=v_{{\rm cm},\parallel}^{2}+v_{{\rm cm},\perp}^{2}.
\end{aligned}
\label{eq:solar-thermal-compact-kappa}
\end{equation}
When $v_{{\rm cm},\perp}\sin\theta_{*}=0$, the outgoing DM speed is independent of the scattering azimuth; $P_{A}^{\rm cap}$ is then unity or zero according to whether its outgoing speed is below or above $v_{{\rm esc},\odot}(R)$.
The radion-mediated Born interaction in Eq.~\eqref{eq:dsigma} gives the differential cross section for the endothermic transition $\chi_{1}A\to\chi_{2}A$,
\begin{equation}
\frac{d\sigma_{12,A}}{d\cos\theta_{*}}
=
\frac{\sigma_{N}^{0}A^{2}\mu_{\chi A}^{2}}{2\mu_{\chi N}^{2}}
\frac{v_{\rm rel}'}{v_{\rm rel}}
F_{A}^{2}(Q)
\left(\frac{m_{r}^{2}}{m_{r}^{2}+Q^{2}}\right)^{2},
\label{eq:solar-thermal-dsigma}
\end{equation}
where $Q^{2}=\mu_{\chi A}^{2}[v_{\rm rel}^{2}+(v_{\rm rel}')^{2}-2v_{\rm rel}v_{\rm rel}'\cos\theta_{*}]$.
The endothermic threshold is imposed through $\mathcal{D}_{A}$, while $P_{A}^{\rm cap}$ selects the $\chi_{2}$ particles that are gravitationally bound immediately after the initial scattering.
In the stationary-nucleus limit $T_{\odot}(R)\to0$, the capture rate reduces to $C_{\odot}^{(0)}$ in Eq.~\eqref{eq:solar_cap_0}.
Although thermal nuclear motion changes the total capture rate by only $0.9\%$, we retain its effects
on the initial orbital distribution of captured $\chi_{2}$ particles when calculating their subsequent evolution and possible re-ejection.

To initialize the orbital calculation, let $\mathcal{Z}$ denote the capture radius, nuclear species, halo speed, thermal nuclear velocity, and outgoing scattering direction.
The normalized differential capture rate defines the probability distribution of the initial orbital states,
\begin{equation}
dP_{\rm cap}(\mathcal{Z})
=
\frac{dC_{\odot}^{(1)}(\mathcal{Z})}{C_{\odot}^{(1)}},
\qquad
\int_{\rm bound}dP_{\rm cap}(\mathcal{Z})=1.
\label{eq:solar-native-initial-distribution}
\end{equation}
We sample the initial orbits according to this distribution, restricting the scattering azimuth to the range satisfying the gravitational binding condition.
The incoming halo direction relative to the local solar radius is taken to be isotropic.
For each captured $\chi_{2}$, we follow its gravitationally bound orbit in the solar potential $\Phi_{\odot}(R)$.
The orbital trajectory determines the residence time at each solar radius, and hence the accumulated nuclear
scattering probability, including excursions outside the Sun where scattering vanishes.

For the reverse exothermic transition $\chi_{2}A\to\chi_{1}A$, energy conservation gives
\begin{equation}
v_{\rm rel}'
=
\sqrt{v_{\rm rel}^{2}+\frac{2\delta}{\mu_{\chi A}}}.
\label{eq:solar-signed-transition}
\end{equation}
The corresponding radion-mediated Born cross section is
\begin{equation}
\frac{d\sigma_{21,A}}{d\cos\theta_{*}}
=
\frac{\sigma_{N}^{0}A^{2}\mu_{\chi A}^{2}}{2\mu_{\chi N}^{2}}
\frac{v_{\rm rel}'}{v_{\rm rel}}
F_{A}^{2}(Q)
\left(\frac{m_{r}^{2}}{m_{r}^{2}+Q^{2}}\right)^{2},
\label{eq:solar-downscattering-dsigma}
\end{equation}
where $Q^{2}=\mu_{\chi A}^{2}[v_{\rm rel}^{2}+(v_{\rm rel}')^{2}-2v_{\rm rel}v_{\rm rel}'\cos\theta_{*}]$ is evaluated using the exothermic outgoing speed.
For a captured $\chi_{2}$ with velocity $\v_{\chi}$,
the local rate of exothermic down-scattering $\chi_{2}A\to\chi_{1}A$ is
\begin{equation}
\gamma_{\chi_{2}\to\chi_{1}}(R,\v_{\chi})
=
\sum_{A}n_{A}(R)
\int d^{3} v_{A}\,f_{A}(\v_{A};R)\,v_{\rm rel}
\int_{-1}^{1}d\cos\theta_{*}\,
\frac{d\sigma_{21,A}}{d\cos\theta_{*}}.
\label{eq:solar-native-collision-rate}
\end{equation}
Along its orbit, the accumulated scattering optical depth and no-collision probability are
\begin{equation}
\mathcal{T}(t_{1},t_{2})
=
\int_{t_{1}}^{t_{2}}dt\,
\gamma_{\chi_{2}\to\chi_{1}}[R(t),\v_{\chi}(t)],
\qquad
P_{0}(t_{1},t_{2})
=
e^{-\mathcal{T}(t_{1},t_{2})}.
\label{eq:solar-native-optical-depth}
\end{equation}
The orbital Monte Carlo samples the first down-scattering time from this nonhomogeneous Poisson process.
For each exothermic down-scattering, we reconstruct the outgoing $\chi_{1}$ velocity $\v_{\chi}'$
from two-body kinematics, sampling the scattering direction according to Eq.~\eqref{eq:solar-downscattering-dsigma}.
The resulting specific orbital energy is
\begin{equation}
\varepsilon_{1}'
=
\frac{1}{2}
\left[
|\v_{\chi}'|^{2}
-v_{{\rm esc},\odot}^{2}(R)
\right].
\label{eq:solar-postcollision-orbit}
\end{equation}
The particle is re-ejected if $\varepsilon_{1}'\geq0$ and remains bound otherwise.
For each simulated initial capture $k$, we record the first down-scattering time $\tau_{21,k}$ and the
specific orbital energy $\varepsilon_{1,k}'$ of the outgoing $\chi_{1}$. 
The cumulative probabilities are estimated by
\begin{equation}
\begin{aligned}
F_{\rm b}(t)
&=
\frac{1}{N_{\rm MC}}
\sum_{k=1}^{N_{\rm MC}}
\Theta(-\varepsilon_{1,k}')
\Theta(t-\tau_{21,k}),
\\
P_{\text{re-ej}}(t)
&=
\frac{1}{N_{\rm MC}}
\sum_{k=1}^{N_{\rm MC}}
\Theta(\varepsilon_{1,k}')
\Theta(t-\tau_{21,k}).
\end{aligned}
\label{eq:solar-retention-average}
\end{equation}
The two fractions sum to unity once all initially captured particles have undergone their first
down-scattering. 
For an approximately constant initial capture rate,
the effective source of retained $\chi_{1}$ particles is
\begin{equation}
C_{\rm core}(t) = C_{\odot}^{(1)}F_{\rm b}(t),
\label{eq:solar-native-delayed-source}
\end{equation}
which incorporates both re-ejection and the delay before the first bound conversion.

\section{Thermal-core population and annihilation}
\label{app:solar-thermalization}

The orbital calculation in Appendix~\ref{app:solar-thermal} determines the retained $\chi_{1}$ source $C_{\rm core}(t)$ after the first exothermic down-scattering and possible re-ejection. We assume that these particles subsequently thermalize according to Eq.~\eqref{eq:solar-core-reference}, with the core populations $N_{1,2}(t)$ and excited fraction $f_{2}(t)$ defined in Eq.~\eqref{eq:solar-populations}. Neglecting corrections of relative order $\delta/m_{\chi}$, the two states share the normalized spatial profile
\begin{equation}
n_{i}(R,t)=N_{i}(t)p_{\chi}(R),
\qquad
\int d^{3}R\,p_{\chi}(R)=1,
\qquad i=1,2.
\label{eq:solar-core-profile-definition}
\end{equation}
For an isothermal population in the solar gravitational potential $\Phi_{\odot}(R)$,
\begin{equation}
p_{\chi}(R)\propto
\exp\left[-\frac{m_{\chi}[\Phi_{\odot}(R)-\Phi_{\odot}(0)]}{T_{\odot,c}}\right].
\label{eq:solar-thermal-profile}
\end{equation}
Approximating the solar core by a constant density $\rho_{c}=150\,\mathrm{g/cm}^{3}$ gives
\begin{equation}
r_{\chi}^{2}=\frac{3T_{\odot,c}}{2\pi G\rho_{c}m_{\chi}},
\label{eq:solar-core-radius}
\end{equation}
and the effective volume
\begin{equation}
V_{\rm eff}
=\left[\int d^{3}R\,p_{\chi}^{2}(R)\right]^{-1}
=(2\pi)^{3/2}r_{\chi}^{3}
\simeq2.07\times10^{26}\,\mathrm{cm}^{3}.
\label{eq:solar-effective-volume}
\end{equation}

The equilibrium excited-state fraction in this thermal core is
\begin{equation}
f_{2}^{\rm th}
\equiv\left.\frac{N_{2}}{N_{\odot}}\right|_{\rm th}
=\left[1+\left(\frac{m_{\chi_{1}}}{m_{\chi_{2}}}\right)^{3/2}
e^{\delta/T_{\odot,c}}\right]^{-1}
\simeq\left(\frac{m_{\chi_{2}}}{m_{\chi_{1}}}\right)^{3/2}
e^{-\delta/T_{\odot,c}},
\label{eq:solar-thermal-fraction}
\end{equation}
which gives $f_{2}^{\rm th}\simeq2.6\times10^{-114}$ at the reference point. To determine how residual excited particles relax toward this equilibrium, we average the local $\chi_{2}\to\chi_{1}$ down-scattering rate in Eq.~\eqref{eq:solar-native-collision-rate} over the thermal-core distribution,
\begin{equation}
\Gamma_{\chi_{2}\to\chi_{1}}
=\int d^{3}R \, p_{\chi}(R)
\int d^{3} v_{\chi}\,
f_{\chi}(\v_{\chi};T_{\odot,c})\,
\gamma_{\chi_{2}\to\chi_{1}}(R,\v_{\chi}),
\label{eq:solar-down-scattering-rate}
\end{equation}
where $f_{\chi}$ is the normalized Maxwell-Boltzmann velocity distribution~\cite{Blennow:2018xwu,DiMauro:2026dqp}. 
Using the BS05(AGS,OP) solar profile~\cite{Bahcall:2004pz}, AGS05 abundances~\cite{Asplund:2004eu}, and the nuclear response adopted in Section~\ref{sec:solar_icecube}, we obtain the conversion rate and the relaxation time
\begin{equation}
\Gamma_{\chi_{2}\to\chi_{1}}
\simeq8.6\times10^{-5}\,\mathrm{s}^{-1},
\qquad
\tau_{\rm conv} \simeq3.2\,\mathrm{hours}.
\label{eq:solar-down-scattering-reference}
\end{equation}
Since the retained particles enter the thermal core as $\chi_{1}$ and thermal excitation is strongly
suppressed for $\delta\gg T_{\odot,c}$, we adopt $f_{2}\simeq0$ in the annihilation calculation.

The annihilation of the retained population is therefore dominated by $\chi_{1}\chi_{1}$ initial states.
Nevertheless, the heavier $\chi_{2}\chi_{2}$ channel contributes virtually through the coupled-channel potential
$V_{S}(r)$ in Eq.~\eqref{eq:solar-sommerfeld-potential}.
The leading annihilation is $p$-wave, corresponding to orbital angular momentum $\ell=1$ of the incoming DM pair
and the Born behavior $\sigma v_{\rm rel}\propto v_{\rm rel}^{2}$.
We therefore introduce the two-component reduced radial wave function
\begin{equation}
\Psi(r)=
\begin{bmatrix}
\psi_{11}(r)\\
\psi_{22}(r)
\end{bmatrix},
\label{eq:solar-sommerfeld-wavefunction}
\end{equation}
which satisfies the coupled radial Schr\"odinger equation
\begin{equation}
\left[
-\frac{1}{m_{\chi}}\frac{d^{2}}{dr^{2}}
+\frac{2}{m_{\chi}r^{2}}
+V_{S}(r)
\right]\Psi(r)
=
\frac{m_{\chi}v_{\rm rel}^{2}}{4}\Psi(r).
\label{eq:solar-sommerfeld-radial}
\end{equation}
The centrifugal term corresponds to $\ell=1$. 
Regularity requires $\psi_{11}(r),\psi_{22}(r)=O(r^{2})$ near the origin.
We normalize the incoming $\chi_{1}\chi_{1}$ wave and impose exponential decay of the $\chi_{2}\chi_{2}$
component when $v_{\rm rel}<\sqrt{8\delta/m_{\chi}}$.
The short-distance $p$-wave annihilation amplitudes are determined by the coefficients of the regular
radial solutions near the origin.
With $k=m_{\chi}v_{\rm rel}/2$, we define
\begin{equation}
d_{i}(v_{\rm rel})
=
\frac{3}{k^{2}}
\lim_{r\to0}\frac{\psi_{ii}(r)}{r^{2}},
\qquad i=1,2,
\label{eq:solar-sommerfeld-amplitudes}
\end{equation}
where the normalization follows from the free
$p$-wave solution
$krj_{1}(kr)\simeq k^{2}r^{2}/3$.
Thus $(d_{1},d_{2})=(1,0)$ for a free incoming
$\chi_{1}\chi_{1}$ state.
Following the coupled-channel Sommerfeld formalism for partial-wave annihilation~\cite{Hryczuk:2011vi,Beneke:2014gja},
the annihilation amplitude of an incoming $\chi_{1}\chi_{1}$ pair is the coherent sum of
the short-distance amplitudes weighted by the corresponding $p$-wave derivative factors,
\begin{equation}
\mathcal{M}_{\Psi\to rr}
=
d_{1}\mathcal{M}_{11\to rr}^{\rm Born}
+
d_{2}\mathcal{M}_{22\to rr}^{\rm Born}.
\label{eq:solar-sommerfeld-amplitude}
\end{equation}
At leading order in $\delta/m_{\chi}$, the two Born amplitudes are equal in the adopted channel
convention. The annihilation cross section therefore becomes
\begin{equation}
\sigma_{11\to rr}v_{\rm rel}
=
b_{rr}v_{\rm rel}^{2}|d_{1}+d_{2}|^{2},
\qquad
S_{1}(v_{\rm rel})=|d_{1}+d_{2}|^{2}.
\label{eq:solar-sommerfeld-annihilation}
\end{equation}
For the Maxwell-Boltzmann distribution of relative
velocities, the thermally averaged Sommerfeld factor is
\begin{equation}
\overline{S}_{1}
=
\frac{1}{12\sqrt{\pi}}
\left(\frac{m_{\chi}}{T_{\odot,c}}\right)^{5/2}
\int_{0}^{\infty}dv_{\rm rel}\,
v_{\rm rel}^{4}
e^{-m_{\chi}v_{\rm rel}^{2}/(4T_{\odot,c})}
S_{1}(v_{\rm rel}).
\label{eq:solar-sommerfeld-average}
\end{equation}
At the reference point $m_{r}=30\,\mathrm{GeV}$, we obtain $\sqrt{\langle v_{\rm rel}^{2}\rangle}\simeq8.83\times10^{-5}$ and $\overline{S}_{1}\simeq71.0$.

\begin{acknowledgments}
R.L. and H.-H.Z. are supported by the National Natural Science Foundation
of China under Grant No.~12275367, the Fundamental Research Funds for the
Central Universities, and the Sun Yat-Sen University Science Foundation.
K.X. is supported by the National Natural Science Foundation
of China and by the State Key Laboratory of Dark Matter Physics.
K.M. is supported by the National Science Foundation under Grant No. PHY-2310497.

\end{acknowledgments}

\bibliographystyle{apsrev4-2-inspire}
\bibliography{references}

\end{document}